\documentclass[
  aps, prb,
  reprint,
  amsmath,
  floatfix,
  ]{revtex4-2}

\usepackage[until=2027-01-01]{blocklatexml}
\usepackage{papertypesetting}

\makeatletter
  \AddToHook{env/table/begin}{%
    \skip@\abovecaptionskip
    \abovecaptionskip\belowcaptionskip
    \belowcaptionskip\skip@
  }
\makeatother

\makeatletter
  \newcommand*\@withperiod[1]{#1.}
  \patchcmd\paragraph{\normalfont\normalsize\itshape}{\normalfont\normalsize\itshape\@withperiod}{}{}
\makeatother

\makeatletter
  \def\bibsection{%
    \expandafter\section\expandafter*\expandafter{\refname}%
    \@nobreaktrue
    \raggedright
  }%
\makeatother

\usepackage{newfloat}
\DeclareFloatingEnvironment{spacerfloat}

\usepackage{figutil}

\makeatletter
  \newenvironment{paperpart}{%
    \clearpage
    \clearpage
  }{%
    \clearpage
    \clearpage
  }

  \newenvironment{mainsetup}{%
    \paperpart
    \graphicspath{{./img/}}%
    \let\addtocontents\@gobbletwo
    \patchcmd\frontmatter@setup{\normalfont}{\normalfont\sffamily}{}{}
    \patchcmd\frontmatter@title@format{\bfseries}{\sffamily\bfseries}{}{}
    \patchcmd\frontmatter@affiliationfont{\it}{\slshape}
    \patchcmd\frontmatter@title@format{\large}{\Large}{}{}
    \patchcmd\section{\bfseries}{\sffamily\bfseries}{}{}
    \patchcmd\subsection{\bfseries}{\sffamily\bfseries}{}{}
    \patchcmd\subsubsection{\itshape}{\sffamily}{}{}
    \patchcmd\@makecaption{\rmfamily}{\sffamily}{}{}
    \pretocmd\@make@capt@title{\begingroup\bfseries}{}{}
    \patchcmd\@make@capt@title{\@caption@fignum@sep}{\@caption@fignum@sep\endgroup}{}{}
    \def\@caption@fignum@sep{: }

    \advance\textheight 5pt\relax
    \def\paper@beforeskip{.7cm \@plus 2ex \@minus .5ex}
    \def\paper@afterskip{.4cm \@plus 1ex \@minus .5ex}
    \patchcmd\section{0.8cm \@plus1ex \@minus .2ex}{\paper@beforeskip}{}{\error}
    \patchcmd\subsection{.8cm \@plus1ex \@minus .2ex}{\paper@beforeskip}{}{\error}
    \patchcmd\subsubsection{.8cm \@plus1ex \@minus .2ex}{\paper@beforeskip}{}{\error}
    \patchcmd\section{0.5cm}{\paper@afterskip}{}{\error}
    \patchcmd\subsection{.5cm}{\paper@afterskip}{}{\error}
    \patchcmd\subsubsection{.5cm}{\paper@afterskip}{}{\error}
    \dbltextfloatsep=16pt plus 6pt minus 4pt\relax
    \dbltextfloatsep=16pt plus 6pt minus 4pt\relax
    \AddToHook{env/table/begin}{\advance\abovecaptionskip 1.8pt\relax}
    \AtBeginDocument{%
      \abovedisplayshortskip=5pt plus 2pt minus 5pt\relax
    }
    \skip\footins=26pt plus 4pt minus 4pt
  }{%
    \onecolumngrid
    \endpaperpart
  }

  \def\@firstoffour#1#2#3#4{#1}
  \let\suppl@lbibitem\@lbibitem
  \patchcmd\suppl@lbibitem
    {\global\advance\c@NAT@ctr\@ne}
    {
      \@ifundefined{b@#2}
        {\global\c@NAT@ctr=-999\relax}
        {\global\c@NAT@ctr=\expandafter\expandafter\expandafter\@firstoffour\csname b@#2\endcsname\relax}%
      \edef\NAT@num{\the\c@NAT@ctr}%
    }
    {}{\error}
  \let\suppl@NAT@wrout\NAT@wrout
  \patchcmd\suppl@NAT@wrout
    {#5}
    {#5\@extra@b@citeb}
    {}{\error}
  \newenvironment{supplsetup}{%
    \paperpart
    \graphicspath{{./suppl_img/}}%
    \appto\maketitle{\onecolumngrid}%
    \let\@oddhead\@empty
    \def\@oddfoot{\hfil\thepage\hfil}%
    \AddToHook{__hyp/target/setname}{%
      \ifHy@localanchorname
        \expandafter\preto
      \else
        \expandafter\gpreto
      \fi
        \@currentHref{suppl.}%
    }%
    \def\@extra@b@citeb{:suppl}%
    \let\@lbibitem\suppl@lbibitem
    \let\NAT@wrout\suppl@NAT@wrout
    \setcounter{page}{1}%
    \setcounter{section}{0}%
    \setcounter{figure}{0}%
    \setcounter{table}{0}%
    \setcounter{equation}{0}%
    \pdfstringdefDisableCommands{%
      \let\,\space
    }%
    \preto\thepage{S\,}%
    \preto\thesection{S\,}%

    \def\theequation@prefix{S}
    \def\refname{Supplemental References}
    \patchcmd\frontmatter@setup{\normalfont}{\normalfont\sffamily}{}{}
    \patchcmd\frontmatter@title@format{\bfseries}{\sffamily\bfseries}{}{}
    \patchcmd\frontmatter@affiliationfont{\it}{\slshape}
    \patchcmd\frontmatter@title@format{\large}{\Large}{}{}
    \patchcmd\section{\small}{}{}{}
    \patchcmd\section{\bfseries}{\sffamily\bfseries}{}{}
    \patchcmd\section{\centering}{\raggedright}{}{}
    \patchcmd\subsection{\small}{}{}{}
    \patchcmd\subsection{\bfseries}{\sffamily\bfseries}{}{}
    \patchcmd\subsection{\centering}{\raggedright}{}{}
    \patchcmd\subsubsection{\small}{}{}{}
    \patchcmd\subsubsection{\itshape}{\sffamily}{}{}
    \patchcmd\subsubsection{\centering}{\raggedright}{}{}
    \patchcmd\@startsection{\@afterindenttrue}{\@afterindentfalse}{}{}
    \patchcmd\@hangfrom@section{\MakeTextUppercase{##3}}{##3}{}{}
    \patchcmd\@hangfroms@section{\MakeTextUppercase{##3}}{##3}{}{}
    \pretocmd\@seccntformat{\expandafter\ifx\csname the##1\endcsname\thesection Supplemental Note \fi}{}{}
    \patchcmd\@makecaption{\rmfamily}{\sffamily}{}{}
    \pretocmd\@make@capt@title{\begingroup\bfseries}{}{}
    \patchcmd\@make@capt@title{\@caption@fignum@sep}{\@caption@fignum@sep\endgroup}{}{}
    \def\@caption@fignum@sep{: }
    \def\paper@beforeskip{.7cm \@plus 2ex \@minus .5ex}
    \def\paper@afterskip{.4cm \@plus 1ex \@minus .5ex}
    \patchcmd\section{0.8cm \@plus1ex \@minus .2ex}{\paper@beforeskip}{}{\error}
    \patchcmd\subsection{.8cm \@plus1ex \@minus .2ex}{\paper@beforeskip}{}{\error}
    \patchcmd\subsubsection{.8cm \@plus1ex \@minus .2ex}{\paper@beforeskip}{}{\error}
    \patchcmd\section{0.5cm}{\paper@afterskip}{}{\error}
    \patchcmd\subsection{.5cm}{\paper@afterskip}{}{\error}
    \patchcmd\subsubsection{.5cm}{\paper@afterskip}{}{\error}
    \AtBeginDocument{%
      \abovedisplayshortskip=5pt plus 2pt minus 5pt\relax
    }
  }{%
    \endpaperpart
  }
\makeatother

\begin{document}

\begin{mainsetup}
\title{Magnetically induced Josephson nano-diodes in\\field-resilient superconducting microwave circuits}

\def\pitaffil{%
  \affiliation{Physikalisches Institut, Center for Quantum Science~(CQ) and \lisaplus, Universität Tübingen, 72076~Tübingen, Germany}
}
\author{Benedikt~Wilde}
\email{benedikt.wilde@uni-tuebingen.de}
\pitaffil
\author{Mohamad~Kazouini}
\pitaffil
\author{Timo~Kern}
\pitaffil
\author{Kevin~Uhl}
\pitaffil
\author{Christoph~Füger}
\pitaffil
\author{Dieter~Koelle}
\pitaffil
\author{Reinhold~Kleiner}
\pitaffil
\author{Daniel~Bothner}
\email{daniel.bothner@uni-tuebingen.de}
\pitaffil

\begin{abstract}
  The development of nonlinear and frequency-tunable superconducting microwave circuits for operation in large magnetic fields is of high relevance for hybrid quantum systems such as spin resonance spectrometers, microwave quantum magnonics, dark matter axion detectors or flux-mediated optomechanics.
  With these exciting perspectives in mind, we investigate niobium-based circuits with integrated nano-constriction quantum interferometers in magnetic in-plane fields of up to several hundred~\unit{\milli\tesla}.
  Our experiments reveal an unexpected and pronounced field-induced asymmetry in the bias-flux response of the circuits, which is demonstrated to originate from a field-induced Josephson-diode effect within the nano-constrictions and which enhances the circuit figures of merit in a magnetic field.
  An intuitive macroscopic Josephson-diode model attributes the effect to inhomogeneous constriction properties and provides us with the diode current-phase relation as a function of the in-plane field.
  Finally, we demonstrate that in the diode-state the circuit Kerr nonlinearity is bimodal in frequency, not only eliminating alternative explanations for the bias-flux-asymmetries but also being potentially useful for quantum circuit applications.
  Overall, our report underlines the potential of niobium nano-constriction circuits for high-field hybrid quantum systems, provides a conceptually simple explanation for the diode effect in superconducting nano-structures, and reveals the untapped potential of combining Josephson nano-diodes with microwave quantum circuits.
\end{abstract}

\makeatletter
  \appto\titleblock@produce{\addvspace{-6\p@}}
\makeatother

\maketitle

\section{Introduction}
Superconducting microwave circuits with integrated nonlinear elements such as Josephson junctions or high-kinetic-inductance nanowires are the workhorse of superconducting and hybrid quantum technologies~\cite{xiang2013, clerk2020, blais2021}.
Lately, there has been growing demand for such circuits being compatible with large magnetic fields for use in field-resilient circuit quantum electrodynamics~\cite{samkharadze2016, kroll2018, kringhoj2021, krause2022}, field-compatible parametric amplifiers~\cite{xu2023, zapata2024, frasca2024}, dispersive magnetometry~\cite{levensonfalk2016}, and hybrid systems with e.g.~micromechanical oscillators~\cite{rodrigues2019, zoepfl2020, schmidt2024}, spin ensembles~\cite{schuster2010, zollitsch2015, ghirri2015}, magnonic oscillators~\cite{huebl2013, li2019, song2025} or topological quantum bits~\cite{nayak2008, hassler2010}.
Some of the envisioned high-field experiments might reveal groundbreaking new insights and technologies regarding quantum gravity, axion dark matter, quantum sensing or topologically protected qubits.
Since the standard material aluminum and the commonly used trilayer superconducting tunnel junctions cannot easily be combined with high magnetic fields~\cite{schneider2019, krause2022}, other circuit materials and junction technologies are currently under intense investigation~\cite{samkharadze2016, kroll2018, kringhoj2021, ghirri2015, luthi2018, kennedy2019, borisov2020, uhl2023}, however, with no technology proven to be superior or dominant yet.

A promising platform towards high-field Josephson microwave devices are niobium (alloy) circuits, which have been demonstrated to possess high coherence in fields up to the tesla regime~\cite{degraaf2012, kwon2018, kroll2019, xu2023}, integrated with nano-constrictions as nonlinear elements, either as single nanowires~\cite{tholen2009, burnett2017, xu2023, uhl2024a} or as superconducting quantum interference devices~(SQUIDs)~\cite{kennedy2019, uhl2024}.
Various implementations of niobium nano-constrictions have been realized in the past~\cite{jamet2001, hasselbach2002, troeman2007, chen2016, kennedy2019, uhl2024, uhl2024a, weber2025}, both in direct-current~(dc) and in microwave platforms, and using both two-dimensional~(\nD 2) and three-dimensional~(\nD 3) constriction types.
Surprisingly, only few experiments, and all in~dc, focused on their high-field characteristics~\cite{chen2010, romans2011, lam2011}, and hence little is known about their high-field microwave properties.
Recently, it has been observed that niobium constrictions can develop a considerable dc superconducting diode~(SD) effect in an external field~\cite{margineda2023}, a nonreciprocal regime with an asymmetry between forward and reverse critical currents that originates from a simultaneous breaking of space-inversion and time-reversal symmetries.
After the early realizations of SDs several decades ago~\cite{zapata1996, krasnov1997, weiss2000, sterck2002, villegas2003, sterck2005}, they have experienced an impressive renaissance during the past few years~\cite{ando2020, wu2022, davydova2022, nadeem2023}, since they promise fascinating possibilities for superconducting electronics, spintronics and quantum technologies \cite{nadeem2023, inglaaynes2025}.
However, reports of integrating intrinsic SDs into microwave circuits are scarce to date~\cite{baumgartner2022}, and many aspects of their sub-critical, inductive properties, their current-phase relations as well as their potential for application in quantum circuits beyond three-wave-mixing~\cite{zorin2016, frattini2017} remain elusive.

\begin{figure*}
  \includegraphics{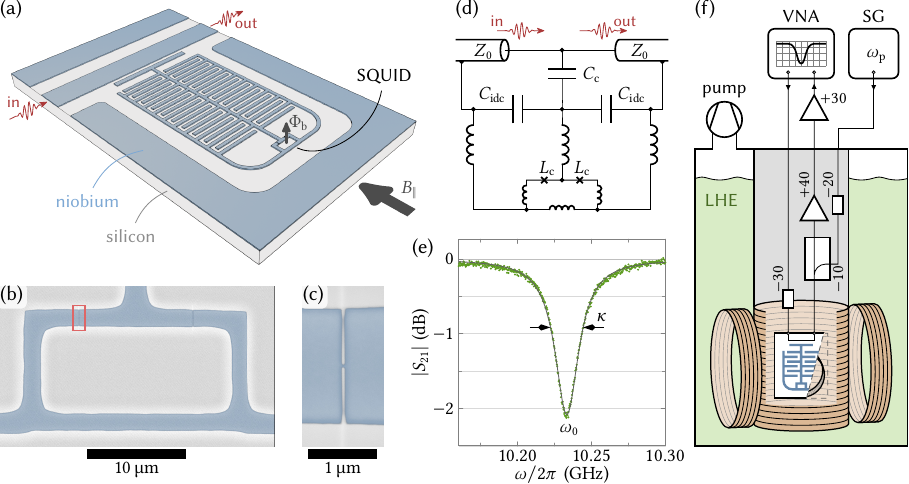}
  \titlecaption{Niobium nano-constriction SQUID circuits for microwave hybrid systems in large magnetic fields}{%
    \sublabel{a}~Rendered schematic of a typical microwave SQUID circuit as used in this work.
    The circuit comprises interdigital capacitors~(IDCs) connected with linear inductors and is patterned from a \qty{100}{\nano\meter}~thick layer of niobium~(blue) on top of an intrinsic silicon substrate~(gray).
    A rectangular-loop SQUID based on ion-beam patterned nano-constrictions is integrated at the bottom of the circuit.
    The circuit is capacitively side-coupled to a coplanar waveguide~(CPW) feedline for driving and readout by means of the scattering matrix element~$S_{21} = S_{\mathrm{out}} / S_{\mathrm{in}}$.
    The SQUID can be flux-biased with a perpendicular flux~$\Phi_{\mathrm b}$ and, independently, a large magnetic in-plane field~$B_\parallel$ can be applied.
    \sublabel{b}~False-color scanning electron microscopy~(SEM) image of a typical SQUID.
    The position of one of the two symmetrically placed nano-constrictions is indicated by a red rectangle.
    \sublabel{c}~Enlarged false-color SEM image of the indicated junction.
    \sublabel{d}~Equivalent circuit of the device, featuring the coupling capacitance~$C_{\mathrm c}$ to the CPW~feedline with characteristic impedance~$Z_0$, the IDC~capacitances~$C_{\mathrm{idc}}$, and the constriction inductances~$L_{\mathrm c}$ (symbolized by crosses).
    The individual linear inductances are not labeled in this diagram for the sake of simplicity and since we will mainly use the effective composite inductance~$L_{\mathrm b}$ (total circuit inductance for $L_{\mathrm c} = 0$).
    For the contributions to the total SQUID loop inductance~$L_\mathrm{loop}$, cf.~Fig.~\ref{fig:flux-response}.
    \sublabel{e}~CPW~transmission~$S_{21}$ vs~probe frequency~$\omega$ for $B_\parallel = 0$, $\Phi_{\mathrm b} = 0$.
    Dots are data, the line is a fit.
    From the fit, we obtain the resonance frequency $\omega_0 = 2 \pi \times \qty{10.233}{\giga\hertz}$ and the linewidth $\kappa = 2 \pi \times \qty{22}{\mega\hertz}$.
    \sublabel{f}~Simplified schematic of the experimental setup.
    The superconducting chip is enclosed in the vacuum compartment of a liquid helium~(LHE) cryostat.
    The microwave input lines are attenuated by \qty{30}{\deci\bel} each to equilibrate the noise of the vector network analyzer~(VNA) and the signal generator~(SG) as closely as possible to the measurement temperature~$T_{\mathrm s} = \qty{2.8}{\kelvin}$ (achieved by pumping on the helium gas) and the output line is equipped with two high-electron-mobility-transistor~(HEMT) amplifiers for maximized signal-to-noise ratio.
    A cylindrical superconducting coil is wrapped directly around the cup of the vacuum compartment for the application of magnetic in-plane fields~$B_\parallel$, and an additional split-coil magnet for compensating out-of-plane components is rigidly attached to the cryostat.
    A small magnetic sample coil is directly attached to the chip housing for the application of~$\Phi_{\mathrm b}$.
    Part of the chip is clipped away in the diagram to reveal the sample coil mounted behind it.
  }
  \label{fig:concept}
  \vspace{-2pt}
\end{figure*}

Here, we investigate the operation of superconducting niobium microwave circuits with integrated nano-constriction quantum interferometers in magnetic in-plane fields.
To guarantee minimized coupling of current-source noise into the SQUIDs, we implement a setup combining a single-current-source \nD 2-vector-magnet with in-situ sample rotation, and characterize the microwave circuits in fields of up to~\qty{300}{\milli\tesla}.
The properties of the circuits turn out to be promising for applications in large magnetic fields, and we observe that the circuit flux responsivity -- one of the most important figures of merit for many applications -- is enhanced by the external field.
At the same time, we find that the characteristic flux-tuning arcs of the microwave SQUID devices become considerably asymmetric with increasing in-plane field, an intriguing effect which points to the constrictions gradually turning into field-induced Josephson diodes~(JDs).
From our data and a macroscopic theoretical JD model, we finally reconstruct the current-phase-relations~(CPRs) of the JD-constrictions and identify spatial inhomogeneities as the likely origin of the diode effect.
Our results illuminate the magnetic-field compatibility of niobium nano-constriction SQUID circuits, and reveal insights into the sub-critical characteristics of field-induced constriction-diodes, detected through their microwave inductance.
The observations presented in this manuscript are of interest for high-field hybrid quantum systems involving superconducting microwave electronics, for focused-ion-beam-patterned nano-devices, for the design and understanding of Josephson diodes, and for diode-enhanced quantum interference circuits.

\section{Results}
\subsection{Devices and setup}\label{sec:concept}
The base layout of our circuits is a superconducting lumped element LC~circuit, which comprises two interdigital capacitors~$C_{\mathrm{idc}}$ with multiple linear inductances that can formally be combined into a single inductor~$L_{\mathrm b}$ (without constriction contributions), cf.~Fig.~\ref{fig:concept}.
The superconducting lines have a width of~\qty{2}{\micro\meter} and the gaps between two capacitor fingers are~\qty{4}{\micro\meter} wide.
At the heart of the circuit is a rectangular loop with inner loop dimensions \qtyproduct{19 x 8}{\micro\meter} and a loop self-inductance $L_{\mathrm{loop}} \approx \qty{44}{\pico\henry}$ including kinetic inductance, cf.~Supplemental Note~\ref{sec:suppl-resonator-params}~\cite{suppl-note}.
By means of a capacitor~$C_{\mathrm c}$, the circuits are side-coupled to a coplanar waveguide~(CPW) feedline with characteristic impedance $Z_0 \approx \qty{50}{\ohm}$ for driving and readout.
Without constrictions, the resonance frequency of the circuits is given by $\omega_{0 \mathrm b} = 1 / \sqrt{L_{\mathrm b} C_{\mathrm{tot}}}$, where~$C_{\mathrm{tot}} = 2 C_{\mathrm{idc}} + C_{\mathrm c}$.

We combine six such circuits along a single microwave CPW feedline on a \qtyproduct{10 x 10}{\milli\meter} microchip for simultaneous characterization and readout.
For a picture of the complete layout and details regarding the individual circuit components and their values, cf.~Supplemental Note~\ref{sec:suppl-chip-layout}~\cite{suppl-note}.
The individual circuit layouts differ solely by their number~$N_{\mathrm{idc}}$ of capacitor fingers, which primarily leads to distinct capacitances, but also to slightly different~$L_{\mathrm b}$, since the capacitor fingers contribute to the total inductance.
The spacing of two neighboring resonance frequencies is roughly~\qty{700}{\mega\hertz}, and the six devices in total cover the frequency range from \qty{8.9}{\giga\hertz} to \qty{12.5}{\giga\hertz}.
The superconducting structures on the chip are patterned from a \qty{\sim 100}{\nano\meter} thick niobium film, deposited on an intrinsic high-resistivity silicon wafer by dc-magnetron sputtering.
All niobium patterning of the main circuits is performed using maskless optical lithography and \ce{SF6} reactive ion etching.
A detailed description of the complete fabrication recipe can be found in Appendix~\ref{sec:meth-device-fabrication}.

After a spectroscopic pre-characterization in liquid helium, three of the six circuits are patterned with a pair of monolithic nano-constrictions, which are placed symmetrically into their SQUID loops~(cf.~Fig.~\ref{fig:concept}); the other three circuits are operated as reference devices.
Patterning of the constrictions is done via local high-precision milling using a focused neon-ion-beam~(\ce{Ne}-FIB), analogous to what is described in Ref.~\cite{uhl2024}.
Two of the three SQUIDs have \nD 3~constrictions (where the constrictions were milled down from the top) and one of them has \nD 2~variants (where the constrictions have the full film height).
In this manuscript, we will focus on the \nD 2~device; we provide analogous results for one of the \nD 3~constriction circuits as well as a comparative discussion in Supplemental Note~\ref{sec:suppl-res4-data}~\cite{suppl-note}.
The constrictions have a length and width of $l \approx w \approx \qty{40}{\nano\meter}$.
After \ce{Ne}-FIB patterning, the chip is mounted and wirebonded to a microwave printed circuit board~(PCB) and enclosed in a radiation-tight copper housing, before it is placed in the vacuum compartment of a liquid helium cryostat.

The experimental setup combines multiple coaxial microwave input/output lines, various dc~twisted-pair copper wires as well as a temperature-control unit consisting of a temperature-diode and a resistive heater for feedback-controlled temperature stability better than~\qty{1}{\milli\kelvin}, cf.~Fig.~\subref{fig:concept}{f} for a simplified schematic and Supplemental Note~\ref{sec:suppl-setup}~\cite{suppl-note} for the complete one.
For magnetic shielding, the entire cryostat is surrounded by a double-layer mu-metal shield at room-temperature.
Two equally attenuated coaxial input lines are used to send a vector network analyzer~(VNA) probe tone and an additional high-power microwave pump tone to the device, respectively, and the output line is equipped with a cryogenic high-electron-mobility-transistor~(HEMT) amplifier as well as a room-temperature HEMT~amplifier.
We insert the pump tone into the device at the probe signal output in order to protect the amplifiers from being saturated by the pump signal.
The chip with a small sample-coil attached to its copper housing for applying SQUID-bias flux~$\Phi_{\mathrm b}$ and the temperature-control components are located in an evacuated cylindrical cup, around which a high-inductance superconducting coil is wrapped for the application of an in-plane magnetic field with $B_\parallel / I_{\mathrm{coil}} = \qty{250}{\milli\tesla\per\ampere}$, with the coil current~$I_{\mathrm{coil}}$.
Thus, the superconducting coil can be immersed directly in liquid helium, while the sample is located in vacuum with variable temperature, maintaining a rigid relative position and orientation between the two.
At the bottom of the cryostat, around the position of the cylindrical sample cup, a second superconducting electromagnet, a split-coil, is installed for the application of an out-of-plane magnetic compensation field $B_\perp / I_{\mathrm{coil}} = \qty{2}{\milli\tesla\per\ampere}$.
Finally, it is possible to pump on the helium compartment of the cryostat to reach sample temperatures~$T_{\mathrm s}$ down to $T_{\mathrm s}^{\mathrm{min}} \sim \qty{2}{\kelvin}$.
Throughout this work, we operate at a temperature~$T_{\mathrm s} = \qty{2.8}{\kelvin}$.

\subsection{Circuit characteristics without in-plane field}\label{sec:flux-response}

\begin{figure*}
  \adjincludegraphics[trim={{\width-\textwidth} 0pt 0pt 0pt}]{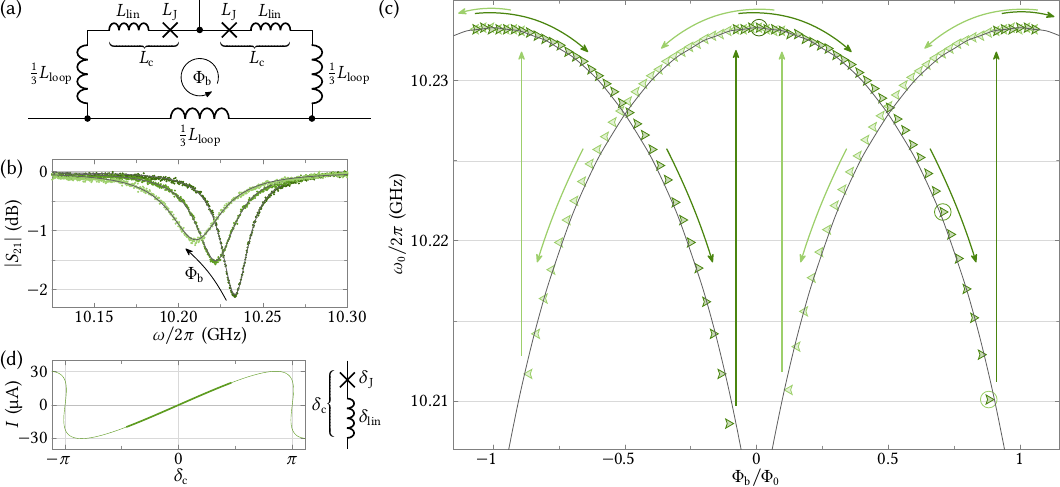}
  \titlecaption{Hysteretic flux response and constriction current-phase relation without magnetic in-plane field}{%
    \sublabel{a}~Circuit schematic of the SQUID.
    The total loop inductance is split into three equal contributions, each carrying~$L_{\mathrm{loop}} / 3$.
    The total constriction inductance~$L_{\mathrm c}$ is described by a series combination of a linear inductance~$L_{\mathrm{lin}}$ and an ideal Josephson inductance~$L_{\mathrm J}$.
    The Josephson part~$L_{\mathrm J}$ can be tuned by applying a bias flux~$\Phi_{\mathrm b}$ through the SQUID.
    \sublabel{b}~Transmission~$S_{21}$ through the circuit for three different bias fluxes $\Phi_{\mathrm b} / \Phi_0 \in \numset{0; 0.71; 0.88}$.
    With increasing $\Phi_{\mathrm b}$ (arrow), the resonance shifts to lower frequencies.
    From fits~(lines) to the data~(dots), we obtain~$\omega_0$ as a function of the bias flux.
    The shift of~$\omega_0$ is due to an increase of~$L_{\mathrm J}$ with~$\Phi_{\mathrm b}$.
    \sublabel{c}~Resonance frequency~$\omega_0$ vs~$\Phi_{\mathrm b} / \Phi_0$ as obtained from the $S_{21}$ fit curves for a larger range of flux values; the presented dataset combines a flux up-sweep (dark kites pointing right) and a flux down-sweep (light kites pointing left), arrows indicate the sweep direction.
    The three encircled data points on the central arc correspond to the resonances shown in panel~\subref*{@}*{b}.
    The resonance frequency modulates with a period of~$\Phi_0$, and the modulation is hysteretic with discontinuous jumps from one flux arc to the neighboring ones at $\Phi_{\mathrm b} / \Phi_0 \sim (n \pm \num{0.9}) \, \Phi_0$ where $n \in \mathbb Z$.
    Jumps are always upwards in frequency and indicated by long vertical arrows.
    Lines are fit curves, from which we obtain the linear inductance $L_{\mathrm{lin}} = \qty{12}{\pico\henry}$ and the sweetspot Josephson inductance $L_{\mathrm J 0} = \qty{11}{\pico\henry}$.
    \sublabel{d}~Inferred current-phase relation of the constriction.
    The total phase~$\delta_{\mathrm c}$ is the sum of a linear phase~$\delta_{\mathrm{lin}}$ and a Josephson phase~$\delta_{\mathrm J}$.
    The critical currents are identical in both current directions and given by $I_0^\pm = \pm \qty{30}{\micro\ampere}$.
    The thick line segment for $\abs{\delta_{\mathrm c}} \leq \num{1.45}$ shows the part of the CPR that corresponds to the experimentally accessible flux arc range between the jumps; the switching currents are~$I_{\mathrm{sw}}^\pm \approx \pm \qty{20}{\micro\ampere}$.
  }
  \label{fig:flux-response}
\end{figure*}

The first experiment we perform is the microwave characterization of the SQUID circuit as a function of bias-flux~$\Phi_{\mathrm b}$ through the SQUID, with the large magnetic coils switched off $B_\parallel = B_\perp = 0$.
This will allow us to quantitatively model the constriction-type Josephson junctions~(cJJs) and to extract their characteristic parameters as a basis for the later experiments and analyses.
To this end, we trace the complex frequency response of the circuit around its resonance frequency by means of the transmission S-parameter $S_{21} = S_{\mathrm{out}} / S_{\mathrm{in}}$ using a~VNA.
We operate in the linear response regime of the circuit, which we ensure by measuring its microwave input-power dependence and then staying around one order of magnitude below the probe powers required for the onset of nonlinearity.
At the sweetspot $\Phi_{\mathrm b} = 0$, the circuit has a resonance frequency $\omega_0 = 2 \pi \times \qty{10.233}{\giga\hertz}$, a total linewidth $\kappa = 2 \pi \times \qty{22}{\mega\hertz}$ and an external linewidth $\kappa_{\mathrm{ext}} = 2 \pi \times \qty{4.7}{\mega\hertz}$, cf.~Fig.~\subref{fig:concept}{e}.

Once $\Phi_{\mathrm b} \neq 0$, we observe a shift of the resonance frequency to lower values and an increase of the internal decay rate, cf.~Fig.~\ref{fig:flux-response}.
On a wider flux scale, the resonance frequency modulates periodically with~$\Phi_{\mathrm b}$, and the period is given by the magnetic flux quantum $\Phi_0 = \qty{2.068e-15}{\tesla\square\meter}$.
The origin of the frequency-shift with bias flux is an increase of the nonlinear constriction inductance~$L_{\mathrm c}$ due to the flux in the SQUID and the resulting dc~current through the cJJs.
We find not only an arc-shaped periodic modulation but also observe a hysteretic flux response of~$\omega_0$; the frequencies obtained during the flux up-sweep partly differ from the ones in the down-sweep and the resonance frequency does not transition smoothly from arc to arc, but rather in discontinuous jumps, as indicated by the vertical arrows in Fig.~\subref{fig:flux-response}{c}.
Adjacent arcs overlap considerably and cross at $\Phi_{\mathrm b} = (n + 1 / 2) \, \Phi_0$ with $n \in \mathbb Z$.
Such flux hysteresis originates from metastable SQUID states that occur when the effective loop inductance is not negligible~\cite{pogorzalek2017, uhl2024}.
The jumps between the arcs occur when the screening current in the SQUID reaches a switching current~$\pm I_{\mathrm{sw}}$ and as a consequence the fluxoid number changes by~$\pm 1$.
Note that $I_{\mathrm{sw}}$ is often considerably smaller than the junction critical current~$I_0$, since the switching can be triggered prematurely by thermal noise, flux instabilities, phase slips or the microwave signals used to probe the resonator.

A practical way to quantitatively model the inductance of each constriction as a function of~$\Phi_{\mathrm b}$ is to separate~$L_{\mathrm c}$ into a linear inductance~$L_{\mathrm{lin}}$ and a flux-dependent Josephson inductance $L_{\mathrm J} = L_{\mathrm J 0} / \cos\delta_{\mathrm J}$~\cite{likharev1979, uhl2024a}.
Here, $L_{\mathrm J 0} = \Phi_0 / 2 \pi I_0$ is the sweetspot Josephson inductance, $I_0$ is the critical constriction current and $\delta_{\mathrm J}$ is the Josephson phase.
Splitting the total cJJ inductance into a linear and a Josephson contribution is equivalent to splitting the total constriction phase drop~$\delta_{\mathrm c}$ as
\begin{equation}
  \delta_{\mathrm c} = \delta_{\mathrm J} + \delta_{\mathrm{lin}}
\end{equation}
where $\delta_{\mathrm{lin}} = 2 \pi L_{\mathrm{lin}} I / \Phi_0$ is the phase drop across the linear inductance.
Using this total phase and the first Josephson relation $I = I_0 \sin\delta_{\mathrm J}$, with the constriction current~$I$, the cJJ current-phase relation can be expressed as the implicit function~\cite{likharev1979}
\begin{equation}\label{eq:cpr-sinlin}
  I(\delta_{\mathrm c}) = I_0 \sin\Bigl( \delta_{\mathrm c} - 2 \pi \, \frac{L_{\mathrm{lin}} I(\delta_{\mathrm c})}{\Phi_0} \Bigr) .
\end{equation}
The total flux~$\Phi$ in the SQUID (without constriction contributions) on the other hand is related to~$\delta_{\mathrm c}$ via $\Phi / \Phi_0 = \delta_{\mathrm c} / \pi$ and to the bias flux via
\begin{equation}\label{eq:flux-relation}
  \Phi = \Phi_{\mathrm b} - L_{\mathrm{loop}} I(\pi \Phi / \Phi_0) .
\end{equation}
Here, we chose the signs such that a small positive bias flux~$\Phi_{\mathrm b}$ leads to a positive ring current~$I$ through the junctions, and assumed the two constrictions in the SQUID to be identical.

The numerical solution of Eqs.~\eqref{eq:cpr-sinlin} and~\eqref{eq:flux-relation} provides us with~$\delta_{\mathrm c}$ as a function of~$\Phi_{\mathrm b}$, which in turn allows us to calculate the constriction inductance for a given~$\Phi_{\mathrm b}$ according to
\begin{equation}\label{eq:constriction-inductance}
  L_{\mathrm c}(\delta_{\mathrm c}) = \frac{\Phi_0}{2 \pi} \Bigl( \diff{I}{\delta_{\mathrm c}} \Bigr)^{-1}
\end{equation}
and to finally fit the flux-tuning of the resonance frequency using
\begin{equation}\label{eq:squidres-frequency}
  \omega_0(\Phi_{\mathrm b}) = \frac{\omega_{0 \mathrm b}}{\sqrt{1 + L_{\mathrm c}(\Phi_{\mathrm b}) / 2 L_{\mathrm b}}} .
\end{equation}
Here, $\omega_{0 \mathrm b} = 2 \pi \times \qty{10.380}{\giga\hertz}$ is the fixed circuit resonance frequency before the cJJ patterning and we use the sweetspot resonance frequency~$\omega_{00} = \omega_0(0)$ to fix the sweetspot constriction inductance $L_{\mathrm c 0}$, while $I_0$ is a free fit parameter.
For the flux response tuning curves shown in Fig.~\subref{fig:flux-response}{c} we obtain $L_{\mathrm c 0} = \qty{23}{\pico\henry}$ and $I_0 = \qty{30}{\micro\ampere}$, which corresponds to $L_{\mathrm{lin}} = L_{\mathrm c 0} - L_{\mathrm J 0} = \qty{12}{\pico\henry}$.

The corresponding constriction CPR is shown in Fig.~\subref{fig:flux-response}{d} and is typical for niobium constrictions.
Instead of a sinusoidal shape, as expected for an ideal Josephson junction, the CPR is considerably forward-skewed which reflects the presence of~$L_{\mathrm{lin}}$.
In a small interval around $\delta_{\mathrm c} = \pi$ the CPR even becomes multi-valued, which indicates that $L_{\mathrm{lin}} / L_{\mathrm c 0} > 1 / 2$.
From considering the part of the CPR that is experimentally accessible before the jump to the next arc occurs, indicated by a thick line in Fig.~\subref{fig:flux-response}{d}, we conclude that the switching is indeed premature and occurs at $\delta_{\mathrm c} \sim \pi / 2$ and $I_{\mathrm{sw}} \approx \qty{20}{\micro\ampere}$.
This is not unusual for the type of constriction we use here~\cite{uhl2024, uhl2024a}, although the exact mechanism behind this discrepancy is not yet fully understood.

In summary, both the microwave circuit and the constrictions are well-behaved and quantitatively modeled at this point and it is time to switch on~$B_\parallel$.

\subsection{Impact of the magnetic in-plane field}\label{sec:field-response}
When investigating in-plane-field operation of superconducting circuits, the magnetic field needs to be aligned with the circuit plane as parallel as possible, in particular when large SQUIDs are involved, but also to avoid Abrikosov vortices in the leads.
One way to achieve parallel alignment would be to operate the vector magnet with two independent current sources and to adjust the split-coil current such that~$B_\perp$ exactly cancels the out-of-plane component of the main-coil field~$B_\parallel$.
The out-of-plane component and the necessary compensation factor can be determined using the SQUID itself.
This standard method, however, leads to the two current sources coupling uncorrelated noise into the SQUID, which is very noise-sensitive due to its large area.
Additionally, independent drifts of the two current sources would pose a severe challenge for effective field alignment and long-term stability.
In order to eliminate the source-related field noise and avoid compensation drifts, we instead operate the two magnets in series, driven by a single current source.
Naturally, the out-of-plane components of the two fields do not coincidentally cancel then from the get-go, and an additional measure has to be taken: rotation of the entire sample assembly, including the in-plane magnet, in the field of the fixed split-coil to an optimal compensation-angle where the out-of-plane component of the total field vanishes.
We estimate the residual misalignment angle between the sample plane and the magnetic field to be better than~\ang{\pm 0.02}.
A detailed description of the alignment procedure is given in Appendix~\ref{sec:meth-field-alignment}.

\begin{figure*}
  \includegraphics{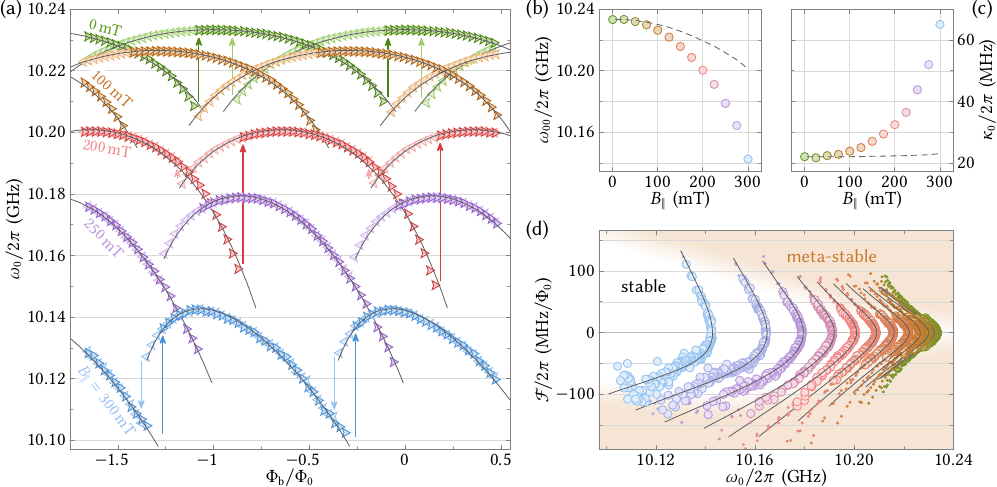}
  \vskip -.4\abovecaptionskip
  \titlecaption{Skewed flux-response and enhanced flux-responsivity in large magnetic in-plane fields}{%
    \sublabel{a}~Bias flux response~$\omega_0(\Phi_{\mathrm b})$ of the circuit for five different magnetic in-plane fields between \qty{0}{\milli\tesla} and~\qty{300}{\milli\tesla}; labels next to the datasets denote~$B_\parallel$.
    Data combine flux up-sweep (dark kites pointing right) and flux down-sweep (light kites pointing left); lines are fits, cf.~main text.
    With increasing $B_\parallel$, sweetspot resonance frequency and flux hysteresis decrease and a tilt of the arcs emerges, an asymmetry with respect to the field-shifting sweetspot flux.
    The arc skewing impacts the discontinuous transitions from one arc to the next, indicated by vertical arrows for $B_\parallel \in \qtyset{0; 200; 300}{\milli\tesla}$.
    For $B_\parallel = 0$, these jumps from the metastable to the stable branch always occur upwards in frequency and are symmetric around the arc crossing points~(ACPs).
    At~\qty{200}{\milli\tesla}, the jumps occurring in the up-sweep are larger compared to the $B_\parallel = 0$ case, which correlates with an increase of the total tuning range $\omega_{0, \mathrm{max}} - \omega_{0, \mathrm{min}}$.
    The jumps in the down-sweep, on the other hand, are significantly diminished in size.
    Still, both jumps are upwards in frequency.
    For the two largest fields, the ACPs disappear and the jumps in the down-sweep are downwards in frequency, meaning that the metastable state has a higher frequency than the stable one.
    \sublabel{b},~\sublabel{c}~Sweetspot resonance frequency~$\omega_{00}$ and linewidth~$\kappa_0$ as a function of~$B_\parallel$.
    The in-plane field reduces the energy gap in the \ce{Nb}~film and the cJJs, increasing the kinetic inductance and the number of thermal quasiparticles, resulting in $\omega_{00}$~decreasing and $\kappa_0$~increasing, respectively.
    Symbols are data extracted from the flux arcs, colors denote the value of~$B_\parallel$, as in all panels.
    The dashed lines show the expected behavior in the absence of the constrictions as inferred from reference circuits.
    Hence, about two thirds of the $\omega_{00}$~decrease and almost all of the $\kappa_0$~increase can be attributed to the constrictions.
    \sublabel{d}~Flux responsivity $\mathcal F = \idell{\omega_0}{\Phi_{\mathrm b}}$ for various~$B_\parallel$ and plotted vs~$\omega_0$ as a figure of merit for sensing and parametric coupling applications.
    Large symbols correspond to values on stable branches, small symbols to those on metastable branches~\cite{metastable-transition-note}, lines are derivatives of the arc fits.
    The stable operation regime is considerably and asymmetrically magnified by~$B_\parallel$.
  }
  \label{fig:field-response}
  \vspace{-2.6pt}
\end{figure*}

With the field aligned, we apply increasing coil currents in steps of $\Delta I_{\mathrm{coil}} = \qty{100}{\milli\ampere}$, which corresponds to field steps $\Delta B_\parallel = \qty{25}{\milli\tesla}$, and for each in-plane field characterize the circuit, up to a maximum field of~\qty{300}{\milli\tesla}.
We implement a field-cooling procedure to further minimize potential noise factors and flux instabilities that can occur with critical states in zero-field-cooling situations, in particular when the alignment is not completely perfect~\cite{brandt1993, ghigo2007,  nulens2023}.
To this end, we set~$I_{\mathrm{coil}}$ to its desired value, then heat the sample in the vacuum compartment to $T \sim \qty{12}{\kelvin}$ for a few seconds, and finally let it thermalize to $T_{\mathrm s} = \qty{2.8}{\kelvin}$ again.
We begin the characterization by sweeping~$\Phi_{\mathrm b}$ using the sample coil and measure $S_{21}$ traces with the VNA for each SQUID flux as described above.
The results are summarized in Fig.~\ref{fig:field-response}.

First, we find that the sweetspot resonance frequency~$\omega_{00}$ (that is, $\omega_0$ at the flux arc maximum) decreases with $B_\parallel$.
Partly, this effect is caused by the in-plane field increasing the penetration depth~$\lambda_{\mathrm L}$ and correspondingly the kinetic inductance of the superconducting film, i.e.~by an increase of~$L_{\mathrm b}$ with~$B_\parallel$.
From simultaneous reference measurements of the cJJ-less circuits on the same chip, however, we find that this effect is by far not sufficient to explain the magnitude of the shift, cf.~Supplemental Note~\ref{sec:suppl-resonator-params}~\cite{suppl-note}.
Hence, the cJJs themselves are impacted considerably by the field and their inductance increases even more strongly than that of the remaining circuit; roughly two thirds of the total sweetspot frequency shift can be attributed to the constrictions and only one third to the remaining film.
Similarly, the sweetspot linewidth~$\kappa_0$ increases with~$B_\parallel$, indicating an increased quasiparticle density due to a field-induced suppression of the superconducting energy gap.
Again, the increase of~$\kappa_0$ is much stronger in the circuits with cJJs than in those without cJJs, in fact by more than one order of magnitude; almost the complete additional loss can be attributed to the constrictions.
Note here that the small increase of the linewidth in the reference circuits in combination with the analysis of the field-alignment procedure (cf.~Appendix~\ref{sec:meth-field-alignment}) eliminates Abrikosov vortices as relevant loss contributors.

A second and quite surprising effect of~$B_\parallel$ is that the shape of the flux arcs is modified.
They become narrower and more and more skewed with increasing~$B_\parallel$, i.e.~asymmetric with respect to their sweetspots.
For the highest fields, neighboring arcs do not even cross anymore.
In conjunction with the skewing, one of the arc branches (the right one for each arc in Fig.~\subref{fig:field-response}{a}) gets extended, while the opposite branch (the left ones in Fig.~\subref{fig:field-response}{a}) becomes shorter.
For $B_\parallel = \qty{200}{\milli\tesla}$ this lengthening/shortening has reached a level where one of the discontinuous transitions between arcs more than doubled in frequency range while the other nearly vanished.
For still larger~$B_\parallel$, the shortening of the left branch is so strong that the corresponding frequency jump is downwards, meaning that the metastable state has a higher resonance frequency than the stable one.
Before explaining this unusual shape of the flux response in detail below, it is useful to briefly discuss how it impacts the circuit characteristics with respect to the envisioned applications.

We find that the field-induced modifications -- especially on the right, longer arc branch -- enhance both the total frequency tuning range $\omega_{0, \mathrm{max}} - \omega_{0, \mathrm{min}}$ and the tuning range between sweetspot and transition to the metastable part of each arc, as compared to the zero-field case.
These enhancements appear favorable for sensing and hybrid systems, where tuning range and the flux responsivity $\mathcal F = \idell{\omega_0}{\Phi_{\mathrm b}}$ are critical parameters that need to be maximized~\cite{rodrigues2019, bothner2021, paradkar2025}, particularly in the stable operation regime.
As can be seen in Fig.~\subref{fig:field-response}{d}, the in-plane field increases the highest stable~$\abs{\mathcal F} / 2 \pi$~\cite{metastable-transition-note} by a factor of more than~\num{5} from \qty[per-mode=symbol]{\sim 25}{\mega\hertz\per\Phizero} at $B_\parallel = 0$ to \qty[per-mode=symbol]{\sim 130}{\mega\hertz\per\Phizero} at high fields.
Due to the asymmetry, the maximized~$\abs{\mathcal F}$ can be reached at different in-plane fields, depending on which side of the skewed arc is chosen, which provides another degree of freedom during operation.
In several of the potential target systems, such as SQUID optomechanics~\cite{rodrigues2019, zoepfl2020, schmidt2024}, the most crucial figure of merit is the cooperativity $\mathcal C \propto \mathcal F^2 / \kappa$, which one might expect to suffer from~$B_\parallel$ due to a field-increased~$\kappa$.
However, the field-enhancement of~$\mathcal F$ is sufficiently large to more than compensate for the increase of~$\kappa$ in this ratio (cf.~also Supplemental Note~\ref{sec:suppl-linewidth}~\cite{suppl-note}), and thus $B_\parallel$ seems to actually be boosting~$\mathcal{C}$.
When operating the device in the \unit{\milli\kelvin}~temperature regime in the future, we expect the impact of $B_\parallel$ on~$\kappa$ to be much smaller, since even the field-reduced superconducting energy gap in the cJJs will not lead to a substantial thermal quasiparticle population, while the benefits of an enhanced~$\mathcal F$ likely persist.

Finally the most pressing question: What causes the unexpected arc skewing?
Various candidates may initially spring to mind, such as SQUID asymmetries or slight flux-dependent sample tilts/rotations in the large in-plane fields.
These first explanations, however, can be ruled out upon further reflection.
Differences in junction properties and SQUID loop asymmetries do not lead to asymmetric arcs, but to symmetric arc modifications~\cite{zorin2016, frattini2017} (and to odd circuit nonlinearities).
Furthermore, we find that the tilt direction of the arcs does not depend on the direction of~$\Phi_{\mathrm b}$ in the experiment; it does, however, invert when $B_\parallel$ is reversed, cf.~Supplemental Note~\ref{sec:suppl-reversed-inpl-field}~\cite{suppl-note}.
Remarkably, the arcs of one of the three SQUID circuits we characterized always tilt opposite to the arcs of the other two, cf.~Supplemental Note~\ref{sec:suppl-res4-data}~\cite{suppl-note}.
This third circuit is rotated by~\ang{180} on the chip (cf.~Supplemental Note~\ref{sec:suppl-chip-layout}~\cite{suppl-note}), which in principle is equivalent to reversing the direction of~$B_\parallel$ but in combination with the other circuits demonstrates that the skewing effect is intrinsic to the devices and cannot be related to chip alignment, mechanical rotation, or residual out-of-plane field effects such as Abrikosov vortices.
Overall, the observations indicate a simultaneous symmetry breaking in both cJJs of the SQUID, induced by~$B_\parallel$, that causes a dependence of both the cJJ inductances (shape of the arc) and the switching currents (jumps between arcs) on the direction of the screening current in the SQUID loop.
In other words, the results suggest that the constrictions are turned into intrinsic superconducting nano-diodes by~$B_\parallel$, with asymmetric CPRs and asymmetric critical currents.

We emphasize that this is not the same as the SQUID as a whole turning into a diode with respect to a transport current flowing across it, which can indeed be caused by SQUID or cJJ asymmetries~\cite{zapata1996, weiss2000, sterck2002, sterck2005, zorin2016, frattini2017}.
The individual-constriction diode-model discussed in the next section will further clarify this important aspect.

\subsection{Field-induced Josephson-diode effect}\label{sec:cpr-response}
To quantitatively demonstrate how a superconducting diode leads to the tilted flux response, the corresponding junction CPRs are of central relevance.
Once the CPRs are known, the constriction inductance and the circuit resonance frequency as a function of~$\Phi_{\mathrm b}$ can be calculated using Eqs.~\eqref{eq:cpr-sinlin}--\eqref{eq:squidres-frequency}.
More precisely, once a model for the CPRs is available, the model parameters can be adjusted to resemble the experimental flux arcs using a fitting routine, and in a second step the CPRs can be calculated using the resulting model parameters.
In contrast to earlier reports of diodes in constrictions~\cite{zhang2022, margineda2023}, we formulate a conceptually simple macroscopic model in terms of only phase, flux and inductances, which does not require sophisticated microscopic details and can be applied to other types of Josephson diodes as well.

\begin{figure*}
  \includegraphics{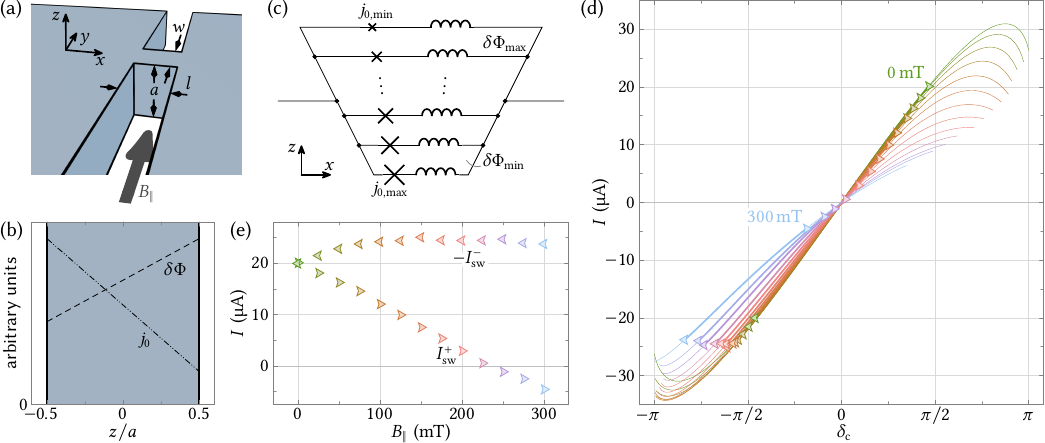}
  \vskip -.5\abovecaptionskip
  \titlecaption{Inhomogeneous nano-constrictions are Josephson-diodes in large magnetic fields}{%
 		\sublabel{a}~Schematic of a \nD 2~nano-constriction defining its width~$w$, length~$l$ and height~$a$ as well as the direction of~$B_\parallel$.
    For our device $a = \qty{100}{\nano\meter}$ and $l = w = \qty{40}{\nano\meter}$.
 		\sublabel{b}~To model the skewed flux response, we assume the critical current density~$j_0$ to linearly decrease from the bottom of the constriction at $z = -a / 2$ to the top at $z = +a / 2$.
    This is consistent with e.g.~surface damage by the \ce{Ne}-FIB process.
    Such a quality gradient is also associated with more magnetic flux per height $\delta\Phi = B_\parallel l_{\mathrm{eff}}$ entering the junction at the top than at the bottom, and we model it as increasing linearly in $z$-direction.
 		\sublabel{c}~Circuit equivalent of the Josephson-diode model consisting of a multi-loop parallel arrangement of infinitesimal constrictions whose inductive Josephson contribution grows with~$z$ due to a decreasing critical current density.
    At the same time, the flux per height~$\delta\Phi$ between the paths increases with~$z$, here illustrated by an increasing loop length.
 		\sublabel{d}~Constriction~CPR~$I(\delta_{\mathrm c})$ as a function of~$B_\parallel$ obtained from fits to the flux arc data (lines in Fig.~\subref{fig:field-response}{a}).
    Thin lines show the full range for which our numerical algorithm yields a result, thick line segments show the parts that correspond to the experimental flux arcs in between the discontinuous jumps, symbols mark the corresponding switching currents/phases (i.e.~the ends of the thick line segments).
    With increasing~$B_\parallel$, the CPRs become increasingly asymmetric, both in slope and in critical current (points of vanishing slope).
    Negative critical currents initially increase in magnitude with~$B_\parallel$ and then slightly decrease for the higher fields, positive critical currents decrease quickly and monotonically.
    Switching currents are~\qty{\sim 10}{\micro\ampere} lower than critical currents.
    As before as well as in the next panel, colors encode~$B_\parallel$.
 		\sublabel{e}~Switching currents $I_{\mathrm{sw}}^+$ and~$-I_{\mathrm{sw}}^-$ as derived from the flux arc discontinuities and the CPR model.
    For $B_\parallel = 0$, positive and negative switching currents are equal in magnitude, but with increasing~$B_\parallel$ the difference between the two values grows, indicating an increasing diode effect.
    Negative switching currents stay larger in magnitude than the initial value for all $B_\parallel > 0$, positive switching currents decrease nearly linearly.
    For the largest fields $B_\parallel \ge \qty{250}{\milli\tesla}$ both switches happen at the same current polarity ($I_{\mathrm{sw}}^+ < 0$), emphasizing that the switching current does not match the critical current~$I_0^+$.
  }
  \label{fig:cpr-response}
  \vspace{-2.1pt}
\end{figure*}

In the presence of a magnetic in-plane field, the entire nano-constriction will be penetrated by the field due to the cJJ dimensions all being smaller than the penetration depth of the Nb film $\lambda_{\mathrm L} \sim \qty{130}{\nano\meter}$.
Most likely, the constriction itself has an even larger penetration depth due to its reduced transition temperature and electron mean free path as compared to the un-milled film~\cite{uhl2024a}.
Now, $B_\parallel$ determines the phase gradient across the constriction in $z$-direction, cf.~Fig.~\ref{fig:cpr-response}, and the total phase drop is
\begin{equation}
	\delta(z) = \delta_0 + \frac{2 \pi}{\Phi_0} \, B_\parallel l_{\mathrm{eff}} z
\end{equation}
with $\delta_0$ the phase drop at $z = 0$ and $l_{\mathrm{eff}}$ the effective length of the cJJ in current direction.
The local current density is then described by the local first Josephson relation
\begin{equation}\label{eq:current-density-simple}
	j(z, \delta_0) = j_0 \sin\Bigl( \delta_0 + \frac{2 \pi}{\Phi_0} \, B_\parallel l_{\mathrm{eff}} z - \frac{2 \pi}{\Phi_0} \, \ell_{\mathrm{lin}} j(z, \delta_0) \Bigr)
\end{equation}
with the critical current density~$j_0$ and the specific linear inductance contribution~$\ell_{\mathrm{lin}}$.
The junction CPR can be obtained from~$j(z, \delta_0)$ by integration over the cJJ cross-section as
\begin{equation}\label{eq:cpr-integral}
	I(\delta_0) = w \int_{-a/2}^{a/2} j(z, \delta_0) \dif z
\end{equation}
where~$w$ is the width of the cJJ.
Note that up to this point, the model basically represents a combination of the standard models for constriction CPRs, as used above for $B_\parallel = 0$~\cite{likharev1979}, and the textbook description of short JJs in a magnetic field.
For $\ell_{\mathrm{lin}} = 0$, the model simply results in a sinusoidal CPR with a field-dependent critical current following the well-known Fraunhofer interference pattern.
However, this combination does not generate a diode~CPR, as the spatial symmetry remains unbroken.

To get the constrictions to turn into nano-diodes, we need at least one of their intrinsic properties to be a function of~$z$, i.e.~we need a constriction with some kind of gradient along the $z$-direction.
Conveniently, this is very likely the experimental situation due to the fabrication method using \ce{Ne}-FIB milling, a process that is known to damage the material in close proximity to the milled parts.
Partly, this damage results from the Gaussian profile of the ion beam, leading to slight milling and/or ion implantation next to the main cuts~\cite{troeman2007, potter2025}, and partly it is related to re-deposition of the milled material onto the adjacent structures, e.g.~the constrictions.
One signature of this damage is the reduced transition temperature compared to the unmilled film~\cite{uhl2024, uhl2024a}.
It is plausible to assume that the surface of the constriction receives more damage from this process than the lower parts, which are somewhat shielded by the surface layer, and that thus a damage-gradient is formed.
A natural question is whether any large gradient is compatible with the length scales of the nano-constriction.
It can be answered by comparing the $z$-extent~$a$ of the constriction with the mean superconducting coherence length~$\overline{\xi}(T_{\mathrm s})$.
The second critical field of the constriction must be $B_{\mathrm c2} > \qty{300}{\milli\tesla}$, and so the coherence length is $\overline{\xi}(T_{\mathrm s}) = \sqrt{\Phi_0 / 2\pi B_{\mathrm c2}} < \qty{34}{\nano\meter}$, indicating $a > 3 \overline\xi$.
Independently, the shape of the CPR in Fig.~\subref{fig:flux-response}{d} suggests $l \gtrsim 3.5 \, \overline\xi$~\cite{likharev1979, uhl2024a} and thus $\overline\xi \lesssim \qty{12}{\nano\meter}$, suggesting $a > 8 \overline\xi$.
Given these lower bounds, we can safely conclude that $a$ is large enough for inhomogeneities to be realized inside the junctions.

We currently do not know the exact details of the constriction non-uniformity; in principle all of the quantities $j_0$, $B_\parallel$, $l_{\mathrm{eff}}$ and~$\ell_{\mathrm{lin}}$ could be a (not necessarily linear) function of~$z$ (and also~$y$, but this is not relevant here).
To minimize complexity and the number fit parameters, we only consider a linear decrease of the critical current density as well as a linear increase of the flux per length with~$z$
\begin{align}
  \label{eq:j0-gradient}
	j_0(z) &= \Bigl( 1 - \frac{2 \epsilon}{a} \, z \Bigr) \, j_0(0) \\
  \label{eq:leff-gradient}
	B_\parallel l_{\mathrm{eff}}(z) &= \Bigl( 1 + \ooalign{$\displaystyle \phantom{\frac{2 \epsilon}{a}}$\cr$\hidewidth\displaystyle \frac{b}{a} \hidewidth$} \, z \Bigr) \, B_\parallel l_{\mathrm{eff}}(0)
\end{align}
with $\epsilon, b \in [0, 1]$.
Either one of the two gradients would be sufficient to obtain a diode, cf.~Supplemental Note~\ref{sec:suppl-cpr-parameters}~\cite{suppl-note}, but the effect is stronger and resembles our experiment more closely if both are included.
For completeness, also the linear contribution~$\ell_{\mathrm{lin}}$ could be considered a function of~$z$, which would likely enhance the diode effect even further, since for $\ell_\mathrm{lin} = 0$ none of the other factors could produce a diodic CPR in the first place; however, we have not implemented it here, since it is not required to reproduce the experimental data.

The numerical solution of the model, fitting it to all experimental flux arcs simultaneously, reveals compelling agreement between theory and measurement, both qualitatively and quantitatively, cf.~the lines in Fig.~\subref{fig:field-response}{a}.
In the fit, we vary the model parameters $j_0(0)$, $\epsilon$, $l_{\mathrm{eff}}(0)$, and $b$, as well as, separately for each in-plane field, $\ell_{\mathrm{lin}}(B_\parallel)$; cf.~Appendix~\ref{sec:meth-arcfit} for a detailed description of the fit routine and Supplemental Note~\ref{sec:suppl-fit-parameters}~\cite{suppl-note} for the resulting fit parameters.
The circuit parameters $\omega_{0 \mathrm b}(B_\parallel)$, $L_{\mathrm b}(B_\parallel)$ and~$L_{\mathrm{loop}}(B_\parallel)$ are inferred from their $B_\parallel = 0$ values and the corresponding field-dependence of the reference circuits, cf.~Supplemental Note~\ref{sec:suppl-resonator-params}~\cite{suppl-note}, and are not varied in the arc fitting routine.
After the arc fit, we numerically calculate the corresponding current-phase relations for all~$B_\parallel$.
Note that with the finite $z$-inhomogeneity and $B_\parallel \neq 0$, the phase~$\delta_0$ in the center of the junction is not necessarily zero when $I = 0$.
Since we are in principle free to choose any~$\delta_{\mathrm c} \coloneq \delta(z_0)$ to describe the CPR (rather than~$\delta_0$, i.e.~$z_0 = 0$) and the SQUID loop current should be zero when no external flux is applied, we choose $z_0$ to be the position where the phase vanishes when no net current flows through the junction, such that for $\delta_{\mathrm c} = 0$ we have $I = 0$ as well as $\Phi = \Phi_{\mathrm b} = 0$, cf.~Eq.~\eqref{eq:flux-relation} and Appendix~\ref{sec:meth-cpr-implementation}.

Figure~\subref{fig:cpr-response}{d} shows the CPRs resulting from the fit, which clearly reveal the field-induced diode effect, both in the corresponding Josephson inductance (inverse CPR slope) and in the critical currents.
The slope of the straight part of the CPRs decreases with~$B_\parallel$, reflecting the increased sweetspot inductance (point of maximum slope on each CPR).
Notably, the point of maximum slope is not at $\delta_{\mathrm c} = 0$ anymore for $B_\parallel \neq 0$, but shifts to negative phases and currents, which is also visible in the sweetspot not being located at $\Phi_{\mathrm b} = n \Phi_0$ for $B_\parallel > 0$ in Fig.~\subref{fig:field-response}{a}.
Furthermore, a clear asymmetry between the critical currents in positive and negative direction develops with increasing~$B_\parallel$.
The magnitude of the negative critical current~$I_0^-$ increases at first, before it starts to gently decline again to a value slightly smaller than the one at $B_\parallel = 0$.
The positive critical current~$I_0^+$ on the other hand immediately and monotonically decreases with the application of~$B_\parallel$, and so does the critical phase.
At the highest fields, we get $\abs{I_0^- / I_0^+} \sim 3$; the exact values are hard to determine, since our numerical algorithm is not able to provide the CPR for all phases, cf.~Appendix~\ref{sec:meth-cpr-implementation} and Supplemental Note~\ref{sec:suppl-cpr-parameters}~\cite{suppl-note}.

The experimental branch switching currents~$I_{\mathrm{sw}}^\pm$ are consistently smaller than~$I_0^\pm$, as has been observed in \ce{Nb}~constrictions before~\cite{, uhl2024a}, which -- as mentioned above -- could have multiple reasons including phase slips, phase diffusion, flux noise or thermal noise.
Nevertheless, $I_{\mathrm{sw}}^\pm$ has the same trend as $I_0^\pm$, just with a nearly constant current offset, which leads to the interesting situation that for the highest~$B_\parallel$ both switching currents have the same polarity; the circulating screening current in the SQUID never changes direction, only magnitude.
Still, the ensemble of positive and negative switching currents resembles the shape of the main bulge of a skewed Fraunhofer interference pattern, which is typical for superconducting Josephson diodes, cf.~e.g.~Ref.~\cite{schmid2025}.
This also allows us to perform a sanity check on the behavior.

Typically in Josephson junctions, the critical current in a magnetic field reaches its first minimum at the field~$B_0$ at which one flux quantum is coupled into the junction.
From the data in Fig.~\subref{fig:cpr-response}{e}, $I_0^+ \approx 0$ seems to happen roughly at around~\qtyrange[range-units=bracket]{350}{400}{\milli\tesla} considering the trend of~$I_{\mathrm{sw}}^+$ and a \qtyrange[range-units=bracket]{10}{15}{\micro\ampere}~offset to~$I_0^+$.
However, due to the presence of the $j_0$\nobreakdash-gradient, a complete disappearance of the critical current is no longer expected (cf.~Supplemental Note~\ref{sec:suppl-cpr-parameters}~\cite{suppl-note}), and indeed the value $B_0 = \Phi_0 / a l_{\mathrm{eff}}(0)$ we find from the fits is only~\qty{305}{\milli\tesla}.
The effective area~$A_{\mathrm{eff}} = al_\mathrm{eff}(0)$ of the cJJ then corresponds to $A_{\mathrm{eff}} = \Phi_0 / B_0 = \qty{6.8e-15}{\meter\squared}$, which for homogeneous field penetration and $a = \qty{100}{\nano\meter}$ implies an effective length $l_{\mathrm{eff}}(0) = \qty{68}{\nano\meter}$.
This value might seem small at first glance, but since the junction is close to or in the nonlocal limit~\cite{ivanchenko1990, rosenthal1991, clem2010} due to the near-complete field penetration of the superconducting film, the theoretical value for the effective junction length is expected to be $l_{\mathrm{eff}} \approx a / 1.5 \approx \qty{67}{\nano\meter}$~\cite{clem2010} rather than $l_{\mathrm{eff}} = l + 2 \lambda_{\mathrm L}$ as one might suspect at first, and thus the agreement is quite good.
For the smaller \nD 3~constriction of a second device with $a = \qty{75}{\nano\meter}$, cf.~Supplemental Note~\ref{sec:suppl-res4-data}~\cite{suppl-note}, analogous considerations yield a slightly larger $B_0 = \qty{361}{\milli\tesla}$, which is consistent with our interpretation.
One should only compare these values with caution, though, as it is well possible for e.g.~$a$ itself to be a function of~$B_\parallel$ due to the top of the constriction possibly having $B_{\mathrm c2} < \qty{300}{\milli\tesla}$, or for the effective length of the nonlocal limit to be modified by the finite constriction length and the gradients.

The presented diode CPR model has an intuitive and useful circuit representation, cf.~Fig.~\subref{fig:cpr-response}{c}, which can be interpreted as a continuous generalization of the diode SQUID-models discussed in Refs.~\cite{fominov2022, souto2022, greco2023}.
It is equivalent to a parallel multi-loop combination of (infinitesimal) constrictions, each being a series arrangement of an ideal Josephson junction and a linear inductance.
With increasing~$z$, the critical currents of the Josephson elements decrease, while the loops between two neighboring constrictions gradually increase in size.
One can now imagine to design constriction arrays patterned into a single wire following this circuit arrangement that show a tailored diode effect at much lower magnetic fields, requiring neither fabrication-based, uncontrolled material inhomogeneities nor more space than a single constriction.
In fact, this approach would not be limited to constrictions, but (with larger footprints) could also be implemented with more standard tunnel junctions, and the necessary linear inductance contribution could be added e.g.~by high kinetic-inductance wiring.

Finally, these considerations provide us with a strategy to evade the diode effect and the associated arc skewing in ion-beam-based nano-constriction circuits in large magnetic fields, should this be desired.
All one needs to do is to align the junctions in a way that the constriction current direction and the in-plane field are parallel to each other, here e.g.~by placing the constrictions in the short SQUID loop sides, which are parallel to both the circuit symmetry axis and~$B_\parallel$ (cf.~Fig.~\ref{fig:concept}).
On the other hand, it is possible to utilize the diode effect in the constrictions for three-wave mixing circuits while avoiding the arc skewing by either using a single constriction or by orienting them perpendicular to the external field but with opposite dc current flow, e.g.~by again placing them in the short SQUID arms but applying the in-plane field in $y$-direction instead of in $z$-direction.

\subsection{Analyzing the CPR derivatives using the Kerr anharmonicity}\label{sec:kerr-response}
The last part of this manuscript is dedicated to the discussion of the Kerr anharmonicity~$\mathcal K$ of the circuit, which is both an important design parameter for applications~\cite{frattini2018, bothner2022, rodrigues2022, zoepfl2023} and a sensitive probe for the device nonlinearity.
Its origin are third and fourth order corrections to the total circuit potential energy~\cite{frattini2018, uhl2024a}, which can originate from a nonlinear kinetic inductance of the superconducting film or from the nonlinear inductance of integrated Josephson elements.
Here, it stems from the constriction nonlinearity.
Large Kerr anharmonicities (hundreds of~\unit{\mega\hertz}) are useful for realizing superconducting qubits, while medium to small anharmonicities (\unit{\kilo\hertz} to~\unit{\mega\hertz}) are desired for high-dynamic-range applications like parametric amplifiers, radiation-pressure systems or magnetometry.

For the current experiment, the main relevance of~$\mathcal K$ arises from it being a sensitive probe for the circuit nonlinearities.
In fact, it is a function of the first three CPR derivatives; for the device under consideration here it can be calculated as (cf.~Appendix~\ref{sec:meth-kerr-theory})
\begin{equation}\label{eq:kerr}
	\mathcal K = -\frac{e^2}{2 \hbar C_{\mathrm{tot}}} \Bigl( \frac{L_{\mathrm{arm}} + L_{\mathrm c}}{2 L} \Bigr)^{\! 3 \,} \frac{3 g_2^2 - g_3 (1 + g_1)}{g_1 (1 + g_1)^4}
\end{equation}
with the elementary charge~$e$, the reduced Planck constant~$\hbar$, $L_{\mathrm{arm}} = L_{\mathrm{loop}} / 3$, $L = L_{\mathrm b} + L_{\mathrm c} / 2$ and the dimensionless CPR~derivatives
\begin{eqnarray}\label{eq:dimensionless-cpr-derivs}
	g_k = \frac{2 \pi}{\Phi_0} \, L_{\mathrm{arm}} \, \del[k]{\delta_{\mathrm c}} I(\delta_{\mathrm c}), \qquad k \in \{1, 2, 3\} .
\end{eqnarray}
Thus, the Kerr anharmonicity is able to provide an independent probe of the sub-critical superconducting diode effect.
While the resonance frequency and the flux arc probe only the constriction inductance, i.e.~the first derivative of the CPR, as well as the switching current, $\mathcal K$~depends on $L_{\mathrm c}$ together with the second and third derivatives in a nontrivial way.
For two points on a point-symmetric (i.e.~non-diode) CPR, one will find all derivatives to agree if the first one does (except for a sign reversal in the even derivatives); not so on a diode CPR, where even for identical slopes the magnitudes of curvature and higher order derivatives can differ by a large amount, cf.~as an intuitive example the positive and negative points of $\idell{I}{\delta_{\mathrm c}} = 0$ in Fig.~\subref{fig:cpr-response}{d} for non-zero~$B_\parallel$.
Hence, if the Kerr anharmonicities differ at identical resonance frequencies on the two sides of a skewed flux arc, it eliminates alternative interpretations of the arc skewing, like chip rotation in the magnetic field or bias-flux lag due to screening currents, and once more confirms an intrinsic origin.
Such an effect could also be of practical interest, since it would allow one to choose between one of two anharmonicities for one and the same resonance frequency and vice versa, depending on the needs of a specific experimental configuration.

\begin{figure*}
  \includegraphics{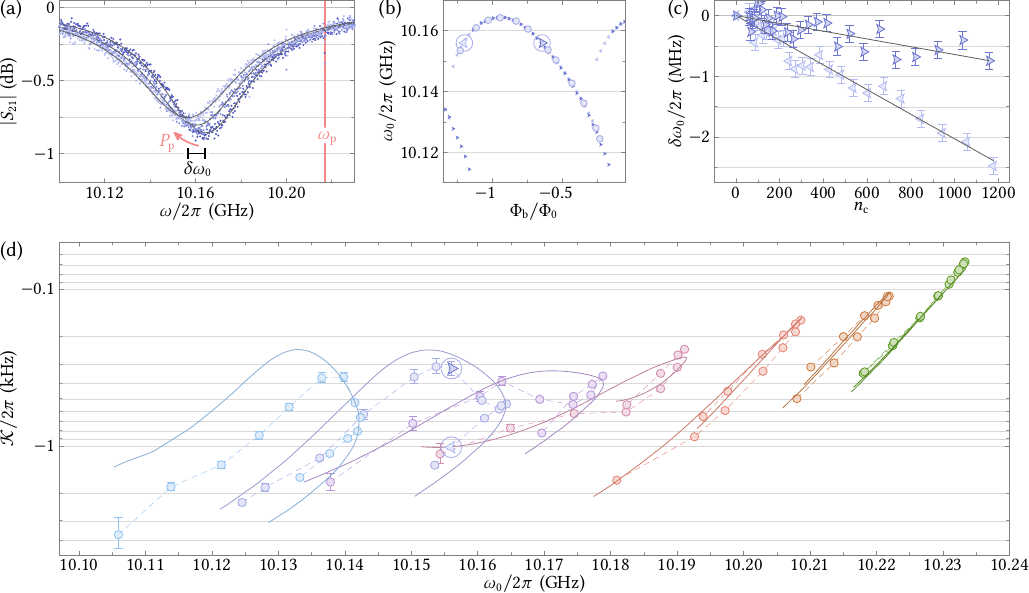}
  \titlecaption{Bimodal Kerr anharmonicity confirms the field-induced Josephson-diode effect and the model current-phase relations}{%
		\sublabel{a}~To determine the Kerr anharmonicity as a function of~$\Phi_{\mathrm b}$ and $B_\parallel$, we implement a two-tone experiment in which we send a pump tone with frequency $\omega_{\mathrm p} \approx \omega_0 + \kappa$ from a signal generator to the circuit and probe the pumped resonance with a small VNA signal for increasing values of pump power~$P_{\mathrm p}$.
    Due to the circuit Kerr anharmonicity originating from the nonlinear current-phase relation, the resonance frequency shifts as a function of pump power by $\domgzero = \omega_{0 \mathrm p} - \omega_0$, where $\omega_{0 \mathrm p}$ is the pump-shifted resonance frequency.
    Symbols are data, lines are fit curves, $B_\parallel = \qty{275}{\milli\tesla}$.
		\sublabel{b}~We conduct this two-tone scheme for multiple points on the flux arc at each~$B_\parallel$, shown here for the arc at $B_\parallel = \qty{275}{\milli\tesla}$.
    Large symbols (disks and kites) show the arc points for which we performed the pump-and-probe experiment, small kites are the $\omega_0$\nobreakdash-values from the flux-response characterization (as shown in Fig.~\subref{fig:field-response}{a} for other in-plane fields).
    Of particular relevance are points with identical~$\omega_0$ on opposite sides of the skewed arc, such as the two encircled ones marked by the left- and right-pointing kite symbols.
		\sublabel{c}~Frequency shift~$\domgzero$ at the two arc points shown as kite symbols in panel~\subref*{@}*{b} vs intracircuit pump photon number~$n_{\mathrm c}$.
    Despite identical resonance frequencies~$\omega_0$, the slopes of the Stark shifts at these arc points differ by a factor~\num{\gtrsim 3}.
    Lines are fits, which provide us with the Kerr anharmonicity~$\mathcal K$.
		\sublabel{d}~Kerr anharmonicity~$\mathcal K$ for various in-plane fields and points on the flux arc vs the corresponding un-pumped resonance frequency~$\omega_0$.
    Symbols are data, connected with dashed lines as guides to the eye, solid lines are fit curves based on the CPRs shown in Fig.~\ref{fig:cpr-response} plus a small polynomial correction, for details see main text and Appendix~\ref{sec:meth-kerrfit}.
    At $B_\parallel = 0$, there is a unique relationship between $\omega_0$ and~$\mathcal K$, but with increasing in-plane field a bimodal distribution of the anharmonicity appears.
    Simultaneously, the overall values of~$\abs{\mathcal K}$ increase with~$B_\parallel$.
    The bimodal distribution confirms a diode CPR and eliminates alternative mechanisms behind the skewed flux responses discussed in Fig.~\ref{fig:field-response}.
    Field values shown are $B_\parallel \in \qtyset{0; 125; 175; 225; 250; 275; 300}{\milli\tesla}$, indicated by color as before.
  }
  \label{fig:kerr-response}
  \vspace{-.1pt}
\end{figure*}

Experimentally, it is straightforward to determine the Kerr anharmonicity, and we apply a well-established two-tone pump-and-probe scheme for it.
The pump is a strong microwave tone from a signal generator with frequency~$\omega_{\mathrm p}$ and on-chip power~$P_{\mathrm p}$.
For each in-plane field we pick multiple values on each flux arc, spaced to cover the complete range of the arc, and then send in the pump tone slightly blue-detuned from the circuit resonance at $\omega_{\mathrm p} \approx \omega_0 + \kappa$.
Again using a weak VNA probe tone, we scan the pumped cavity resonance at various pump powers~$P_{\mathrm p}$ and subsequently determine the pump-induced resonance frequency shift (ac~Stark shift)
\begin{equation}
	\domgzero = \omega_{0 \mathrm p} - \omega_0
\end{equation}
as a function of~$P_{\mathrm p}$, where $\omega_{0 \mathrm p}$ is the resonance frequency at pump power~$P_{\mathrm p}$.
Combined with the intracircuit pump photon number~$n_{\mathrm c}$, which we obtain by analytical calculation in combination with a fit estimate for the pump line attenuation, cf.~Appendices~\ref{sec:meth-twotone-response} and~\ref{sec:meth-kerrfit}, the pump-induced frequency shift can be modeled by
\begin{equation}
	\domgzero = \Delta_{\mathrm p} - \sqrt{ (\Delta_{\mathrm p} - \mathcal K n_{\mathrm c} ) ( \Delta_{\mathrm p} - 3 \mathcal K n_{\mathrm c}) - (\kappa_{\mathrm p} - \kappa)^2 / 16 }
\end{equation}
with $\Delta_{\mathrm p} = \omega_{\mathrm p} - \omega_0$ and the pump-broadened linewidth $\kappa_{\mathrm p}$, cf.~Fig.~\subref{fig:kerr-response}{a} and~\subref{fig:kerr-response}*{c}.
The result of determining $\mathcal{K}$ via this procedure is shown and discussed in Fig.~\subref{fig:kerr-response}{d}, additional fit parameters are given in Supplemental Note~\ref{sec:suppl-fit-parameters}~\cite{suppl-note}.

The Kerr anharmonicity is clearly bimodal.
For large~$B_\parallel$ -- the regime of the strongest diode effect in the CPR and the strongest skewing of the flux response -- the experimental values for~$\mathcal K$ at a single resonance frequency (i.e.~identical~$L_{\mathrm c}$) differ by a factor up to~\num{\gtrsim 4}, depending on the side of the arc or correspondingly of the CPR.
For lower fields the difference is smaller but still significant with a factor~\num{\sim 1.5}.
Only for~$B_\parallel = 0$ the bimodality disappears completely, which is also very much expected.
In terms of absolute values of~$\mathcal K$, we find values between $\qty{\sim 70}{\hertz}$ and~$\qty{\sim 4}{\kilo\hertz}$, i.e.~we are deep in the low-anharmonicity regime $\mathcal K / \kappa \sim (\numrange{e-5}{e-7})$, which is ideal for most of our envisioned target devices like SQUID optomechanics and magnetometry, even if at lower temperatures $\kappa$ will decrease by one to two orders of magnitude.

Finally, we find that the experimental values for~$\mathcal K$ not only confirm a diode effect, but are also consistent with the independently obtained current-phase relations in Fig.~\ref{fig:cpr-response}.
Since tiny changes of the CPR in the experimentally accessible regime can have a strong impact on~$\mathcal K$ through the second and third derivatives, we allow small modifications~$\Delta I(\delta_{\mathrm c})$ to the CPRs in this step to better match the experimental Kerr constants.
These micro-adjustments are implemented as a high-order polynomial, added on top of the existing CPR.
To guarantee minimal deviation from the established CPR, we fit both Kerr and flux arcs simultaneously in this procedure.
The resulting CPRs deviate by less than~\qty{50}{\nano\ampere} from the original ones; a detailed discussion of the modified CPR fits can be found in Appendix~\ref{sec:meth-kerrfit} and in Supplemental Note~\ref{sec:suppl-poly-correction}~\cite{suppl-note}.
The lines shown in Fig.~\subref{fig:kerr-response}{d} are derived from these micro-modified CPRs and the agreement to the experimental values is very good, considering the amount of uncertainties such as frequency-dependent values for the pump attenuation, the external decay rate of the circuit towards the pump input, the gradient functions for~$j_0$ and~$B_\parallel l_{\mathrm{eff}}$, gradients in $\ell_{\mathrm{lin}}$ and~$w$ or SQUID asymmetries due to the two constrictions not being exactly identical.

\section{Discussion}
In this work, we investigated niobium quantum interference microwave circuits in magnetic in-plane fields up to~$\qty{300}{\milli\tesla}$, and observed a field-induced superconducting diode effect in the monolithic interferometer nano-constrictions.
The diode effect leads to a strong skewing of the circuit flux response, which we were able to fully reproduce based on a simple macroscopic constriction model.
The model revealed that inhomogeneities in the constriction itself are likely the origin of the diode effect.
Since gradients in e.g.~film quality, film stress, critical current densities or grain sizes are ubiquitous in non-epitaxial superconducting thin films and superconducting nano-structures, the insights gained here are likely relevant to any experiment involving the superconducting or Josephson diode effect.
In combination with inductive sensing of the CPR-derivative, i.e.~the constriction inductance, our model furthermore enabled us to reveal the asymmetric constriction current-phase relation as a function of magnetic in-plane field.
Finally, we were able to eliminate alternative mechanisms behind the experimental findings by measuring the Kerr anharmonicity of the circuit which showed a strong bimodal distribution as a function of circuit resonance frequency and is completely compatible with the presented diode model.

From an application-oriented perspective, the circuits show promising characteristics for hybrid quantum systems, sensing and metrology in magnetic in-plane fields up to~$\qty{300}{\milli\tesla}$ and possibly beyond; frequency tuning range and flux responsivity are both enhanced by the in-plane field and the resulting diode effect, making \ce{Nb}~constriction devices a highly competitive technology for magnetic field applications.
Furthermore, the simple circuit model we developed for the constriction diodes is easily applicable for future designs of yet unexplored constriction- or barrier-junction-array diodes with small footprints but large CPR asymmetries, which could be used to study and exploit diode-based enhancements of microwave circuits even at low magnetic fields.
Finally, we expect the bimodal distribution of the Kerr anharmonicity and the flux responsivity, i.e.~a certain decoupling of these two quantities from the circuit resonance frequency, to have useful implications for SQUID-based radiation-pressure systems and magnetometry.

In summary, our results constitute a promising starting point for the experimental combination of superconducting microwave circuits and superconducting diodes, in particular towards enhanced quantum circuits with yet unavailable functionalities, and towards a deeper understanding of the SD~effect thanks to the possibility to inductively characterize the diode CPR and its derivatives using microwave measurements.
We are convinced that a wide range of research fields may benefit from the findings presented here, including SQUID optomechanics for investigating the interplay between quantum physics and gravity, photon-pressure circuits (radio-frequency up-converters) for superconducting axion dark-matter detectors, quantum-limited parametric amplifiers with tunable nonlinearities, or dispersive magnetometry for high-bandwidth SQUID microscopy.

\begin{acknowledgments}
  The authors thank Markus Turad, Ronny Löffler (instrument scientists of the core facility \lisaplus), Christoph Back and Christoph Kalkuhl for technical support.
  This research received funding from the Deutsche Forschungsgemeinschaft~(DFG) via grant numbers 490939971 (\mbox{BO~6068/1-1}) and 511315638 (\mbox{BO~6068/2-1}) and from the Vector Stiftung via project number \mbox{P2023-0201}.
  \initials{MK}~gratefully acknowledges financial support by the Studienstiftung des Deutschen Volkes.
  We also acknowledge support by the COST actions NANOCOHYBRI (CA16218) and SUPERQUMAP (CA21144).
\end{acknowledgments}

\appendix

\section{Device fabrication and preparation}\label{sec:meth-device-fabrication}
The fabrication and preparation of the superconducting chip is executed in six steps, five of which regard the actual fabrication.
The steps are individually described below.
Note that the fabrication is much more sophisticated than necessary for SQUID circuits, since the chip also contains micromechanical elements in some resonators, though not in the one presented in the main text.
Despite the mechanical elements thus being completely irrelevant for the experiments and results presented here, we describe the entire chip fabrication for completeness and reproducibility.

\paragraph*{Step 1: Sacrificial layer patterning}
The fabrication starts with the patterning of the sacrificial layer for mechanical beams on the chip.
To this end, a \qtyproduct{26 x 26}{\milli\meter}~chip of high-resistivity ($\rho > \qty{10}{\kilo\ohm\centi\meter}$ at room~temperature) intrinsic silicon with a nominal thickness of~\qty{525}{\micro\meter} is covered with the adhesion promoter~\brandcode{AR\,300-80} as well as the photo-resist~\brandcode{ma-N\,1405} by spin-coating (resist thickness~\qty{\sim 0.5}{\micro\meter}).
The resist is then patterned by means of maskless photolithography ($\lambda_{\mathrm{litho}} = \qty{365}{\nano\meter}$) and developed using the developer~\brandcode{ma-D\,533/S}.
Since the resist making up the sacrificial layer is sensitive to the solvents usually used to clean samples between fabrication steps (acetone, isopropanol), no such cleaning can be performed from here on out.

\paragraph*{Step 2: Niobium deposition and patterning}
Before depositing the conductive niobium layer, the sample is subjected to plasma ashing in an oxygen plasma to remove possible resist and adhesion-promoter residues on the silicon surface.
Next, \qty{300}{\nano\meter} of niobium are deposited by dc-magnetron sputtering.
Optical lithography follows as before, but with the thicker positive resist~\brandcode{ma-P\,1215} (thickness~\qty{\sim 1.5}{\micro\meter}) and the developer~\brandcode{ma-D\,331/S}.
The pattern is etched into the \ce{Nb}~film using reactive ion etching with~\ce{SF6} at an angle of~\ang{40} to the chip surface in order to ensure consistent etching around the mechanical beams, in particular on the sidewalls of the sacrifical-layer patches.
During etching, the chip is constantly rotated to ensure homogeneous exposure.

\paragraph*{Step 3: Dicing and release}
Before removing all resist on the chip and thus releasing the sensitive mechanical beams, the sample is diced into the final pieces of \qtyproduct{10 x 10}{\milli\meter}.
The resist is then removed using oxygen plasma etching.

\paragraph*{Step 4: Global etching for niobium thinning}
The rather thick \qty{300}{\nano\meter} niobium layer helped ensure mechanical stability as well as electrical contact at the edges of the mechanical beams but for the final device we target a film thickness of only~\qty{100}{\nano\meter}.
In this step, the film thickness is reduced to the final value, again using \ce{SF6} reactive ion etching, this time perpendicular to the chip surface for directed etching from the top.
At this point, the chip is mounted as described in Step~\hyperref[sec:meth-fabrication-final-mounting]{6} and pre-characterized before the constriction-type Josephson junctions are added in the next step.

\paragraph*{Step 5: Constriction cutting}
Each resonator contains a loop in the inductive part of the circuit where the constriction junctions are placed after pre-characterization, converting it into a SQUID (cf.~Fig.~\ref{fig:concept}).
This is achieved using a neon ion microscope~(NIM) which is capable of high-precision milling with a focused ion beam~(FIB) of nano-scale spot size.
For the \nD 2~junctions, two \qty{\sim 40}{\nano\meter}~wide rectangles are cut into the wire from both sides, leaving a remaining constriction of around the same width between them.
For the \nD 3~junctions, the constriction is additionally milled from the top with a lower dose, leading to a reduced conductor thickness.
Ion acceleration voltages and ion doses can be found in Supplemental Table~\ref{tab:suppl-resonator-parameters}~\cite{suppl-note}.

\paragraph*{Step 6: Final mounting}\label{sec:meth-fabrication-final-mounting}
After fabrication is complete, the chip is mounted into a radiation tight copper housing with a printed circuit board~(PCB) surrounding the chip, where it is wire-bonded to microwave feedlines and ground.
The PCB contains coplanar waveguide feedlines leading to SMP~connectors, where microwave cables are attached during measurement.
After mounting into the measurement setup, the characterization of the device is performed.

\section{Single-tone circuit response}\label{sec:meth-S21-fitting}
The ideal transmission-response function of a high-$Q$ parallel RLC~circuit side-coupled to a feedline with characteristic impedance~$Z_0$ by a coupling capacitance~$C_{\mathrm c}$ is given by
\begin{equation}\label{eq:S21-ideal}
  S_{21}^{\mathrm{ideal}} = 1 - \frac{\kappa_{\mathrm{ext}}}{\kappa + 2 \i (\omega - \omega_0)}
\end{equation}
with the angular excitation frequency~$\omega$ and the resonance frequency $\omega_0 = 1 / \sqrt{L C_{\mathrm{tot}}}$.
The total capacitance is given by $C_{\mathrm{tot}} = C + C_{\mathrm c}$, $L$ and~$C = 2C_\mathrm{idc}$ are inductance and capacitance of the uncoupled circuit, respectively.
The internal and external decay rates (linewidths) $\kappa_{\mathrm{int}}$ and~$\kappa_{\mathrm{ext}}$, are given by
\begin{align}
  \kappa_{\mathrm{int}} &= \frac{1}{R (C + C_{\mathrm c})} \\
  \kappa_{\mathrm{ext}} &= \frac{\omega_0^2 C_{\mathrm c}^2 Z_0}{2 (C + C_{\mathrm c})}
\end{align}
and the total decay rate by $\kappa = \kappa_{\mathrm{int}} + \kappa_{\mathrm{ext}}$.
The effective resistance~$R$ accounts for all internal losses of the circuit, such as resistive, dielectric or radiative losses.

Due to the cabling and all the microwave components in between the vector network analyzer and the circuit, the ideal reflection is not what we measure, though.
To take frequency-dependent attenuation, the electrical cable length and possible interferences (e.g.~parasitic transmission around the chip) into account, we model the actual measurement signal as
\begin{equation}\label{eq:S21-real}
  S_{21}^{\mathrm{real}} = (a_0 + a_1 \omega + a_2 \omega^2) \e^{\i (\phi_0 + \phi_1 \omega)} \Bigl( 1 - \frac{\kappa_{\mathrm{ext}} \e^{\i \theta}}{\kappa + 2 \i (\omega - \omega_0)} \Bigr) .
\end{equation}
The factors $a_0$, $a_1$, $a_2$, $\phi_0$, $\phi_1$ and $\theta$ are real-valued fit parameters.
During our automated data fitting routine we first remove the absorption resonance from the dataset (leaving a gap in the $S_{21}$-dataset) and fit the remaining $S_{21}$-response with the background function
\begin{equation}\label{eq:S21-bg}
  S_{21}^{\mathrm{bg}} = (a_0 + a_1 \omega + a_2 \omega^2) \e^{\i (\phi_0 + \phi_1 \omega)} .
\end{equation}
As a result, we obtain preliminary values for $a_0$, $a_1$, $a_2$, $\phi_0$ and $\phi_1$.
Then, we calculate $S_{21}^{\mathrm{real}} / S_{21}^{\mathrm{bg}}$ for the complete dataset and fit the resulting data with
\begin{equation}
  S_{21}^\theta = 1 - \frac{\kappa_{\mathrm{ext}} \e^{\i \theta}}{\kappa + 2 \i (\omega - \omega_0)}
\end{equation}
from which we obtain a preliminary set of values for $\omega_0$, $\kappa$, $\kappa_{\mathrm{ext}}$ and $\theta$.
Finally, we use all the preliminary values for $a_0$, $a_1$, $a_2$, $\phi_1$, $\phi_2$, $\omega_0$, $\kappa$, $\kappa_{\mathrm{ext}}$ and $\theta$ as starting parameters to re-fit the original dataset using Eq.~\eqref{eq:S21-real}.
All the $S_{21}$-datasets in the manuscript figures as well as the corresponding fit curves have been completely background-corrected (cf.~next section) and we have additionally removed the interference angle~$\theta$.

\section{Two-step background-correction}\label{sec:meth-bg-correction}
In order to achieve good results with the fitting procedure described in the previous section, the background of the data must already be quite uniform, as assumed in Eq.~\eqref{eq:S21-bg}, which is not usually the case in the experiment due to interfering resonances in the measurement setup in combination with our large resonance linewidths.
Measuring the (approximate) background directly allows us to first divide each $S_{21}$~measurement by this measured background before in a second step fitting the resulting signal as described above.

We remove the resonance signal from the measurement by heating the sample to~\qty{5}{\kelvin}, which leaves the measurement background largely unaffected.
At this temperature, the circuit resonances are completely absent due to the reduced transition-temperature of the nano-constrictions~\cite{uhl2024}, while the on-chip coplanar waveguide feedline ($T_{\mathrm c} \approx \qty{9}{\kelvin}$) remains superconducting with nearly unchanged characteristics.
Then, we record $S_{21}$~traces for each resonator frequency window with identical settings to those we use for the measurements at~\qty{2.8}{\kelvin}.
To reduce noise in the background signal, we take \num{20}~traces and average them point-wise.
This procedure was repeated each day of the measurement series (four in total) with very consistent results, validating the stability of the measurement setup.
To even further reduce noise, we use the mean of these four sets of measured backgrounds for the final evaluation.

\section{Field alignment by rotation}\label{sec:meth-field-alignment}
In order to align the applied field with the sample plane, we employ an arrangement as shown in Fig.~\subref{fig:meth-field-alignment}{a}.
The large main coil is designed to apply a strong field~$B_\parallel$ parallel to the chip surface, however, due to unavoidable assembly inaccuracies, it contains a small out-of-plane component~$B_\parallel^z$.
To compensate this out-of-plane component, a split-coil is mounted in the cryostat in such a way that it creates a field~$B_\perp$ orthogonal to~$B_\parallel$.
At the same current, the compensation coils generate a much smaller field at the sample location than the main coil $B_\perp / B_\parallel = \qty{0.8}{\percent}$, as it only needs to account for the small out-of-plane component of~$B_\parallel$.
However, the actual magnitude of~$B_\parallel^z$ depends on the exact alignment between the main coil and the sample and cannot be precisely controlled, so the fields of the two coils need to be coordinated such that their out-of-plane components cancel out~$B_\parallel^z + B_\perp^z = 0$.
Note that the application of SQUID bias flux~$\Phi_{\mathrm b}$ is completely independent of~$B_\parallel$ and~$B_\perp$ and accomplished via the small sample coil attached directly to the sample housing.

\begin{figure}
  \raisebox{0pt}[\height][0pt]{%
    \begin{minipage}[b][\textheight]{\columnwidth}
      \includegraphics[trim={0pt 0pt 0pt .29cm}]{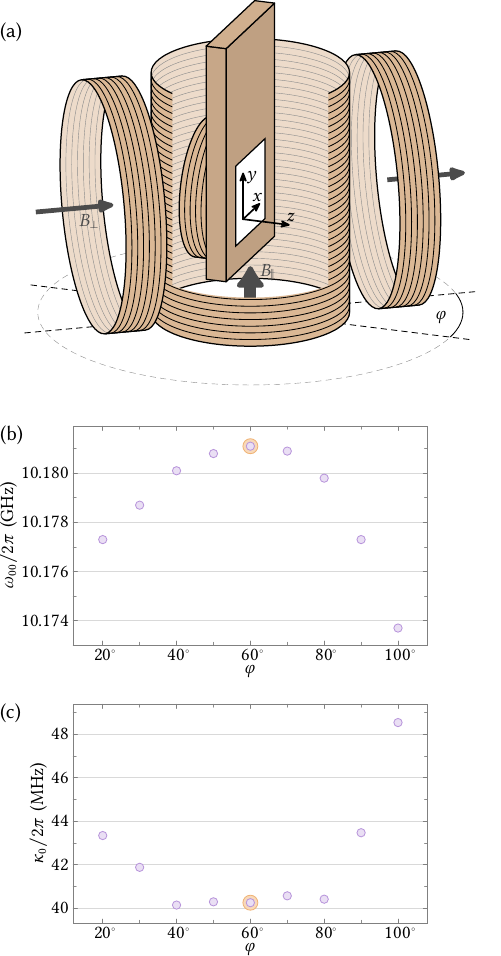}
      \vfill
      \titlecaption{Alignment of the in-plane field by rotation}{%
        \sublabel{a}~Magnet arrangement in the cryostat.
        Directly attached to the chip housing is a small sample coil for applying~$\Phi_{\mathrm b}$, which is not involved in the application of the in-plane field.
        A large coil generating~$B_\parallel$ is wrapped on the outside of the vacuum compartment that chip and sample coil reside in.
        A weaker split-coil magnet applying~$B_\perp$ is rigidly attached to the cryostat, that is operated in series with the large main coil and used to compensate out-of-plane components of~$B_\parallel$.
        To balance the compensation of the out-of-plane components of $B_\parallel$ and~$B_\perp$, the vacuum compartment together with the chip assembly can be rotated inside the cryostat, adjusting their angle~$\varphi$ with respect to the symmetry axis of the split coil.
        \sublabel{b}~Sweetspot resonance frequencies~$\omega_{00}$ vs~rotation angle~$\varphi$ at an applied field of $B_\parallel = \qty{250}{\milli\tesla}$.
        The orange highlighted maximum at $\varphi = \ang{60}$ marks the ideal field alignment~$\varphi_0$; all experiments were performed at that angle.
        \sublabel{c}~Sweetspot linewidth~$\kappa_0$ vs~rotation angle~$\varphi$ extracted from the same resonance traces as the frequencies shown in~\subref*{@}*{b}, again with the ideal angle~$\varphi_0$ highlighted orange.
        Note that the sweetspot linewidth remains constant for $\varphi \in [\ang{40}, \ang{80}]$, only increasing outside that range due to the presence of Abrikosov vortices.
        \label{fig:meth-field-alignment}%
      }
    \end{minipage}%
  }%
\end{figure}

Rather than using two separate current sources for the two coils generating $B_\parallel$ and $B_\perp$, which would entail uncorrelated current drifts as well as noise, and pose a severe challenge for effective field compensation, we operate the two coils in series, using a single in-house-built, low-noise, battery-powered current source.
While this fixes the ratio~$B_\perp / B_\parallel$ between the fields generated by the coils, we can control the ratio of their out-of-plane components~$B_\perp^z / B_\parallel^z$ by rotating the vacuum cup with the attached sample assembly inside the cryostat and thus adjusting the angle~$\varphi$ between that assembly and the compensation coils.
Quantitatively, $B_\parallel^z$ is constant with respect to~$\varphi$ while $B_\perp$ rotates around the $y$-axis and thus~$B_\perp^z = B_\perp \cos\varphi$.
Here, we neglect small misalignments between chip, rotational axis and split-coil, but while including inaccuracies would complicate the previous expression (and thus subsequent ones), perfect $B_\parallel^z$~compensation remains possible.
We want to operate the setup with the optimal compensation angle~$\varphi_0$ at which the out-of-plane components of the two fields cancel exactly $B_\perp^z + B_\parallel^z = (\cos\varphi - \cos\varphi_0) B_\perp = 0$.

In order to find~$\varphi_0$, we apply a constant coil current $I_{\mathrm{coil}} = \qty{1}{\ampere}$ (i.e.~$B_\parallel = \qty{250}{\milli\tesla}$ and $B_\perp = \qty{2}{\milli\tesla}$) and, for different~$\varphi$, sweep the flux~$\Phi_{\mathrm b}$ through the SQUID loop using the sample coil attached directly to the chip housing, recording the flux response.
The sweetspot frequencies of the resulting flux arcs are presented in Fig.~\subref{fig:meth-field-alignment}{b}.
Similarly to the in-plane field, cf.~Fig.~\ref{fig:field-response}, a magnetic out-of-plane field suppresses the resonance frequency~$\omega_0$ and increases the linewidth~$\kappa$ of the resonator.
Unlike the vortex-free Meissner response induced by a magnetic in-plane field, however, an out-of-plane field introduces vortices into the superconducting film as well as much larger screening currents due to demagnetization effects, causing extra losses and frequency shifts at much smaller field values~\cite{bothner2011, bothner2012a}, which in turn provide a sensitive probe for the out-of-plane field component.
Comparing the sweetspot resonance frequencies~$\omega_{00}(\varphi)$, we expect to find a minimum at $\varphi = 0$ (where $B_\perp^z / B_\parallel^z < -1$ is minimal), maxima at $\varphi = \pm \varphi_0$ (where $B_\perp^z / B_\parallel^z = -1$) and the global minimum at $\varphi = \pi$ (where $B_\perp^z / B_\parallel^z > 0$ is maximal).
The sweetspot linewidths~$\kappa_0(\varphi)$ correspondingly show opposite extrema at the same angles, but here the minimum at the optimum angle forms a plateau, with the linewidth remaining constant inside an interval surrounding~$\varphi_0$ and only increasing outside that range.
This is in line with all flux being expelled from the superconducting film of the resonator below some critical field~$B_{\mathrm m}$, with magnetic vortices appearing and increasing the linewidth only for larger fields~\cite{stan2004, song2009}.
To avoid hysteresis in the magnetic response due to Bean-like flux gradients / critical states~\cite{brandt1993, bothner2012} and to minimize the risk of flux avalanches and flux jumps~\cite{ghigo2007, nulens2023} when microwave signals are applied, we perform field-cooling before each $\Phi_{\mathrm b}$~sweep and in particular for each new angle $\varphi$, heating the sample to~\qty{\sim 12}{\kelvin} and then cooling back down to~\qty{2.8}{\kelvin}, as described in Sec.~\ref{sec:field-response}.

Using this procedure, we find $\varphi_0 = \ang{60}$, implying $\abs{B_\parallel^z} / B_\parallel = \qty{0.4}{\percent}$ and that the misalignment between the main coil field and the sample surface is~\ang{\sim 0.2}.
Considering a possible uncertainty in the optimal compensation angle of up to~\ang{3} due to the limited angle resolution of the experiment, we can estimate the residual out-of-plane field to be $\abs{(B_\parallel^z + B_\perp^z) / B_\parallel} < \qty{0.04}{\percent}$, which is equivalent to a residual misalignment on the order of~\ang{0.02}.
The additional orthogonal in-plane component~$B_\perp^x$ introduced by the compensation coils at $\varphi_0 = \ang{60}$ is~\qty{0.7}{\percent} of~$B_\parallel$, corresponding to a rotation of~\ang{\sim 0.4}, i.e.~it is small, on the order of the assembly inaccuracies that are present in any case.
The sweetspot linewidth remains constant at its minimum value in the interval from \ang{40} to~\ang{80}, i.e.~$\varphi_0 \pm \ang{20}$, once more confirming the choice of compensation angle.
Calculating the total out-of-plane field at these angles we estimate the threshold field for vortex penetration to be $B_{\mathrm m} \approx \qty{0.7}{\milli\tesla}$.
This value is in line with the threshold fields reported in Ref.~\cite{stan2004} for our strip width of~\qty{\sim 2}{\micro\meter} and serves to emphasize the quality of our field alignment.

These values again illustrate the necessity for a precise and stable out-of-plane field compensation as we achieve it with this alignment procedure.
At our largest field value $B_\parallel = \qty{300}{\milli\tesla}$ the magnitude of the compensated out-of-plane component is $\abs{B_\parallel^z} = \abs{B_\perp^z} = \qty{1.2}{\milli\tesla}$, which without compensation would correspond to \num{\sim 120} flux quanta in the resonator SQUID loop.
If the main coil and the compensation coils were driven independently and each of them drifted by just~\qty{0.1}{\percent}, this could add up to a $\Phi_{\mathrm b}$~drift of almost a quarter flux quantum, making reproducible flux tuning measurements impossible.
Driving the main coil and the compensation coils in series neatly avoids this issue, keeping $B_\parallel^z / B_\perp^z$ constant such that any driving current drifts affect only the compensated out-of-plane field $B_\parallel^z + B_\perp^z$, not its individual components.
Regarding vortex penetration, we note that the un-compensated out-of-plane field is greater than~$B_{\mathrm m}$, thus introducing vortices into the resonator strips that cause additional losses, whereas the compensated out-of-plane field component will be~\qty{< 0.12}{\milli\tesla} even in the worst-case scenario (\ang{3}~off the optimum compensation angle), well below the vortex penetration threshold.

\section{Numerical algorithm to calculate the CPR}\label{sec:meth-cpr-implementation}
So as to avoid constant factors of no consequence to the numerical solution of the CPR, we introduce the normalized variables $\tilde z = z / a$, $\tilde j = j / j_0(0)$, $\delta_B = 2 \pi B_\parallel l_{\mathrm{eff}}(0) a / \Phi_0$, $\delta_\ell = 2 \pi j_0(0) \ell_{\mathrm{lin}} / \Phi_0$, $I_{00} = w a j_0(0)$ and $\tilde I = I / I_{00}$, and rewrite Eqs.~\eqref{eq:current-density-simple} through~\eqref{eq:leff-gradient} more concisely as
\begin{gather}
  \label{eq:current-density-normalized}
  \tilde j(\tilde z, \delta_0) = (1 - 2 \epsilon \tilde z) \sin\bigl( \delta_0 + \delta_B (1 + b \tilde z) \tilde z - \delta_\ell \tilde j(\tilde z, \delta_0) \bigr) \\
  \label{eq:cpr-integral-normalized}
  \tilde I(\delta_0) = \int_{-1/2}^{1/2} \tilde j(\tilde z, \delta_0) \dif{\tilde z} .
\end{gather}
In order to find solutions to the implicit (normalized) current density equation, we treat~$\tilde j$ as an independent variable, define
\begin{equation}
  \mathcal J \coloneq (1 - 2 \epsilon \tilde z) \sin\bigl( \delta_0 + \delta_B (1 + b \tilde z) \tilde z - \delta_\ell \tilde j \bigr) - \tilde j
\end{equation}
and solve $\mathcal J(\tilde j) = 0$ for a given set of coordinates~$(\tilde z, \delta_0)$ and model parameters~$(\epsilon, b, \delta_B, \delta_\ell)$ using standard root-finding algorithms.
The matter is complicated by the fact that~$\mathcal J(\tilde j)$ may have multiple roots, and we need to find the correct one; that is, we need to find consistent roots for different $(\tilde z, \delta_0)$.
Luckily, for $(\tilde z, \delta_0) = (0, 0)$ the problem is trivially solved by $\tilde j = 0$ and we can trace the continuously changing root from this starting value to any desired coordinates.

Naturally, doing this from scratch every time a current density needs to be computed would be very computationally expensive as well as inefficient, since it would trace the same values over and over again, in particular as we need to integrate over~$\tilde z$ to obtain a CPR value, i.e.~we always need the current density on the entire interval $\tilde z \in [-1/2, 1/2]$.
Hence, instead of re-doing the calculation for each set of coordinates, we scan the relevant coordinate space $(\tilde z, \delta_0) \in [-1/2, 1/2] \xtimes [-\pi, \pi]$ the first time a set of model parameters is used, overscanning it slightly to avoid edge effects, thus obtaining the complete~$\tilde j(\tilde z, \delta_0)$, compute the integral~$\tilde I(\delta_0)$ and cache the result.
We then use fifth-degree splines in order to extract values of~$\tilde I$ and its derivatives for arbitrary~$\delta_0$ from the cached sampling of~$\tilde I(\delta_0)$, not needing to solve $\mathcal J = 0$ more than once for each set of model parameters.
In detail, starting from $(\tilde z, \delta_0) = (0, 0)$, we scan~$\tilde z$ in both directions in steps of~\num{0.01} and then, for all of these solutions together, scan~$\delta_0$ in both directions in steps of~\num{0.05}.
The step sizes were determined by manual experimentation and allow for decent computation time without a significant reduction in accuracy.

A final wrinkle of our constriction model is that, for a given set of model parameters, the current density may not be well-defined for all sets of coordinates, particularly for large $\epsilon$ and~$\delta_\ell$ which result in regions of multiple $\mathcal J(\tilde j)$~roots, similar to the multi-valued CPR of a homogeneous constriction resulting from $L_{\mathrm{lin}} > L_{\mathrm J0}$.
In these cases, the tracked root of~$\mathcal J(\tilde j)$ may disappear at some~$(\tilde z, \delta_0)$ when leaving a region of multiple roots, leaving only roots that cannot be reached from~$(0, 0)$ in a continuous manner as outlined above.
Our algorithm tries to detect these points by comparing the found root~$\tilde j$ to the expected value, estimated using the gradient of~$\tilde j(\tilde z, \delta_0)$, and discarding it if the discrepancy is larger than~\num{0.01}, which is why not all of the CPRs shown in Fig.~\subref{fig:cpr-response}{d} fill the entire interval~$\delta_0 \in [-\pi, \pi]$.
This detection is not perfect, though, -- especially in areas where the gradient grows very large -- and, together with the unavoidable imprecision associated with numerical computations, can lead to noticeable inaccuracies at the edge of the CPR.
These inaccuracies can often be alleviated by choosing different step sizes for the scanning of the coordinate space (which we do for the CPR curves presented in Fig.~\subref{fig:cpr-response}{d}), but for our data evaluation they are conveniently inconsequential, as the experimentally accessible part of the CPR lies in its center, comfortably far from the numerical edge.
Supplemental Note~\ref{sec:suppl-cpr-parameters}~\cite{suppl-note} provides a detailed discussion of the effect each of the model parameters has on the current density distribution and the resulting CPR, also illustrating the appearance of the numerical edge of the solution.

As a final step in the algorithm, we numerically determine the phase~$\tilde\delta_0$ at which $\tilde I(\delta_0 = \tilde\delta_0) = 0$, thus obtaining the shifted function $\tilde I(\delta_{\mathrm c})$ with the nominal constriction phase $\delta_{\mathrm c} = \delta_0 - \tilde\delta_0$.

\section{Numerical algorithm to fit the flux arcs}\label{sec:meth-arcfit}
We need to implement the function~$\omega_0(\Phi_{\mathrm b})$ using the CPR~$I(\delta_{\mathrm c}) = I_{00} \, \tilde I(\delta_{\mathrm c})$ described in the previous section and fit it to the experimental data, varying the model parameters.
To this end, we numerically invert Eq.~\eqref{eq:flux-relation} to get~$\delta_{\mathrm c}(\Phi_{\mathrm b})$, thus obtaining~$L_{\mathrm c}(\Phi_{\mathrm b})$ and finally~$\omega_0(\Phi_{\mathrm b})$ directly from Eqs.~\eqref{eq:constriction-inductance} and~\eqref{eq:squidres-frequency}.
However, this function only describes a single flux arc as a function of~$\Phi_{\mathrm b}$ while the experimental data consist of multiple flux arcs as a function of the bias coil current~$I_{\mathrm b}$.
Hence, we first need to convert this current into the bias flux
\begin{equation}
  \frac{\Phi_{\mathrm b}}{\Phi_0} = \frac{I_{\mathrm b}}{I_{\mathrm b 0}} - \frac{\Delta\Phi_{\mathrm b}}{\Phi_0}
\end{equation}
with the current~$I_{\mathrm b 0}$ coupling one flux quantum into the SQUID and a flux offset~$\Delta\Phi_{\mathrm b}$, both a priori unknown.
We also need to shift all flux arcs on top of each other to be able to fit them using our~$\omega_0(\Phi_{\mathrm b})$.

To this end, we determine the flux arc each data point belongs to, using the discontinuous jumps to detect where the circuit jumps from one arc to the next/previous one.
Due to the strong distortion of the arcs and the jumps between them at high in-plane fields, including the inversion of the jump direction, detection is more challenging there and some manual intervention is necessary to ensure a flawless separation of the data into individual flux arcs.
Next, we determine the sweetspot~$I_{\mathrm b}$ of each arc by calculating the derivative~$\idell{\omega_0}{I_{\mathrm b}}$ in the vicinity of the arc maximum, fitting a linear function to it and extracting its zero.
This allows us to extract~$I_{\mathrm b 0}$ from the spacing of the sweetspot currents, and to define the shifted flux $\Phi_{\mathrm b}' = \Phi_{\mathrm b} - n \Phi_0$ with the flux arc number~$n$, effectively shifting all arcs on top of the first one.
Finally, we are able to fit~$\omega_0(\Phi_{\mathrm b}')$, varying~$\Delta\Phi_{\mathrm b}$ as well as the model parameters $I_{00}$, $\epsilon$, $b$, $\delta_B$ and~$\delta_\ell$.

However, doing this for each in-plane field separately would not lead to consistent model parameters, most of which should not depend on~$B_\parallel$ or, in the case of~$\delta_B \propto B_\parallel$, do so in a known manner.
To account for this, we concatenate the flux arc data for all in-plane fields and fit them simultaneously, using a single set of the parameters $I_{00}$, $\epsilon$, $b$ and~$\delta_B / B_\parallel$, allowing only $\Delta\Phi_{\mathrm b}$ and~$\delta_\ell$ to vary independently for each~$B_\parallel$.
Allowing the latter two quantities to vary with in-plane field is also physically reasonable, since the flux offset $\Delta\Phi_{\mathrm b}$ will depend on e.g. the exact alignment details between chip surface and in-plane field or single trapped flux quanta in the vicinity of the SQUID, and the specific inductance~$\ell_{\mathrm {lin}}$ is a kinetic inductance, which by nature is sensitive to magnetic fields.

In the simpler case of the CPR given by Eq.~\eqref{eq:cpr-sinlin} and discussed in the context of Fig.~\ref{fig:flux-response}, no such concatenation is necessary, as we consider only a single field~$B_\parallel = 0$.
Additionally, we know the sweetspots to be located at $\Phi_{\mathrm b}' = 0$, so we can determine~$\Delta\Phi_{\mathrm b}$ directly from the sweetspot currents and need only vary~$I_0$ in the fit.

\section{Theory of the Kerr anharmonicity}\label{sec:meth-kerr-theory}
In order to derive the Kerr anharmonicity of the system, we write its total energy as a fourth order Taylor approximation
\begin{equation}
  U_{\mathrm{tot}}(\delta_{\mathrm{ac}}) = \sum_{k = 0}^4 \frac{c_k}{k!} \delta_{\negverythinspace\mathrm{ac}}^k + \Olandau(\delta_{\negverythinspace\mathrm{ac}}^5)
\end{equation}
in terms of some ac~phase difference~$\delta_{\mathrm{ac}}$ from the equilibrium state, where $c_k = \del[k]{\delta_{\mathrm{ac}}} U_{\mathrm{tot}} \evalwith[\big]{\delta_{\mathrm{ac}} = 0}$.
We are interested in the dynamics around an equilibrium state at $\delta_{\mathrm{ac}} = 0$, thus $c_1 = 0$.
Furthermore, we choose $c_0 = 0$ without loss of generality.
The Kerr anharmonicity is then given by~\cite{frattini2018}
\begin{equation}\label{eq:kerr-from-coefficients}
  \mathcal K = \frac{e^2}{2 \hbar C_{\mathrm{tot}}} \frac{c_4}{c_2} .
\end{equation}
Thus, we need to determine the SQUID resonator potential as a function of the total resonator phase~$\delta_{\mathrm{tot}}$, find the equilibrium phase and calculate the relevant derivatives.

\begin{figure}
  \includegraphics{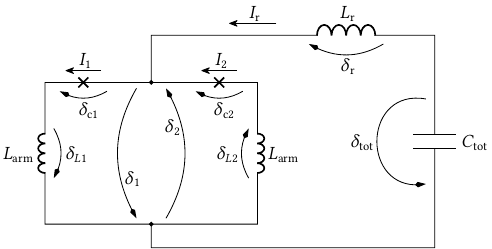}
  \titlecaption{Parameters of a SQUID resonator circuit}{%
    Quantities characterizing the components, including the phase differences across them, are indicated in the diagram.
    Note that the orientation of the currents and phase differences in the SQUID arms is chosen such that a ring current in the SQUID contributes equally to both arms while a bias current through it contributes with opposite signs.
  }
  \label{fig:meth-kerr-circuit}
\end{figure}

We consider a simplified circuit model as shown in Fig.~\ref{fig:meth-kerr-circuit}, where all relevant quantities for the calculation are defined.
Compared with the circuit shown in Fig.~\subref{fig:concept}{d}, we use a more basic representation of the LC-resonator, and we neglect the inductive effect of the bottom arm in Fig.~\subref{fig:flux-response}{a}, using $L_{\mathrm{arm}} = L_{\mathrm{loop}} / 3$ and $L_{\mathrm r} = L_{\mathrm b} - L_{\mathrm{arm}} / 2$.
We assume the junctions to behave identically with respect to current flowing along the SQUID ring, i.e.~$I_1 = I(\delta_{\mathrm c1})$ and $I_2 = I(\delta_{\mathrm c2})$ with the same CPR~$I(\delta)$.
The total circuit energy is
\begin{equation}\label{eq:squidres-Utot}
  U_{\mathrm{tot}} = E_{\mathrm c}(\delta_{\mathrm c1}) + E_{\mathrm c}(\delta_{\mathrm c2}) + \frac{E_{\mathrm{arm}}}{2} (\delta_{L1}^2 + \delta_{L2}^2) + \frac{E_{\mathrm r}}{2} \delta_{\mathrm r}^2
\end{equation}
with the constriction energy
\begin{equation}
  E_{\mathrm c}(\delta) = \frac{\Phi_0}{2 \pi} \int_0^\delta I(\delta') \dif{\delta'}
\end{equation}
and the inductive energies
\begin{align}
  E_{\mathrm{arm}} &= \Bigl( \frac{\Phi_0}{2 \pi} \Bigr)^2 \frac{1}{L_{\mathrm{arm}}} &
  E_{\mathrm r} &= \Bigl( \frac{\Phi_0}{2 \pi} \Bigr)^2 \frac{1}{L_{\mathrm r}} .
\end{align}
Current conservation dictates (cf.~Fig.~\ref{fig:meth-kerr-circuit})
\begin{gather}
  \begin{align}
    \delta_{L1} &= \frac{\Phi_0}{2 \pi} \frac{I(\delta_{\mathrm c1})}{E_{\mathrm{arm}}} &
    \delta_{L2} &= \frac{\Phi_0}{2 \pi} \frac{I(\delta_{\mathrm c2})}{E_{\mathrm{arm}}}
  \end{align} \\
  \delta_{\mathrm r} = \frac{L_{\mathrm r}}{L_{\mathrm{arm}}} (\delta_{L1} - \delta_{L2})
\end{gather}
and additionally the total SQUID flux, comprising the contributions $\delta_1$ and~$\delta_2$ of the two SQUID arms, must compensate the externally applied flux~$\delta_{\mathrm{ext}}$
\begin{equation}
  \delta_1 + \delta_2 + \delta_{\mathrm{ext}} = 0 .
\end{equation}
This means that of the five phases in Eq.~\eqref{eq:squidres-Utot} only one is actually free, or in other words, we can write all of them as a function of the free total phase~$\delta_{\mathrm{tot}}$, and the energy as~$U_{\mathrm{tot}}(\delta_{\mathrm{tot}})$.

Now what remains to be done is to calculate the first four derivatives~$\del[k]{\delta_{\mathrm{tot}}} U_{\mathrm{tot}}$ and evaluate them in the equilibrium state, where no current flows through the resonator inductance, i.e.~only a ring current in the SQUID is present.
This implies $\delta_{\mathrm r} = 0$, $I_1 = I_2$ and thus $\delta_{\mathrm c1} = \delta_{\mathrm c2} \eqcolon \delta_{\mathrm c}$.
Using the dimensionless CPR derivatives (also given in Eq.~\eqref{eq:dimensionless-cpr-derivs} in an alternative form)
\begin{equation}
  g_k
  = \diff[k]{\delta_{L1}}{\delta_{\mathrm c1}} \evalwith{\delta_{\mathrm c1} = \delta_{\mathrm c}}
  = \diff[k]{\delta_{L2}}{\delta_{\mathrm c2}} \evalwith{\delta_{\mathrm c2} = \delta_{\mathrm c}}
  = \frac{\Phi_0}{2 \pi} \frac{1}{E_{\mathrm{arm}}} \diff[k]I\delta(\delta_{\mathrm c})
\end{equation}
and the inductance participation ratio
\begin{equation}
  p
  = \Bigl( 1 + \frac{2 E_{\mathrm{arm}}}{E_{\mathrm r}} \frac{g_1}{1 + g_1} \Bigr)^{-1}
  = \frac{(L_{\mathrm{arm}} + L_{\mathrm c}) / 2}{L_{\mathrm r} + (L_{\mathrm{arm}} + L_{\mathrm c}) / 2}
\end{equation}
we find, after some algebra,
\begin{align}
  c_2 &= 2 p E_{\mathrm{arm}} \frac{g_1}{1 + g_1}
  = \Bigl( \frac{\Phi_0}{2 \pi} \Bigr)^2 \frac{1}{L_{\mathrm r} + (L_{\mathrm{arm}} + L_{\mathrm c}) / 2} \\[1ex]
  c_4 &= -2 p^4 E_{\mathrm{arm}} \frac{3 g_2^2 - g_3 (1 + g_1)}{(1 + g_1)^5}
\end{align}
as well as $c_3 = 0$ due to symmetry.
Finally, plugging these results into Eq.~\eqref{eq:kerr-from-coefficients}, we obtain~$\mathcal K$ as given in Eq.~\eqref{eq:kerr}.

\section{Two-tone circuit response}\label{sec:meth-twotone-response}
\paragraph*{General considerations}
We model the classical intracavity field~$\alpha(t)$ of the SQUID circuits with Kerr nonlinearity and nonlinear damping using the equation of motion~\cite{uhl2023, uhl2024}
\begin{equation}\label{eq:kerr-eom-general}
  \dot\alpha = \Bigl( \i \verythinspace (\negverythinspace \omega_0 + \mathcal K \abs\alpha^2) - \frac{\kappa + \kappa_{\mathrm{nl}} \abs\alpha^2}{2} \Bigr) \, \alpha + \i \sqrt{\frac{\kappa_{\mathrm{ext}}}{2}} S_{\mathrm{in}} .
\end{equation}
Here, $\omega_0$~is the cavity resonance frequency, $\mathcal K$~is the Kerr nonlinearity (frequency shift per photon), $\kappa$~is the bare total linewidth, $\kappa_{\mathrm{nl}}$~is the nonlinear damping constant, $\kappa_{\mathrm{ext}}$~is the external linewidth and $S_{\mathrm{in}}$ is the input field.
The intracavity field is normalized such that $\abs\alpha^2$ corresponds to the intra-cavity photon number and $\abs{S_{\mathrm{in}}}^2$ to the input photon flux on the coplanar waveguide feedline.

The solution of this equation of motion significantly depends on the pump power and on the number of tones sent to the cavity.
However, given this solution~$\alpha$, the circuit output field will always be given by~\cite{gardiner1985, chen2022b}
\begin{align}
  S_{\mathrm{out}} &= S_{\mathrm{in}} + \i \sqrt{\frac{\kappa_{\mathrm{ext}}}{2}} \, \alpha \\
\shortintertext{and the ideal transmission response function thus by}
  S_{21}^{\mathrm{ideal}} &= 1 + \i \sqrt{\frac{\kappa_{\mathrm{ext}}}{2}} \frac{\alpha}{S_{\mathrm{in}}} .
\end{align}

\paragraph*{The nonlinear single-tone regime}
We start by setting the input field to a single pump-tone $S_{\mathrm{in, st}} = S_{\mathrm p} \e^{\i \omega_{\mathrm p} t}$ with the frequency~$\omega_{\mathrm p}$ and a complex-valued amplitude~$S_{\mathrm p}$.
For the intracavity field, we make the ansatz $\alpha = \alpha_{\mathrm p} \e^{\i \omega_{\mathrm p} t}$ with real-valued~$\alpha_{\mathrm p}$; the phase delay between input and response is encoded in~$\arg(S_{\mathrm p})$.
The equation of motion now reads
\begin{equation}\label{eq:kerr-eom-singletone}
  \i \omega_{\mathrm p} \alpha_{\mathrm p} = \Bigl( \i \verythinspace (\negverythinspace \omega_0 + \mathcal K \alpha_{\mathrm p}^2) - \frac{\kappa + \kappa_{\mathrm{nl}} \alpha_{\mathrm p}^2}{2} \Bigr) \, \alpha_{\mathrm p} + \i \sqrt{\frac{\kappa_{\mathrm{ext}}}{2}} S_{\mathrm p} .
\end{equation}
Introducing the intracircuit pump photon number~$n_{\mathrm c} = \alpha_{\mathrm p}^2$, the pump photon flux~$n_{\mathrm p} = \abs{S_{\mathrm p}}^2 = P_{\mathrm p} / \hbar \omega_{\mathrm p}$ and the detuning $\Delta_{\mathrm p} = \omega_{\mathrm p} - \omega_0$ between the pump and the bare cavity resonance, we obtain
\begin{equation}\label{eq:kerr-eom-singletone-rearranged}
  \Bigl( \Delta_{\mathrm p} - \mathcal K n_{\mathrm c} - \i \frac{\kappa + \kappa_{\mathrm{nl}} n_{\mathrm c}}{2} \Bigr) \, \alpha_{\mathrm p} = \sqrt{\frac{\kappa_{\mathrm{ext}}}{2}} S_{\mathrm p}
\end{equation}
or, taking the absolute square on both sides,
\begin{equation}\label{eq:kerr-characteristic-polynomial}
  \biggl( (\Delta_{\mathrm p} - \mathcal K n_{\mathrm c})^2 + \Bigl( \frac{\kappa + \kappa_{\mathrm{nl}} n_{\mathrm c}}{2} \Bigr)^2 \biggr) \, n_{\mathrm c} = \frac{\kappa_{\mathrm{ext}}}{2} n_{\mathrm p} .
\end{equation}
Finding the real-valued roots of this characteristic third-degree polynomial yields the physical solutions for the amplitude~$\alpha_{\mathrm p}$; in the case of three real-valued roots the highest and lowest amplitudes are the stable states, though there is only one such root for our $\mathcal K < 0$ and $\Delta_{\mathrm p} > 0$.
The phase delay between input and response can be read from Eq.~\eqref{eq:kerr-eom-singletone-rearranged} and is
\begin{equation}
  \arg(S_{\mathrm p}) = \atanii\Bigl( - \frac{\kappa + \kappa_{\mathrm{nl}} n_{\mathrm c}}{2}, \Delta_{\mathrm p} - \mathcal K n_{\mathrm c} \Bigr) .
\end{equation}

Having determined the complete complex field solution, we can calculate the transmission
\begin{equation}
  S_{21, \mathrm{st}}^{\mathrm{ideal}} = 1 + \i \sqrt{\frac{\kappa_{\mathrm{ext}}}{2}} \frac{\alpha_{\mathrm p}}{S_{\mathrm p}} ,
\end{equation}
which for low pump powers (using $\mathcal K = \kappa_{\mathrm{nl}} = 0$) corresponds to Eq.~\eqref{eq:S21-ideal}.
Note that we do not use these equations for any data analysis in this manuscript, but we include them as a prerequisite for the calculation concerning the two-tone regime.

\paragraph*{The linearized two-tone regime}
In addition to the strong pump-tone at frequency~$\omega_{\mathrm p}$ with fixed power $P_{\mathrm p} = n_{\mathrm p} \hbar \omega_{\mathrm p}$, we now insert a weak probe signal at the scanning frequency~$\omega_{\mathrm s}$.
We write the total input as $S_{\mathrm{in, tt}} = S_{\mathrm p} \e^{\i \omega_{\mathrm p} t} + S_{\mathrm s} \e^{\i \omega_{\mathrm s} t}$ with another complex field amplitude~$S_{\mathrm s}$.
For the intracavity field, we use the ansatz $\alpha = \smash{ \bigl( \alpha_{\mathrm p} + \alpha_{\mathrm s}(t) \bigr) \e^{\i \omega_{\mathrm p} t} }$ with a complex and time-dependent~$\alpha_{\mathrm s}(t)$ and obtain the equation of motion
\begin{equation}\label{eq:kerr-eom-twotone}
  \scalemuskips{.5}
  \mathclap{
  \begin{aligned}[b]
    \i \omega_{\mathrm p} (\alpha_{\mathrm p} &+ \alpha_{\mathrm s}) + \dot\alpha_{\mathrm s} \\
    = {}& \i \Bigl( \omega_0 + \mathcal K \bigl( \alpha_{\mathrm p}^2 + \alpha_{\mathrm p} (\alpha_{\mathrm s} + \alpha_{\mathrm s}^\ast) + \abs{\alpha_{\mathrm s}}^2 \bigr) \! \Bigr) (\alpha_{\mathrm p} + \alpha_{\mathrm s}) \\
        &- \Bigl( \frac\kappa 2 + \frac{\kappa_{\mathrm{nl}}}{2} \bigl( \alpha_{\mathrm p}^2 + \alpha_{\mathrm p} (\alpha_{\mathrm s} + \alpha_{\mathrm s}^\ast) + \abs{\alpha_{\mathrm s}}^2 \bigr) \! \Bigr) (\alpha_{\mathrm p} + \alpha_{\mathrm s}) \\
      &+ \i \sqrt{\frac{\kappa_{\mathrm{ext}}}{2}} (S_{\mathrm p} + S_{\mathrm s} \e^{\i \Omega_{\mathrm s} t})
  \end{aligned}
  }
\end{equation}
where we introduced the probe frequency with respect to the pump~$\Omega_{\mathrm s} = \omega_{\mathrm s} - \omega_{\mathrm p}$.

We now perform the linearization by dropping all terms not linear or constant in the small quantity~$\alpha_{\mathrm s}$ and get
\begin{equation}\label{eq:kerr-eom-twotone-linearized}
  \mathclap{
  \begin{aligned}[b]
    \i \omega_{\mathrm p} (\alpha_{\mathrm p} &+ \alpha_{\mathrm s}) + \dot\alpha_{\mathrm s} \\
    = {}& \Bigl( \i \verythinspace (\negverythinspace \omega_0 + \mathcal K \alpha_{\mathrm p}^2) - \frac{\kappa + \kappa_{\mathrm{nl}} \alpha_{\mathrm p}^2}{2} \Bigr) (\alpha_{\mathrm p} + \alpha_{\mathrm s}) \\
      &+ \Bigl( \i \mathcal K - \frac{\kappa_{\mathrm{nl}}}{2} \Bigr) \, (\alpha_{\mathrm s} + \alpha_{\mathrm s}^\ast) \alpha_{\mathrm p}^2 \\
      &+ \i \sqrt{\frac{\kappa_{\mathrm{ext}}}{2}} (S_{\mathrm p} + S_{\mathrm s} \e^{\i \Omega_{\mathrm s} t}) .
  \end{aligned}
  }
\end{equation}
The time-independent terms are identical to Eq.~\eqref{eq:kerr-eom-singletone} of the single-tone case and allow us to determine $\alpha_{\mathrm p}$ and~$n_{\mathrm c}$ as before.
Fourier transforming the remaining terms with the transform variable~$\Omega$ (frequency relative to the pump frequency) yields
\begin{equation}
  \frac{\hat\alpha_{\mathrm s}}{\chi_\alpha}
  = \Bigl( \i \mathcal K - \frac{\kappa_{\mathrm{nl}}}{2} \Bigr) n_{\mathrm c} \hat\alpha_{\mathrm s}^\rats + \i \sqrt{\frac{\kappa_{\mathrm{ext}}}{2}} \hat S_{\mathrm s}
\end{equation}
where $\hat\alpha_{\mathrm s}(\Omega)$ is the Fourier transform of~$\alpha_{\mathrm s}(t)$, $\hat S_{\mathrm s}(\Omega)$ is a Dirac~peak at $\Omega = \Omega_{\mathrm s}$ and we introduced the counter-rotating complex conjugate $\hat\alpha_{\mathrm s}^\rats(\Omega) = \hat\alpha_{\mathrm s}^\ast(-\Omega)$, such that $\hat\alpha_{\mathrm s}^\rats(\Omega)$ is the Fourier transform of~$\alpha_{\mathrm s}^\ast(t)$.
Furthermore, we defined
\begin{equation}\label{eq:kerr-twotone-chipr}
  \chi_\alpha(\Omega) = \Bigl( \i \verythinspace (\Delta_{\mathrm p} - 2 \mathcal K n_{\mathrm c} + \Omega) + \frac{\kappa + 2 \kappa_{\mathrm{nl}} n_{\mathrm c}}{2} \Bigr)^{-1} .
\end{equation}
Substituting the equivalent relation for~$\hat\alpha_{\mathrm s}^\rats$, we find
\begingroup\belowdisplayskip=\abovedisplayshortskip
\begin{equation}\label{eq:kerr-twotone-solution}
  \hat\alpha_{\mathrm s}
  = \i \chi_{\mathrm s} \sqrt{\frac{\kappa_{\mathrm{ext}}}{2}} \biggl(
      \hat S_{\mathrm s}
      - \Bigl( \i \mathcal K - \frac{\kappa_{\mathrm{nl}}}{2} \Bigr) n_{\mathrm c} \chi_\alpha^\rats \hat S_{\mathrm s}^\rats
    \biggr)
\end{equation}
\endgroup
with
\begin{equation}\label{eq:kerr-twotone-chig}
  \chi_{\mathrm s} = \frac{\chi_\alpha}{ 1 - (\mathcal K^2 + \kappa_{\mathrm{nl}}^2 / 4) n_{\mathrm c}^2 \chi_\alpha^{} \chi_\alpha^\rats} .
\end{equation}
The VNA measures only the circuit response at $\Omega = \Omega_{\mathrm s}$, so we need not consider the second, counter-rotating term in Eq.~\eqref{eq:kerr-twotone-solution} and finally find the transmission response at the probe frequency
\begin{equation}
  S_{21, \mathrm{tt}}^{\mathrm{ideal}} = 1 - \frac{\kappa_{\mathrm{ext}}}{2} \chi_{\mathrm s} .
\end{equation}

\paragraph*{The pumped Kerr modes}
To find the resonance frequencies of the susceptibility~$\chi_{\mathrm s}$, we determine the complex frequencies~$\tilde\Omega_{0 \mathrm p}$ for which $1 / \chi_{\mathrm s} = 0$.
The real part of such a solution is the resonance frequency $\Omega_{0 \mathrm p} = \Re(\tilde\Omega_{0 \mathrm p}) = \omega_{0 \mathrm p} - \omega_{\mathrm p}$ while the imaginary part corresponds to half the mode linewidth $\kappa_{\mathrm p} = 2 \Im(\tilde\Omega_{0 \mathrm p})$.
Setting the denominator from Eq.~\eqref{eq:kerr-twotone-chig} to zero leads to
\begin{equation}
  \scalemuskips{.8}
  \tilde\Omega_{0 \mathrm p}^\pm \! = \i \frac{\kappa + 2 \kappa_{\mathrm{nl}} n_c}{2} \pm \sqrt{ (\Delta_{\mathrm p} - \mathcal K n_{\mathrm c}) (\Delta_{\mathrm p} - 3 \mathcal K n_{\mathrm c}) - \frac{\kappa_{\mathrm{nl}}^2 n_{\mathrm c}^2}{4} } .
\end{equation}
The radicand is always positive for our experimental parameters, so the system has two resonances
\begin{equation}
  \omega_{0 \mathrm p}^\pm = \omega_{\mathrm p} \pm \sqrt{ (\Delta_{\mathrm p} - \mathcal K n_{\mathrm c}) (\Delta_{\mathrm p} - 3 \mathcal K n_{\mathrm c}) - \frac{\kappa_{\mathrm{nl}}^2 n_{\mathrm c}^2}{4} }
\end{equation}
split symmetrically around the pump frequency, though we observe only~$\omega_{0 \mathrm p} = \omega_{0 \mathrm p}^-$ due to our experimental parameters.

In the experiment, we measure the shift of this mode with respect to the un-pumped resonance frequency $\delta\negverythinspace\omega_0 = \omega_{0 \mathrm p} - \omega_0$ which is given by
\begin{equation}\label{eq:kerr-frequency-shift}
  \domgzero = \Delta_{\mathrm p} - \sqrt{ (\Delta_{\mathrm p} - \mathcal K n_{\mathrm c}) (\Delta_{\mathrm p} - 3 \mathcal K n_{\mathrm c}) - \frac{\kappa_{\mathrm{nl}}^2 n_{\mathrm c}^2}{4} } .
\end{equation}
Besides the shifted resonance frequency, we also extract the pumped linewidth from the transmission data
\begin{equation}
  \kappa_{\mathrm p} = \kappa + 2 \kappa_{\mathrm{nl}} n_{\mathrm c}
\end{equation}
and so the only unknown value remaining in Eq.~\eqref{eq:kerr-frequency-shift} is the product~$\mathcal K n_{\mathrm c}$.
Given~$n_{\mathrm c}$, we can thus determine~$\mathcal K$ by fitting that relation to the experimental data.

\paragraph*{Calculating the photon number}
We now solve Eq.~\eqref{eq:kerr-frequency-shift} for $\Delta_{\mathrm p} - \mathcal K n_{\mathrm c}$, recalling that $\Omega_{0 \mathrm p} = \omega_{0 \mathrm p} - \omega_{\mathrm p} = \domgzero - \Delta_{\mathrm p}$, and get
\begin{equation}
  \Delta_{\mathrm p} - \mathcal K n_{\mathrm c} = \frac 13 \biggl( \Delta_{\mathrm p} \pm \sqrt{ \Delta_{\mathrm p}^2 + 3 \Bigl( \frac{\kappa_{\mathrm{nl}}^2 n_{\mathrm c}^2}{4} + \Omega_{0 \mathrm p}^2 \Bigr) } \, \biggr).
\end{equation}
Note that only the positive solution is relevant for our purposes.
In the experiment, we measure or set all of the values on the right-hand side of this equation, i.e.~we have direct experimental access to $\tilde\Delta \coloneq \Delta_{\mathrm p} - \mathcal K n_{\mathrm c}$.
The characteristic polynomial Eq.~\eqref{eq:kerr-characteristic-polynomial} contains the same quantity and so we can extract~$n_{\mathrm c}$ from it using
\begin{equation}
  n_{\mathrm c} = \frac{2 P_{\mathrm p}}{\hbar \omega_{\mathrm p}} \frac{\kappa_{\mathrm{ext}}}{(\kappa + \kappa_{\mathrm{nl}} n_{\mathrm c})^2 + \tilde\Delta^2}.
\end{equation}
Again, all values on the right-hand side are directly accessible in the experiment, except that the pump power~$P_{\mathrm{SG}}$ is set at the signal generator and reaches the chip attenuated by some unknown factor $P_{\mathrm p} = \zeta P_{\mathrm{SG}}$.
We can, however, determine~$n_{\mathrm c} / \zeta$ from the experimental data and, since only the product~$\mathcal K n_{\mathrm c}$ appears in Eq.~\eqref{eq:kerr-frequency-shift}, use that value to find~$\zeta \mathcal K$ from a fit to the experimental $\domgzero$~data.
By comparing the result with the $\mathcal K$\nobreakdash-values predicted by the CPR model we can later extract a value for~$\zeta$.

\section{Numerical algorithm to fit \texorpdfstring{$\mathcal K$}{K} with modified CPRs}\label{sec:meth-kerrfit}
In a first step, we determine~$\zeta \mathcal K$ for each two-tone measurement as outlined above by extracting $\domgzero$ and~$n_{\mathrm c} / \zeta$ from the measurement data for each power and then fitting the function $\domgzero(n_{\mathrm c} / \zeta)$ given by Eq.~\eqref{eq:kerr-frequency-shift} to those data under variation of~$\zeta \mathcal K$.
From the CPRs resulting from the flux arc fits, we calculate the theoretically predicted~$\mathcal K_{\mathrm{theo}}$ and by dividing the two values obtain $\zeta \sim \num{4e-5}$, corresponding to \qty{44}{\deci\bel} of attenuation, consistent with the expected attenuation in our setup (\qty{30}{\deci\bel} of explicit attenuation, cf.~Fig.~\subref{fig:concept}{f} or Supplemental Fig.~\ref{fig:suppl-setup}~\cite{suppl-note}, and \qty{\sim 14}{\deci\bel} from cables and connectors).

Since the Kerr anharmonicity depends on the second- and third-order derivatives of the CPR, it is extremely sensitive to small errors in the theoretical CPR and indeed the experimental data match the model prediction only roughly, cf.~Supplemental Note~\ref{sec:suppl-poly-correction}~\cite{suppl-note}.
As a consequence, we do not get a constant~$\zeta$ for all flux bias points and magnetic in-plane fields, though the qualitative features of the Kerr response to bias flux response are reproduced nicely by the model.
Possible reasons include imperfections in the model (e.g.~due to the assumed linear gradients of $j_0$ and~$B_\parallel l_{\mathrm{eff}}$ in the junction) as well as as possible pump-frequency dependence of the attenuation~$\zeta$.
To account for these unknowns and demonstrate that a good match between theory and data can be achieved using imperceptible changes to the CPR, we modify the CPR to $I(\delta_{\mathrm c}) + \Delta I(\delta_{\mathrm c})$ using a small polynomial correction
\begin{equation}
  \Delta I = J \, (q_1 \tilde\delta_{\mathrm c} + q_3 \tilde\delta_{\mathrm c}^3 + q_8 \tilde\delta_{\mathrm c}^8 + q_9 \tilde\delta_{\mathrm c}^9)
\end{equation}
where $J = \max(\del{\delta_{\mathrm c}} I) = \del{\delta_{\mathrm c}} I(\delta_{\mathrm c 0})$ is the maximum slope of the original CPR (corresponding to the flux arc sweetspot) and $\tilde\delta_{\mathrm c} = \delta_c - \delta_{\mathrm c 0}$ is the constriction phase relative to that point.
The inclusion of~$J$ serves to keep the free parameters in an intermediate value range and to thus avoid possible numerical problems when varying them in a fitting routine, and it also eases interpretation of the~$q_k$ by making them encode the relative change to the junction inductance, rather than an absolute CPR offset.
Additionally, we consider a~$\zeta$ that is linear in pump frequency
\begin{equation}
  \zeta(\omega_{\mathrm p}) = \zeta_0 \bigl( 1 + \zeta_1 \, (\omega_{\mathrm p} - \omega_{\mathrm p}^{\mathrm{ref}}) \bigr)
\end{equation}
with a constant $\smash{\omega_{\mathrm p}^{\mathrm{ref}}} = 2 \pi \times \qty{10.25}{\giga\hertz}$ which is chosen near the pump frequency used at the zero-field arc sweetspot, such that~$\zeta_0$ is approximately the attenuation at that point.
This is useful both for an intuitive interpretation of the fit parameters and for the fit routine itself, as changes in~$\zeta_1$ do not cause large changes in the absolute value of~$\zeta$ in the relevant $\omega_{\mathrm p}$~region, only changing its slope.

Now, we fit the theoretical values to the measured Kerr anharmonicity and to the flux arcs simultaneously, varying $\zeta_0$ and~$\zeta_1$ (affecting the experimental~$\mathcal K(\Phi_{\mathrm b})$) as well as, individually for each field, $q_1$, $q_3$, $q_8$ and~$q_9$ (affecting~$\mathcal K_{\mathrm{theo}}(\Phi_{\mathrm b})$ and, to a lesser extent, the flux-arc fits~$\omega_0(\Phi_{\mathrm b})$).
Since there are many more flux-arc points than two-tone measurements, and their numerical value is several orders of magnitude larger, we need to weight the flux arc residuals in the fit such that they do not overwhelm the Kerr data (or vice versa).
We find that a relative weight of~\num{e-4} strikes a nice balance between the two datasets, except for the highest fields, where we use slightly larger weights (\num{2.5e-4} at~\qty{275}{\milli\tesla} and \num{5e-4} at~\qty{300}{\milli\tesla}).
The discrepancy between the measured and the predicted Kerr response is largest near the edge of the flux arc, where the uncertainty of the experimental values is also greatest, cf.~the error bars in Fig.~\subref{fig:kerr-response}{d} or Supplemental Fig.~\subref{fig:suppl-poly-correction}{c}.
We take this fact into account by weighting the Kerr residuals in the fit with the factor $\smash{(\sigmaK^{\mathrm{min}} / \sigmaK)^{\frac 13}}$, where $\sigmaK$ is the fit uncertainty reported by the $\zeta \mathcal K$~fit used to determine the value in question and $\sigmaK^{\mathrm{min}}$ is its minimum value at the same in-plane field.
The exponent~$\frac 13$ was again determined by experimentation and found to strike a good balance, reducing the impact of the uncertain points without neglecting them completely.

Varying all of the fit parameters at the same time does not result in good fit convergence, so we employ a multi-step process.
In a first step, we set $\zeta_1 = 0$ and vary the other parameters (including $\zeta_0$) individually for each~$B_\parallel$.
Using starting values for $\zeta_0$ and~$\zeta_1$ that result in $\zeta(\omega_{\mathrm p})$ on the order of those just found, we vary only those two parameters in a second fit for all fields simultaneously, keeping the polynomial coefficients~$q_k$ constant at the values from the first fit.
Finally, in the third step, we vary the coefficients~$q_k$ once more, at each field individually, keeping $\zeta_0$ and~$\zeta_1$ constant at the values from the previous fit.

\section*{Data availability}
All data presented in this paper and the Supplemental Material, including raw data, as well as the corresponding processing scripts will be made publicly available on the repository Zenodo with the identifier~\doi{10.5281/zenodo.17093328}.

\section*{Author contributions}
\initials{BW}~conducted the experiments, performed data processing and analysis, implemented the algorithms for fitting and numerical solutions, prepared the figures, wrote the first draft of the manuscript and contributed to sample design and theory development.
\initials{MK}~contributed to the experiments.
\initials{TK}~designed and fabricated the device and contributed to the experimental setup.
\initials{KU}~contributed to design and installation of the vector magnet and to sample fabrication.
\initials{CF}~contributed to design and installation of the vector magnet.
\initials{DK}~and \initials{RK} contributed to funding acquisition and participated in scientific discussions.
\initials{DB}~conceived the experiment, supervised all parts of the project, performed funding acquisition, developed the theoretical framework and wrote the first draft of the manuscript.
All authors discussed the results and conclusions, and contributed to manuscript revisions.

\section*{Competing interests}
The authors declare no competing interests.

\end{mainsetup}

\begin{supplsetup}
\title{Supplemental Material for:\\Magnetically induced Josephson nano-diodes in\\field-resilient superconducting microwave circuits}

\author{Benedikt~Wilde}
\author{Mohamad~Kazouini}
\author{Timo~Kern}
\author{Kevin~Uhl}
\author{Christoph~Füger}
\author{Dieter~Koelle}
\author{Reinhold~Kleiner}
\author{Daniel~Bothner}

\maketitle

\tableofcontents
\clearpage

\nocite{uhl2024}

\section{Complete experimental setup}\label{sec:suppl-setup}

\begin{figure}[b]
  \begin{minipage}[b]{.6\textwidth}
    \titlecaption{Schematic of the experimental setup}{%
      The superconducting chip is mounted inside a radiation-tight copper sample holder, which in turn is placed inside the vacuum compartment of a liquid helium~(LHE) cryostat.
      A temperature-sensing diode mounted directly on the copper sample box, in combination with a resistive heater placed close by in the dipstick assembly, allows temperature control with a stability $\Delta T_{\mathrm s} < \qty{1}{\milli\kelvin}$ using PID~feedback control provided by a temperature controller.
      Pumping on the helium gas allows setting the measurement temperature to $T_{\mathrm s} = \qty{2.8}{\kelvin}$.
      Coaxial microwave cables connect the vector network analyzer~(VNA) as well as, via a directional coupler, a signal generator~(SG) to the sample, with both devices sharing the \qty{10}{\mega\hertz}~reference-clock signal provided by the SG.
      The input lines both of VNA and of~SG are attenuated by~\qty{30}{\deci\bel} at cryogenic temperatures to reduce the input noise to approximately the noise level at the measurement temperature.
      The output line is equipped with two high-electron-mobility-transistor~(HEMT) amplifiers for maximized signal-to-noise ratio, one at cryogenic temperatures and one at room temperature.
      To prevent the amplifiers from saturating, the pump signal is inserted into the device at the probe signal output and an attenuator is placed at the input of the room-temperature amplifier.
      Another attenuator is placed at the SG output to shift its power output range, with the power~$P_{\mathrm{SG}}$ of the pump signal with frequency~$\omega_{\mathrm p}$ defined as the shifted power.
      Numbers near the attenuators, the directional coupler and the amplifiers are given in units of~\unit{\deci\bel}.
      For the application of the magnetic in-plane field, a large cylindrical superconducting coil is wrapped directly around the cup of the vacuum compartment, connected in series with a split-coil magnet that is rigidly attached to the cryostat and used to compensate out-of-plane components of the main coil field~$B_\parallel$.
      Part of the sample assembly is clipped away in the diagram to reveal the small magnetic sample coil mounted behind it.
      Two low-noise dc~current sources provide the driving currents $I_{\mathrm b}$ for the sample coil and~$I_{\mathrm{coil}}$ for the in-plane field coils, and $I_{\mathrm b}$ is additionally low-pass filtered at cryogenic temperatures with a cutoff frequency of~\qty{\sim 3}{\kilo\hertz}.
      The entire cryostat is surrounded by a double-layer mu-metal shield for magnetic screening.
      \label{fig:suppl-setup}%
    }
  \end{minipage}%
  \hfill
  \begin{minipage}[b]{.4\textwidth}\raggedleft
    \includegraphics{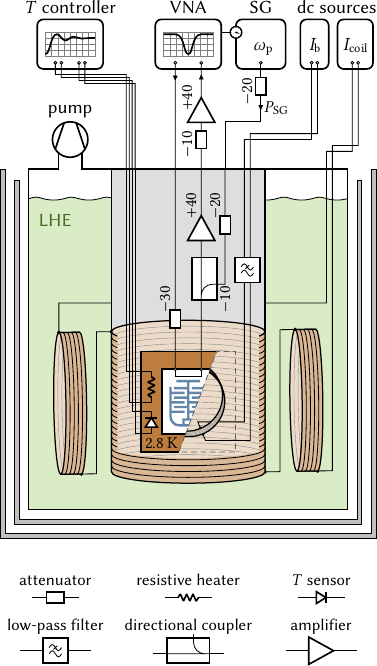}
  \end{minipage}
\end{figure}

\begin{spacerfloat}[b]
  \rule{0pt}{.5cm}
\end{spacerfloat}

A schematic of the complete experimental setup is shown in Supplemental Fig.~\ref{fig:suppl-setup} and described there.
We use in-house-built, low-noise, battery-powered dc~current sources to drive the different magnetic coils in the setup.
Driving the compensation coils in series with the main coil ensures that any current drifts or noise contributions from the current source are present equally in all of them, thus not affecting the compensation, i.e.~the orientation of the total field.
To minimize the coupling of external fields as well as wire cross-talk, all dc~wires inside the vacuum compartment are installed as twisted pairs and those outside the cryostat are shielded.

The microwave cables in the cryostat dipstick are coaxial stainless steel cables for low thermal conductivity.
At our measurement frequencies of~$\qty{\sim 10}{\giga\hertz}$ we expect~\qty{\sim 8}{\deci\bel} of attenuation from the \qty{\sim 1.5}{\meter}~stainless-steel cable and~\qty{\sim 3}{\deci\bel} from the \qty{2}{\meter}~room-temperature cable in each input/output line.
Additionally attenuating the input signals by~\qty{30}{\deci\bel} at cryogenic temperatures ensures that room-temperature noise from the vector network analyzer~(VNA) and the signal generator~(SG) is reduced to a level close to that of the sample temperature $T_{\mathrm s} = \qty{2.8}{\kelvin}$.

Since the temperature-sensing diode used to control~$T_{\mathrm s}$ is placed directly on the copper sample box for optimal thermal contact, it is also exposed to the strong magnetic field~$B_\parallel$ applied to the sample, affecting its temperature response and thus the temperature measured by the $T$~controller.
Even though the orientation of the diode in the field is chosen to minimize this effect, it is still considerable, reducing the temperature reading at the measurement temperature by more than~\qty{0.5}{\kelvin} at $B_\parallel = \qty{300}{\milli\tesla}$ compared to the zero-field value.
In order to keep the temperature constant despite this shift, we set the temperature to~\qty{2.8}{\kelvin} at zero field and interrupt the PID~feedback temperature control while ramping the coil current~$I_{\mathrm{coil}}$, instead keeping the power supplied to the heating resistor constant.
After the target current has been reached, we use the shifted temperature reading as the new temperature set-point and resume the PID~control.
We verify the integrity of this process using a second temperature diode placed outside the large $B_\parallel$~coil in the sample assembly, which does not experience as high fields and is thus less affected, as well as by checking the set-point that results from it after ramping the field back to zero.
Typically, the discrepancy from the original set-point after a full measurement day comprising several field-ramps is only a few~\unit{\milli\kelvin}, mostly driven by the fact that the heating power of the $T$~controller cannot be set with the same precision as that used by the device during PID~control.

\clearpage

\vspace*{0pt plus 0fil}

\section{Full chip layout}\label{sec:suppl-chip-layout}

\begin{figure}
  \includegraphics{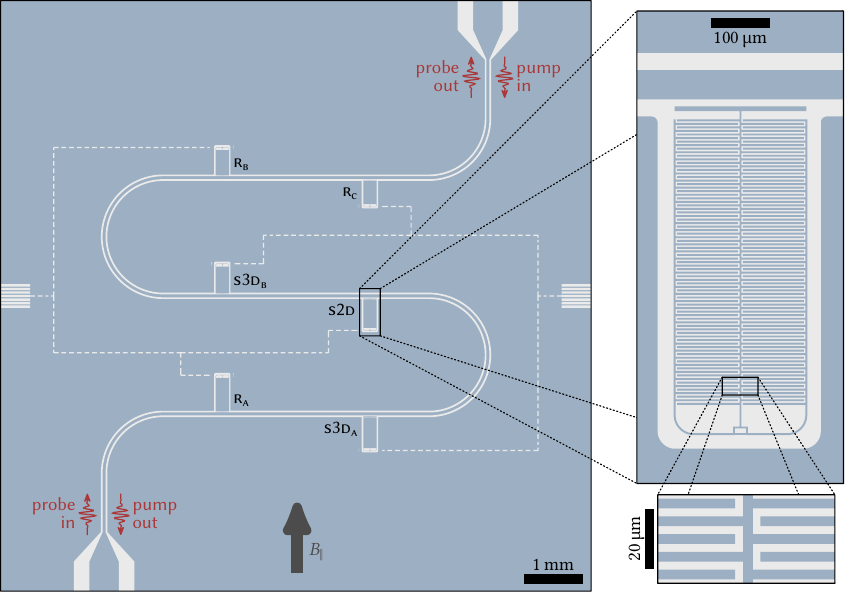}
  \titlecaption{Full chip layout}{%
    The microwave is guided through a meandering coplanar waveguide~(CPW) alongside which six side-coupled LC-resonators are placed.
    Each resonator consists of a capacitive plate for coupling to the CPW, interdigital capacitors~(IDCs) and inductive wires, all arranged symmetrically around the central axis.
    At the core of each resonator is a loop that can be turned into a SQUID acting as a flux-tunable inductance by cutting constrictions into its two arms using a neon ion microscope~(NIM), cf.~Fig.~\ref{fig:concept}.
    This is done for the resonators \res 2, \res 3 and~\res 4, while \res 1, \res 5 and~\res 6 remain un-altered and serve as reference resonators.
    \res 3~is the resonator discussed in the main text, data for~\res 4 are shown in Supplemental Note~\ref{sec:suppl-res4-data}.
    The dashed lines leading to the resonators from the chip edges are markers used for focusing, orientation and accurate movement in the~NIM, such that unnecessary irradiation of circuit structures can be avoided.
  }
  \label{fig:suppl-chip-layout}
\end{figure}

The full chip layout is depicted in Supplemental Fig.~\ref{fig:suppl-chip-layout}.
The six resonator designs differ solely in the number~$N_{\mathrm{idc}}$ of capacitor fingers in each IDC.
Additional differences are introduced during fabrication, namely through cutting constriction-type Josephson junctions into the loop at the center of some resonators, turning them into SQUID resonators, as well as by integrating a mechanical beam into the long side of the loop for some resonators.
The parameters used for each resonator are summarized in Supplemental Table~\ref{tab:suppl-resonator-parameters}.
Note that all \ce{Nb}~structures on the chip are~\qty{\sim 1}{\micro\meter} smaller than in the layout depicted in Supplemental Fig.~\ref{fig:suppl-chip-layout} due to over-exposure of the photo-resist during fabrication, e.g.~the interdigital capacitor~(IDC) lines are~\qty{2}{\micro\meter} wide with gaps of~\qty{4}{\micro\meter} between them.
The true, fabricated device dimensions were extracted from scanning electron microscope~(SEM) images and used in all simulations for determining device parameters.

In the next section, we will use the reference resonators to infer certain parameters of the SQUID resonators that we do not have direct access to.
Since the reference resonators \res 1 and~\res 5 are closest to the SQUID resonators in resonance frequency (cf.~Supplemental Table~\ref{tab:suppl-resonator-parameters}), we only use those two for that purpose, not considering~\res 6 as a reference.

\begin{table}[p]
  \titlecaption{Key resonator parameters}{%
    We show the number~$N_{\mathrm{idc}}$ of capacitor fingers, whether the resonator SQUID loop contains a mechanical beam, as well as the neon-ion acceleration voltage~$V_{\mathrm{acc}}$ and beam doses $D_1$ for cutting the constrictions and $D_2$ for thinning them down.
    In addition to these fabrication parameters, we show the resonance frequencies $\omega_{0 \mathrm b}$ and~$\omega_0$ before and after cutting the constrictions, respectively, the resonance linewidth~$\kappa$, the inductances of the resonator before cutting the junctions~$L_{\mathrm b}$, of the constrictions~$L_{\mathrm c}$ and of the SQUID loop~$L_{\mathrm{loop}}$, as well as the total inductance~$L$ and capacitance~$C_{\mathrm{tot}}$ of the resonator.
    For the SQUID resonators, the bias-flux dependent values $\omega_0$, $\kappa$ and~$L_{\mathrm c}$ are given at the sweetspot.
    The dimensions used for simulating the loop inductances~$L_{\mathrm{loop}}$ (as well as~$L_{\mathrm b}$) were extracted from SEM images of the fabricated circuits, which is why their values vary slightly despite the SQUID ring layouts being identical.
    \label{tab:suppl-resonator-parameters}%
  }
  \sisetup{per-mode=symbol}
  \def\extraspace{\hspace{.8em}}
  \begin{tabular}[b]{l !{\extraspace} S[table-format=2] c S[table-format=2] S[table-format=5] S[table-format=4] !{\extraspace} *{2}{S[table-format=2.3]} S[table-format=2] S[table-format=3] *{2}{S[table-format=2]} S[table-format=3] S[table-format=3]}\toprule
    {} & {$N_{\mathrm{idc}}$} & mech. & {$V_{\mathrm{acc}}$} & {$D_1$} & {$D_2$} & {$\omega_{0 \mathrm b} / 2 \pi$} & {$\omega_0 / 2 \pi$} & {$\kappa / 2 \pi$} & {$L_{\mathrm b}$} & {$L_{\mathrm c}$} & {$L_{\mathrm{loop}}$} & {$L$} & {$C_{\mathrm{tot}}$} \\
    {} & {} & {} & {in \unit{\kilo\volt}} & \multicolumn{2}{c !{\extraspace}}{in \unit{\ions\per\square\nano\meter}} & {in \unit{\giga\hertz}} & {in \unit{\giga\hertz}} & {in \unit{\mega\hertz}} & {in \unit{\pico\henry}} & {in \unit{\pico\henry}} & {in \unit{\pico\henry}} & {in \unit{\pico\henry}} & {in \unit{\femto\farad}} \\\midrule
    \res 1 & 94 & no & {--} & {--} & {--} & 8.937 & 8.937 & 8 & 441 & {--} & {--} & 441 & 719 \\
    \res 2 & 87 & yes & 25 & 18000 & 1000 & 9.654 & 9.475 & 28 & 418 & 32 & 45 & 433 & 651 \\
    \res 3 & 81 & no & 25 & 18000 & 0 & 10.380 & 10.233 & 22 & 397 & 23 & 44 & 409 & 592 \\
    \res 4 & 75 & yes & 25 & 18000 & 2000 & 10.983 & 10.774 & 44 & 377 & 29 & 45 & 392 & 557 \\
    \res 5 & 70 & no & {--} & {--} & {--} & 11.641 & 11.641 & 11 & 360 & {--} & {--} & 360 & 519 \\
    \res 6 & 65 & yes & {--} & {--} & {--} & 12.532 & 12.532 & 10 & 343 & {--} & {--} & 343 & 470 \\\bottomrule
  \end{tabular}
\end{table}

\begin{figure}[p]
  \includegraphics{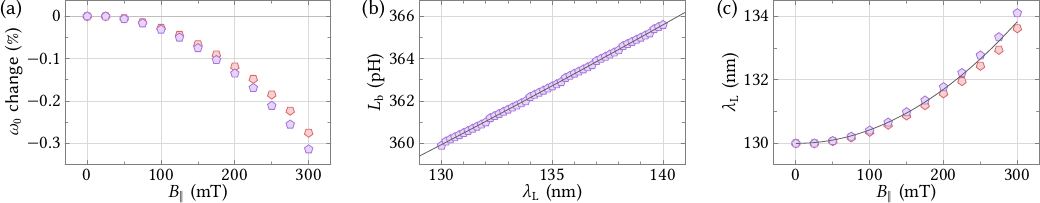}
  \titlecaption{Magnetic in-plane field response of the superconductor penetration depth}{%
    Symbols are data for the two relevant reference resonators \plotref{plt:res1}~\res 1 and~\plotref{plt:res5}~\res 5.
    \sublabel{a}~Relative change of the resonance frequency~$\omega_0$ with~$B_\parallel$.
    \sublabel{b}~Simulated resonator inductance~$L_{\mathrm b}(\lambda_{\mathrm L})$ without constrictions, shown here for \res 5.
    Symbols are simulated data points, the gray line is a fit.
    \sublabel{c}~$\lambda_{\mathrm L}(B_\parallel)$ inferred from the simulation results together with the resonance frequency shift.
    The gray line is a fit to the data of both resonators.
  }
  \label{fig:suppl-lmbd-calibration}
\end{figure}

\begin{figure}[p]
  \includegraphics{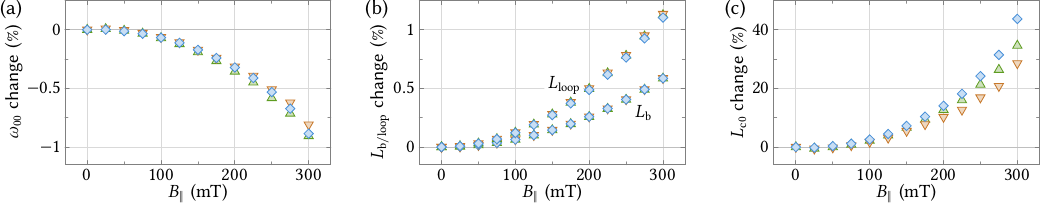}
  \titlecaption{Magnetic in-plane field response of the SQUID~resonator inductances}{%
    Symbols are data for the three SQUID~resonators \plotref{plt:res2}~\res 2, \plotref{plt:res3}~\res 3 and~\plotref{plt:res4}~\res 4.
    \sublabel{a}~Relative change of the sweetspot resonance frequency~$\omega_{00}$ with~$B_\parallel$.
    \sublabel{b}~Relative change of the inductance~$L_{\mathrm b}$ of the SQUID resonators and of their SQUID loop inductances~$L_{\mathrm{loop}}$ as calculated from the simulated $\lambda_{\mathrm L}$-dependence together with~$\lambda_{\mathrm L}(B_\parallel)$ (cf.~Supplemental Fig.~\subref{fig:suppl-lmbd-calibration}{c}).
    Since all values are calculated using the same~$\lambda_{\mathrm L}(B_\parallel)$, they are almost identical and the symbols mostly cover each other.
    Small visible deviations in~$L_{\mathrm{loop}}$ are due to the simulated dimensions being extracted from SEM images of the SQUID rings, thus carrying small differences between the circuits over to the simulation.
    \sublabel{c}~Relative change of the constriction inductance~$L_{\mathrm c0}$ at the flux arc sweetspot for the SQUID resonators as inferred from the resonance frequency shifts in combination with~$L_{\mathrm b}(B_\parallel)$.
  }
  \label{fig:suppl-sweetspot-L}
\end{figure}

\clearpage

\section{Determination of the resonator parameters}\label{sec:suppl-resonator-params}
The resonance frequency of our resonators is given by
\begin{equation}\label{eq:resonance-frequency}
  \omega_0 = \frac{1}{\sqrt{C_{\mathrm{tot}} L}}
\end{equation}
with the total capacitance~$C_{\mathrm{tot}}$ and inductance~$L$ of the resonator.
For the SQUID resonators, $L = L_{\mathrm b}$ before cutting the constrictions and $L = L_{\mathrm b} + L_{\mathrm c} / 2$ after, cf.~the main manuscript.
For the reference resonators there are no constrictions, so $L = L_{\mathrm b}$ always.

Applying an in-plane magnetic field~$B_\parallel$ partly suppresses the superconducting state of the resonator material, increasing its penetration depth~$\lambda_{\mathrm L}$ and thus its kinetic inductance, causing a downwards shift in resonance frequency, as shown in Supplemental Fig.~\subref{fig:suppl-lmbd-calibration}{a} for the reference resonators and in Supplemental Fig.~\subref{fig:suppl-sweetspot-L}{a} for the SQUID resonators.
Assuming the resonator capacitance to remain unaffected by the applied field, we obtain the in-plane-field dependence of the resonator inductance
\begin{equation}\label{eq:L-B-dependence}
  L(B_\parallel) = \Bigl( \frac{\omega_0(0)}{\omega_0(B_\parallel)} \Bigr)^2 \, L(0) .
\end{equation}

In order to obtain the field-dependent penetration depth~$\lambda_{\mathrm L}(B_\parallel)$, we simulate the inductance~$L_{\mathrm b}$ of each circuit geometry using the software \software{3D-MLSI}~\cite{khapaev2001a}, varying the penetration depth, and, in order to remove simulation output resolution artifacts as well as for convenient interpolation, fit the results using~\cite{gubin2005}
\begin{equation}
  L_{\mathrm b}(\lambda_{\mathrm L}) = L_{\mathrm b}^{\mathrm{geo}} + \lambda_{\mathrm L} L_{\mathrm b}^\ast \coth(d / \lambda_{\mathrm L})
\end{equation}
with the film thickness~$d = \qty{100}{\nano\meter}$ and fit parameters $L_{\mathrm b}^{\mathrm{geo}}$ and~$L_{\mathrm b}^\ast$, as shown in Supplemental Fig.~\subref{fig:suppl-lmbd-calibration}{b}.
Assuming a zero-field value $\lambda_{\mathrm L}(0) = \qty{130}{\nano\meter}$ typical for our films at low temperatures $T \lesssim \qty{4}{\kelvin}$, we find~$L_{\mathrm b}(0)$ for all circuits from these simulations.
For the reference circuits, we thus obtain~$L_{\mathrm b}(B_\parallel)$ using Eq.~\eqref{eq:L-B-dependence} and, again from the simulation data, $\lambda_{\mathrm L}(B_\parallel)$.
The result of this calculation is shown in Supplemental Fig.~\subref{fig:suppl-lmbd-calibration}{c}.
Due to the inaccuracies involved, the resulting values are not exactly identical for both resonators.
We fit the expected behavior $\lambda_{\mathrm L}(B_\parallel) = \bigl( 1 + (B_\parallel / B^\ast)^2 \bigr) \lambda_{\mathrm L}(0)$~\cite{healey2008} to these data for \res 1 and \res 5 and use the resulting fit values for subsequent calculations.
The fit parameter~$B^\ast$ introduced here can be interpreted as a characteristic magnetic field, but has no deeper meaning for the further analysis.

Now that we have obtained~$\lambda_{\mathrm L}(B_\parallel)$, we can calculate the field-dependent inductance~$L_{\mathrm b}(B_\parallel)$ of the SQUID resonators as well as their loop inductances~$L_{\mathrm{loop}}(B_\parallel)$ from the simulated $L_{\mathrm b}(\lambda_{\mathrm L})$ and~$L_{\mathrm{loop}}(\lambda_{\mathrm L})$.
The resulting field-induced change of both inductances is shown in Supplemental Fig.~\subref{fig:suppl-sweetspot-L}{b}.

For the SQUID resonators, we also need to consider the constriction inductance~$L_{\mathrm c}$ that contributes to the total inductance.
Indeed, comparing Supplemental Figs.~\subref{fig:suppl-lmbd-calibration}{a} and~\subref{fig:suppl-sweetspot-L}{a} reveals that the sweetspot resonance frequency~$\omega_{00}$ of the SQUID resonators reacts considerably stronger to~$B_\parallel$ than the reference resonator frequency~$\omega_0$, highlighting the additional effect of the sweetspot constriction inductance~$L_{\mathrm c0}$.
Using the resonance frequency~$\omega_{0 \mathrm b}$ of the SQUID circuits before cutting the junctions (i.e.~without~$L_{\mathrm c}$), we find
\begin{gather}
  \omega_{00}
  = \frac{1}{\sqrt{ C_{\mathrm{tot}} (L_{\mathrm b} + L_{\mathrm c 0} / 2) }}
  = \frac{\omega_{0 \mathrm b}}{\sqrt{ 1 + L_{\mathrm c 0} / 2 L_{\mathrm b} }} \\[1ex]
  \mathllap{\implies\qquad}
  L_{\mathrm c 0}(B_\parallel)
  = 2 L_{\mathrm b}(B_\parallel) \, \biggl( \! \Bigl( \frac{\omega_{0 \mathrm b}(B_\parallel)}{\omega_{00}(B_\parallel)} \Bigr)^2 - 1 \biggr) .
\end{gather}

Before constriction cutting, the resonators were not characterized at the measurement temperature~\qty{2.8}{\kelvin} of the main experiment.
Hence, we have to infer~$\omega_{0 \mathrm b}$ at that temperature from other data.
Assuming that the capacitance~$C_{\mathrm{tot}}$ remains unaffected by temperature and that any resonance frequency shift is due to a change in~$L$ caused by a changing~$\lambda_{\mathrm L}$ (as for the field-induced shift discussed above), the ratio $\omega_{0 \mathrm b} / \omega_0^{\res 1}$ must be constant with temperature for each SQUID resonator.
Thus, using the measured ratio at~\qty{4.2}{\kelvin} as well as $\omega_0^{\res 1}(0)$ at~\qty{2.8}{\kelvin}, we can calculate~$\omega_{0 \mathrm b}(0)$.
To improve accuracy, we do this both using \res 1 and using~\res 5 as a reference, obtaining values differing by~\qty{\sim 0.05}{\percent}, and use the average.
Analogously to Eq.~\eqref{eq:L-B-dependence}, we now obtain the field dependence of this frequency
\begin{equation}
  \omega_{0 \mathrm b}(B_\parallel) = \omega_{0 \mathrm b}(0) \sqrt{\frac{L_{\mathrm b}(0)}{L_{\mathrm b}(B_\parallel)}} ,
\end{equation}
finally allowing us to calculate~$L_{\mathrm c 0}(B_\parallel)$.
The resulting field-induced change is depicted in Supplemental Fig.~\subref{fig:suppl-sweetspot-L}{c}.

Having determined all resonator inductances, we can calculate the total capacitance of each circuit from Eq.~\eqref{eq:resonance-frequency}.
All of the zero-field parameter values resulting from this procedure are summarized in Supplemental Table~\ref{tab:suppl-resonator-parameters}.

\clearpage

\section{The resonator linewidth}\label{sec:suppl-linewidth}

\begin{figure}[b]
  \begin{minipage}[b]{.5\textwidth}
    \rlap{\adjincludegraphics[trim={{\width-2\textwidth} 1.9pt 0pt 0pt}]{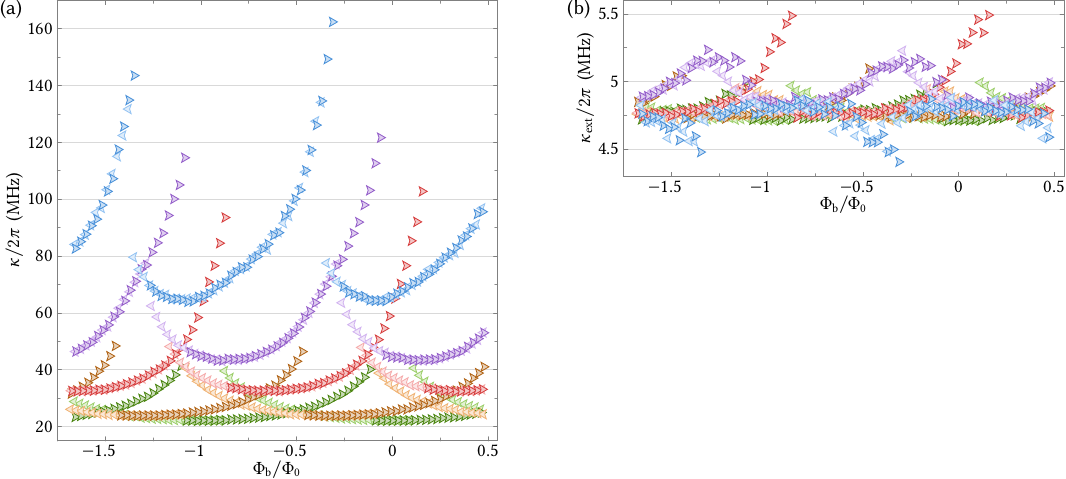}}
  \end{minipage}\hfill
  \begin{minipage}[b]{.46\textwidth}
    \titlecaption{Linewidth flux-response in large magnetic in-plane fields}{%
      \sublabel{a}~Total linewidth~$\kappa(\Phi_{\mathrm{b}})$ and \sublabel{b}~external linewidth~$\kappa_{\mathrm{ext}}(\Phi_{\mathrm b})$ corresponding to the frequency flux-response data~$\omega_0(\Phi_{\mathrm b})$ presented in Fig.~\subref{fig:field-response}{a}.
      Again, the in-plane field values are $B_\parallel \in \qtyset{0; 100; 200; 250; 300}{\milli\tesla}$ (from green to blue) and the pointing direction of the kite symbols indicates the flux-sweep direction.
      We observe a linewidth behavior matching that of the resonance frequencies, with $\kappa$ increasing when $\omega_0$ decreases and a corresponding skewing of the flux arcs appearing.
      $\kappa_{\mathrm{ext}}$ remains largely unaffected by the field, only showing a small, field-dependent bias-flux response away from the sweetspot.
      \label{fig:suppl-flux-linewidth}%
    }
  \end{minipage}
\end{figure}

\begin{spacerfloat}[b]
  \rule{0pt}{6cm}
\end{spacerfloat}

For completeness, we present the linewidths of resonator~\res 3 here, which we did not discuss in detail in the main manuscript.
As already mentioned there, the resonator is in the undercoupled regime, with an external linewidth $\kappa_{\mathrm{ext}} \sim 2\pi\times\qty{5}{\mega\hertz}$ and a total linewidth $\kappa > 4 \kappa_{\mathrm{ext}}$ dominated by the internal linewidth.
Supplemental Fig.~\ref{fig:suppl-flux-linewidth} summarizes the linewidth flux-response~$\kappa(\Phi_{\mathrm b})$ and~$\kappa_{\mathrm{ext}}(\Phi_{\mathrm b})$ for different~$B_\parallel$.
Each data point was extracted from the same fit as the corresponding resonance frequency point in Fig.~\subref{fig:field-response}{a}.
As expected, the linewidth (panel~\subref{fig:suppl-flux-linewidth}*{a}) reacts to both flux and field in an inverse fashion compared to the resonance frequency, increasing where $\omega_0$ decreases, but with a similar shift and skewing present.
The magnetic field suppresses the superconducting energy gap, leading to an increased number of quasiparticles and thus to higher losses, particularly in the constrictions (cf.~Fig.~\subref{fig:field-response}{c}).
These losses contribute to the internal linewidth while leaving the external linewidth (panel~\subref{fig:suppl-flux-linewidth}*{b}) largely unaffected, which only shows a small field-dependent bias-flux response (most likely caused by a frequency-dependence of~$\kappa_{\mathrm{ext}}$ or a misattribution of~$\kappa_{\mathrm{int}}$-contributions to~$\kappa_{\mathrm{ext}}$ by the fit, not by an actual field-dependence).
Bias-flux induced losses (caused by the circulating current) are also present in the internal linewidth, of course, and they are enhanced by the applied~$B_\parallel$, with both the sweetspot linewidth and the linewidth tuning increasing by a similar value, at $B_\parallel = \qty{300}{\milli\tesla}$ by~\qty{\sim 50}{\mega\hertz}.
We note that in principle the apparent linewidth may also be broadened by flux-noise effects when the circuit is biased away from the sweetspot, though experience with similar devices leads us to expect that this is not the case here~\cite{kazouini2026}.

\clearpage

\section{The effect of each CPR model parameter}\label{sec:suppl-cpr-parameters}
To illustrate the effect of the different CPR model parameters on the current density distribution inside the junction as well as the resulting CPR, we will now have a look at the solution obtained using the CPR algorithm described in Appendix~\ref{sec:meth-cpr-implementation} for different sets of model parameters.
Throughout, we will use the normalized parameters $(\epsilon, b, \delta_B, \delta_\ell)$ and the corresponding normalized current density~$\tilde j(\tilde z, \delta_0)$ and normalized CPR~$\tilde I(\delta_0)$.
Note that we consider $\tilde j$ and~$\tilde I$ as functions of the phase~$\delta_0$ in the center of the junction, not of the shifted phase $\delta_{\mathrm c} = \delta_0 - \tilde\delta_0$ which is used elsewhere.
This simplifies the interpretation of the plots shown, as one fewer step is involved in the calculation, as well as revealing the value of~$\tilde\delta_0$ as the zero of~$\tilde I(\delta_0)$ near the origin in the CPR plots.
For reference, let us repeat Eqs.~\eqref{eq:current-density-normalized} and~\eqref{eq:cpr-integral-normalized} from Appendix~\ref{sec:meth-cpr-implementation} here
\begin{gather}
  \label{eq:repeated-current-density-normalized}
  \tag{\ref*{eq:current-density-normalized}}
  \tilde j(\tilde z, \delta_0) = (1 - 2 \epsilon \tilde z) \sin\bigl( \delta_0 + \delta_B (1 + b \tilde z) \tilde z - \delta_\ell \tilde j(\tilde z, \delta_0) \bigr) \\
  \label{eq:repeated-cpr-integral-normalized}
  \tag{\ref*{eq:cpr-integral-normalized}}
  \tilde I(\delta_0) = \int_{-1/2}^{1/2} \tilde j(\tilde z) \dif{\tilde z} .
\end{gather}

\begin{figure}
  \includegraphics{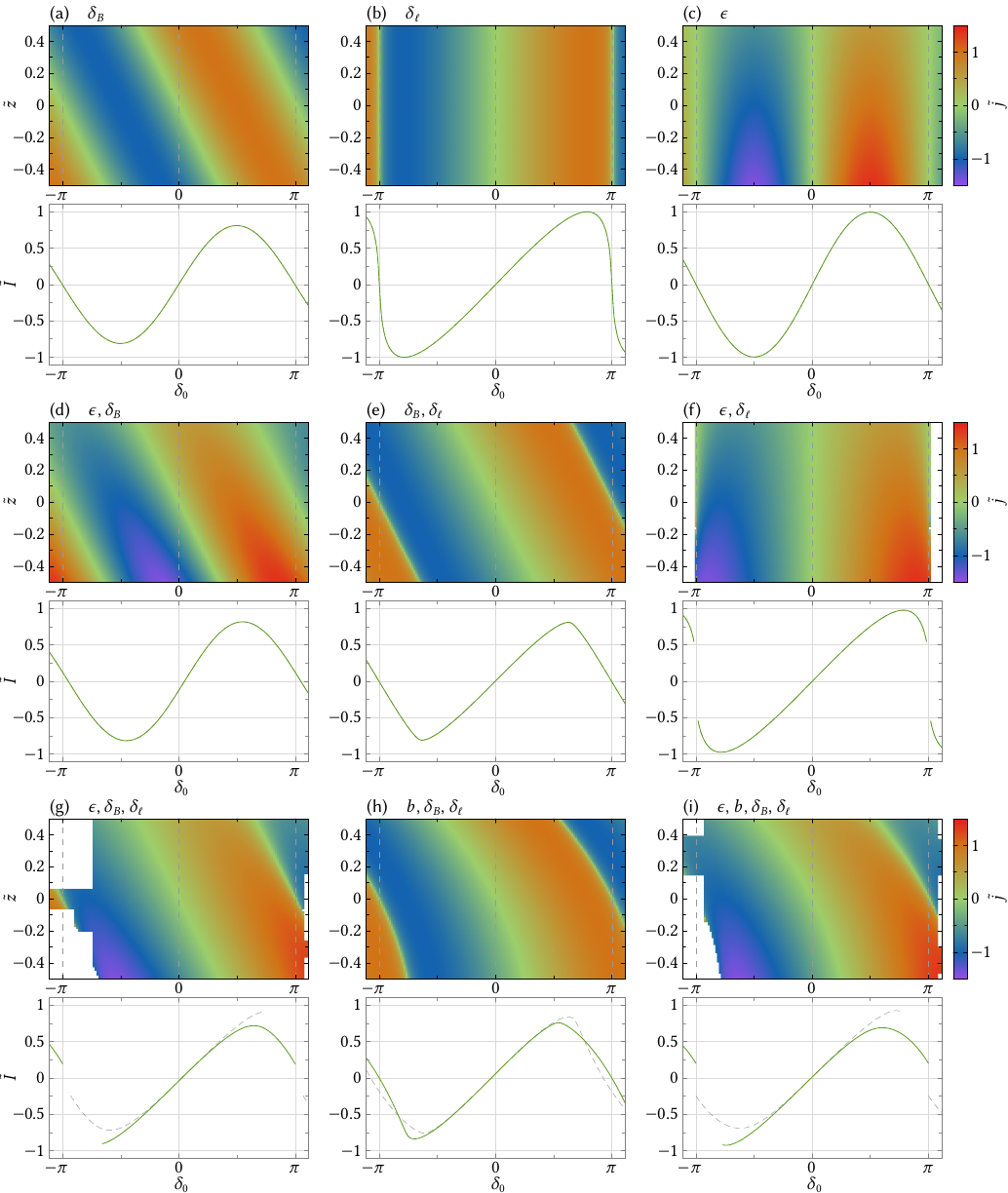}
  \titlecaption{Effect of the CPR model parameters on the current density distribution and the resulting CPR}{%
    Each panel comprises a color plot of~$\tilde j(\tilde z, \delta_0)$ at the top and a plot of the CPR~$\tilde I(\delta_0)$ at the bottom, gray dashed curves correspond to the CPR mirrored about its zero.
    The parameters that are used (i.e.~not set to zero) for each panel are mentioned at the top.
    When mentioned, the parameter values $\epsilon = \num{0.4}$, $b = \num{0.6}$, $\delta_B = \num{2.2}$ and $\delta_\ell = \num{0.9}$ are used.
    Each plot has the same axis limits and color map as other plots showing the same quantity, and both the current density and the CPR plots share the same $x$-axis such that each $\tilde I$\nobreakdash-value corresponds to the integral of the $\tilde j$ values vertically above it.
    For a detailed description of the effect of each parameter, see Supplemental Note~\ref{sec:suppl-cpr-parameters}.
  }
  \label{fig:suppl-cpr-parameters}
\end{figure}

In Supplemental Fig.~\ref{fig:suppl-cpr-parameters}, both of these functions are plotted for various sets of model parameter values, with each parameter being either zero or taking its designated value $\epsilon = \num{0.4}$, $b = \num{0.6}$, $\delta_B = \num{2.2}$ and $\delta_\ell = \num{0.9}$.
Each panel of the figure comprises two plots, a $j(\tilde z, \delta_0)$~color plot at the top and an $\tilde I(\delta_0)$~plot at the bottom.
They share the same $x$-axis such that each CPR value in the bottom panel is the integral of the current density values vertically above it.
The color map used for the $\tilde j$~plots reaches its most orange/blue color for $\tilde j = \pm 1$, changing towards red/violet only for $\abs{\tilde j} > 1$.
The trivial solutions $\smash{\tilde j\bigl( \tilde z, n \pi - \delta_B (1 + b \tilde z) \tilde z \bigr)} = 0$ with $n \in \mathbb Z$ are visible as green stripes in the color plots (when not obscured by a non-trivial solution, e.g.~at low~$\tilde z$ in panel~\subref{fig:suppl-cpr-parameters}*{f}), with the center of the plot marking the solution $\tilde j(0, 0) = 0$ that is used as starting point in the algorithm, cf.~Appendix~\ref{sec:meth-cpr-implementation}.

In the first three panels (first row) we explore the effect of just one of the three model contributions; the magnetic field~($\delta_B$, for now with $b = 0$), the linear inductance~($\delta_\ell$) and the critical current density gradient~($\epsilon$).
If all model parameters were zero (not shown), the current density would be described by a simple sine $\tilde j = \sin\delta_0$ independent of~$\tilde z$, i.e.~the color plot would be vertically invariant and the CPR thus identical~$\tilde I = \tilde j$.
The effect of~$\delta_B$ (panel~\subref{fig:suppl-cpr-parameters}*{a}) is to tilt the $\tilde z$\nobreakdash-dependence of the current density, with the slope of the tilt (green stripes in the plot) given by~$\delta_B$.
In the integral, this leads to a sine with a reduced amplitude $\sin(\delta_B / 2) / (\delta_B / 2)$ giving rise to the well-known Fraunhofer pattern of critical currents.
The depicted $\delta_B$-value corresponds to a junction with $B_\parallel / B_0 = \delta_B / 2 \pi = \num{0.35}$ of a flux quantum coupled into it by the in-plane field.
We already discussed the effect of a linear inductance contribution~$\delta_\ell$ (panel~\subref{fig:suppl-cpr-parameters}*{b}) in the context of Fig.~\ref{fig:flux-response}; it leads to a skewed sine profile with unchanged critical currents, maintaining~$\tilde I = \tilde j$.
The depicted $\delta_\ell$-value corresponds to a linear inductance contribution $L_{\mathrm{lin}} / L_{\mathrm J0} = \delta_\ell = \num{0.9}$.
Finally, the effect of~$\epsilon$ (panel~\subref{fig:suppl-cpr-parameters}*{c}) is a $\tilde z$\nobreakdash-dependent scaling factor, with the amplitude of the sine profile being larger than~\num{1} for negative~$\tilde z$ and smaller for positive~$\tilde z$.
In the integral, this linear gradient drops out and the CPR remains unaffected.

\looseness=-1
The next three panels (second row) depict the consequence of using a combination of two of these three parameters.
In the current density plots, the result is simply a combination of the individual effects of the involved parameters; a $\tilde z$~tilt for~$\delta_B$, skewed $\delta_0$\nobreakdash-dependence for~$\delta_\ell$ and a $\tilde z$\nobreakdash-dependence of the amplitude for~$\epsilon$.
In the CPR, on the other hand, the result is less obvious.

Including $\epsilon$ and~$\delta_B$ (panel~\subref{fig:suppl-cpr-parameters}*{d}) yields a CPR that is still a sine with reduced amplitude, but with its zero~$\tilde\delta_0$ shifted away from $\delta_0 = 0$.
This behavior applies more generally; $\delta_B$ together with a $\tilde z$\nobreakdash-gradient leads to a horizontally shifted~$\tilde I(\delta_0)$.
We stress that this CPR is still completely symmetrical around its zero, the shift only indicates that $\delta_0 \neq 0$ for $\tilde I = 0$.
Looking at the current density plot, we can now understand why the CPR does not become zero when $\delta_B = 2 \pi$ (i.e.~$B_\parallel = B_0$); in this case, the $\tilde z$~tilt reaches exactly one period over the height of the junction, i.e.~the sine~$\tilde j(\delta_0)$ is shifted half a period to the left at $\tilde z = 1 / 2$ and half a period to the right at $\tilde z = -1 / 2$.
Thus, calculating~$I(\delta_0)$ we integrate over all phases of the sine, but in the presence of a non-zero~$\epsilon$ the amplitude is not constant for all~$\tilde z$ and thus the integral is not zero.

In the CPR for $\delta_B$ together with~$\delta_\ell$ (panel~\subref{fig:suppl-cpr-parameters}*{e}), we see a distorted version of the reduced-amplitude CPR caused by~$\delta_B$, but unlike before, due to the $\tilde z$~tilt in~$\tilde j$, the skewing introduced by~$\delta_\ell$ affects each part of the CPR in a non-trivial way.

Finally, the combination of $\epsilon$ and~$\delta_\ell$ (panel~\subref{fig:suppl-cpr-parameters}*{f}) produces a CPR similar to the skewed sine created by just~$\delta_\ell$, but not quite identical.
The gradient in critical current density enters into the inductive term in Eq.~\eqref{eq:repeated-current-density-normalized} and thus changes the shape of the $\delta_0$\nobreakdash-dependence of~$\tilde j$ besides just its amplitude.
In particular, for negative~$\tilde z$, where the critical current density and thus the inductive contribution are amplified, $\tilde j(\delta_0)$ may cease being single-valued, similar as for $\delta_\ell > 1$, cf.~Fig.~\subref{fig:flux-response}{d}.
In this case, the root that is tracked by the CPR algorithm disappears upon exiting the region of multiple solutions.
Since this happens only locally, unlike in the case without~$\epsilon$, the dynamics in the junction are no longer obvious.
As mentioned in Appendix~\ref{sec:meth-cpr-implementation}, the algorithm discards the $\tilde j$~solution in this case, which happens for the first time at the transition between a unique $\tilde j(\delta_0)$ (large~$\tilde z$) and a multi-valued relation (small~$\tilde z$), marked by a missing pixel in the current density plot near $(\tilde z, \delta_0) \approx (\num{-0.16}, \pm \pi)$.
Due to the way the algorithm is implemented, it may not find solutions for any~$\tilde z$ (or large parts of the scanning interval) once any part of the solution has been discarded, which is why $\tilde j(\tilde z, \delta_0)$ is missing for all~$\tilde z$ after the missing pixel.
As the CPR~$\tilde I(\delta_0)$ is undefined once the current density~$\tilde j(\tilde z, \delta_0)$ is undefined for any~$\tilde z$, this is of no consequence to the final CPR solution, though, which is also illustrated by the CPR plot stopping at the last value $\abs{\delta_0} = \num{3.10}$ where the $\tilde j$~solution exists for all~$\tilde z$.

\looseness=-1
Now, we combine all three parameters, obtaining a current density distribution featuring $\tilde z$~tilt, skewed $\delta_0$\nobreakdash-dependence as well as a $\tilde z$\nobreakdash-dependent amplitude (panel~\subref{fig:suppl-cpr-parameters}*{g}).
This finally leads to new behavior in the CPR, namely to an asymmetric profile with different positive and negative critical currents~$I_0^\pm$; a Josephson diode.
A gray dashed curve marks the CPR mirrored about its zero, making any asymmetries apparent.
The tilt introduced by~$\epsilon$ also tilts the locus of the disappearing tracked root and thus the shape of the edge of the solution found by the algorithm, as can be seen at the bottom left edge of the current density plot.
For larger magnetic fields (larger~$\delta_B$), the slope of the tilt will increase and the solution will disappear closer to $\delta_0 = 0$, different model parameters, namely large~$\delta_\ell > 1$, may also cause it to disappear earlier for positive~$\delta_0$.
We note that the algorithm does find solutions for $\delta_0 > \pi$ in this example that could in principle be used to calculate the CPR in a small interval above $\delta_0 = (2 n + 1) \pi$.
This is not implemented, though, as it is not relevant for the evaluation at hand, and so the CPR stops at $\delta_0 = \pi$.

Using a gradient in the magnetic contribution given by~$b$ instead of the critical current density gradient given by~$\epsilon$ (panel~\subref{fig:suppl-cpr-parameters}*{h}) also leads to an asymmetric CPR, albeit with a less pronounced asymmetry, particularly in the region of small~$\delta_0$ most relevant in the experiment.
In the current density plot, $b$~modifies the effect of~$\delta_B$ from a linear tilt of the $\tilde z$\nobreakdash-dependence to a parabolic one (cf.~again green stripes in the $\tilde j$~plots).
Combining both gradients, i.e.~using all four model parameters (panel~\subref{fig:suppl-cpr-parameters}*{i}), leads to a combined diode effect featuring both asymmetry contributions.
Note that the shift~$\tilde\delta_0$ of the CPR zero induced by~$b$ (in combination with~$\delta_B$) opposes that induced by~$\epsilon$, so the total shift is rather small here.
This is the form of the CPR model that was used in evaluating the data presented in the main manuscript, obtaining the CPRs shown in Fig.~\subref{fig:cpr-response}{d}.
Note that a large~$\delta_\ell > 1$ leads to the solution disappearing earlier for positive values of~$\delta_{\mathrm c}$ and later for negative values, which is the case for the CPRs at high fields shown there, cf.~$\delta_\ell$\nobreakdash-values in Supplemental Fig.~\ref{fig:suppl-fit-inductance}.
For the \res 4~data presented in Supplemental Note~\ref{sec:suppl-res4-data}, the magnetic contribution gradient was not used, i.e.~$b = 0$ (panel~\subref{fig:suppl-cpr-parameters}*{g}).

We note for completeness that using both gradients without the linear inductance contribution (i.e.~using $\epsilon$, $b$ and~$\delta_B$) does not lead to a diode CPR.
The resulting current density is similar to that shown in panel~\subref{fig:suppl-cpr-parameters}*{d} but with a parabolic tilt rather than a linear one, and the resulting CPR is also alike, with the only notable difference being the additional shift to the left due to~$b$.

\section{Polynomial correction of the CPR}\label{sec:suppl-poly-correction}

\begin{figure}
  \includegraphics{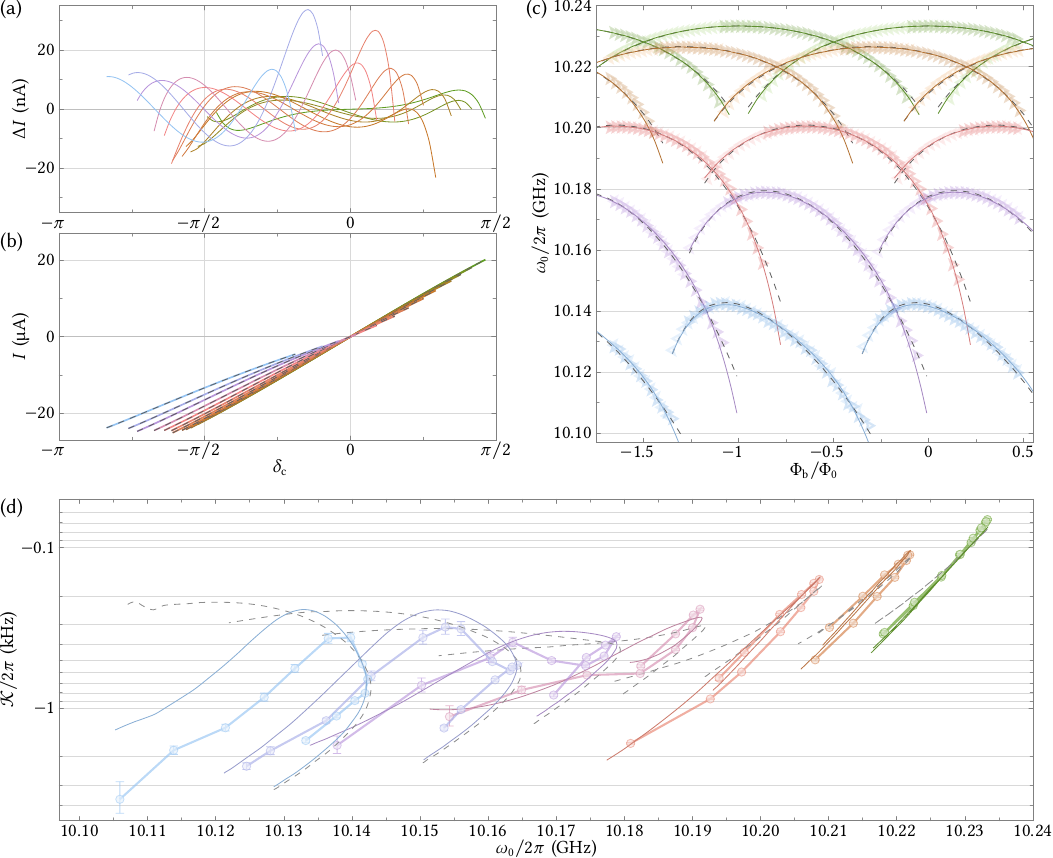}
  \titlecaption{An imperceptibly small polynomial CPR correction considerably improves the Kerr fit}{%
    \sublabel{a}~Polynomial correction $\Delta I(\delta_{\mathrm c})$ used to modify the CPR for the Kerr analysis, with each curve from $B_\parallel = 0$~(green) to \qty{300}{\milli\tesla}~(blue) corresponding to the line of the same color in Fig.~\subref{fig:cpr-response}{d}, though here we only show the part corresponding to the thick segment there, i.e.~the part that corresponds to the experimental flux arcs, since this is the part of the correction that participates in the fits.
    Inside this region, its value stays within a few tens of~\unit{\nano\ampere} for all fields, about three orders of magnitude below the critical current (though the function diverges outside the fit range, as any polynomial fit does).
    \sublabel{b}~Corresponding current-phase relations with and without the polynomial correction; dashed gray lines correspond to the CPR model fits~$I(\delta_{\mathrm c})$ and solid, colored lines show the modified CPRs~$I(\delta_{\mathrm c}) + \Delta I(\delta_{\mathrm c})$.
    Due to the tininess of the modifications, no difference between the curves is visible, the dashed lines seem to lie exactly on top of the colored lines.
    \sublabel{c}~Flux-response fits~$\omega_0(\Phi_{\mathrm b})$ with and without the polynomial correction.
    Kite symbols in the background show the measurement data also presented in Fig.~\subref{fig:field-response}{a}, dashed gray lines correspond to the CPR model fits (using~$I(\delta_{\mathrm c})$) to the data that are also shown there and solid, colored lines are the fits using the modified CPR~$I(\delta_{\mathrm c}) + \Delta I(\delta_{\mathrm c})$.
    As in Fig.~\subref{fig:field-response}{a}, we show data for $B_\parallel \in \qtyset{0; 100; 200; 250; 300}{\milli\tesla}$.
    Unlike in a plot directly comparing the original and the modified CPR, the difference between the two functions is visible here, as the resonance frequency depends on the CPR derivative and is thus more sensitive to small changes.
    The fit quality, however, remains largely unchanged, with the solid line matching the data better in some places and slightly worse in others, the largest differences being located near the edge of the flux arc data, close to the divergence of the polynomial offset.
    \sublabel{d}~Kerr anharmonicity~$\mathcal K$ with and without the polynomial correction.
    As before, dashed, gray lines correspond to the value derived from~$I(\delta_{\mathrm c})$ while solid, colored lines are from $I(\delta_{\mathrm c}) + \Delta I(\delta_{\mathrm c})$, data in the background are the same as presented in Fig.~\subref{fig:kerr-response}{d}, comprising $B_\parallel \in \qtyset{0; 125; 175; 225; 250; 275; 300}{\milli\tesla}$.
    Since~$\mathcal K$ depends on still higher CPR derivatives than~$\omega_0$, it is modified much more strongly by~$\Delta I$, resulting in deviations of up to one order of magnitude.
    These modifications greatly improve the agreement between the theory curve and the measurement data, demonstrating that even minuscule uncertainty in the CPR is sufficient to explain the mismatch in the Kerr data, though all qualitative features of the anharmonicity response are present even before the modification.
  }
  \label{fig:suppl-poly-correction}
\end{figure}

As described in Appendix~\ref{sec:meth-kerrfit}, we modify the CPR using a small polynomial correction~$\Delta I(\delta_{\mathrm c})$ for the fit of the Kerr anharmonicity.
In Supplemental Fig.~\ref{fig:suppl-poly-correction}, we detail the magnitude of this correction and its effect on the CPR as well as the resonance frequency and Kerr response predicted by it.
While the correction is several orders of magnitude smaller than the constriction critical current -- and thus imperceptibly small when considering the CPR directly --, the CPR derivative, and thus the resulting resonance-frequency flux-response, is visibly affected.
The effect is still small, though, and does not significantly impact the agreement with the measurement data.
Higher CPR derivatives are affected even more strongly, and thus the effect of the correction on the Kerr anharmonicity is considerable.
We can greatly improve the match between modeled and observed $\mathcal K$~response using the correction, demonstrating that the present mismatch can be explained by minuscule disparities in the~CPR.
Such discrepancies might have a multitude of causes, like the gradients described by $\epsilon$ and~$b$ having more complicated shapes than the simple linear dependence we used in the CPR model (cf.~Eqs.~\eqref{eq:j0-gradient} and~\eqref{eq:leff-gradient}), additional gradients in other directions or parameters, non-trivial frequency dependencies of the pump attenuation~$\zeta(\omega_{\mathrm p})$ or SQUID asymmetries due to small differences between the two constrictions.
Not all of these can be addressed by the small correction~$\Delta I(\delta_{\mathrm c})$, so some discrepancy remains, but overall the agreement of the modified CPR with the experimental data is very good.
All qualitative features of the asymmetric Kerr response are already present in the un-modified CPR, affirming the fundamental suitability of our CPR model.

We note that some numerical inaccuracies are visible as wiggles at the edge of the un-corrected $\mathcal K$ curves (dashed gray lines) in Fig.~\subref{fig:suppl-poly-correction}{d} for the highest fields, as discussed in Appendix~\ref{sec:meth-cpr-implementation}.
Naturally, these inaccuracies are still present with the polynomial correction, but due to the logarithmic scale of the plot axis and the much larger~$\abs{\mathcal K}$ values they are not visible in the corresponding curve.

\section{Flux-tuning, current-phase relation and Kerr data of sample \res 4}\label{sec:suppl-res4-data}
Besides the resonator~\res 3 discussed in the main manuscript, we also characterized the other two SQUID resonators \res 2 and~\res 4.
Notably, the orientation of~\res 4 on the chip is inverted with respect to that of~\res 3, cf.~Supplemental Fig.~\ref{fig:suppl-chip-layout}, so any in-plane field couples flux into the constrictions, relative to the SQUID loop, with opposite polarity.
Supplemental Fig.~\ref{fig:suppl-res4-field-response} shows the field-dependent flux-response of~\res 4 and indeed the skewing occurs in the opposite direction compared to to~\res 3, cf.~Fig.~\ref{fig:field-response}.
Since the junctions in~\res 4 are implemented as \nD 3~constrictions, i.e.~they have been thinned down from the top to~\qty{\sim 75}{\nano\meter} (estimated from the NIM doses), a weaker response to~$B_\parallel$ is expected, and indeed this is what we find.
All qualitative features discussed before are still present, though, albeit with inverted flux orientation.
We note that the other \nD 3~constriction resonator~\res 2 shows a similar diode effect but with the arcs skewing in the same direction as those of~\res 3, confirming that the inverted skewing is indeed due to the orientation of the resonator with respect to the field rather than some effect related to the \nD 3~constrictions.
The field response of the sweetspot resonance frequency and linewidth of~\res 4, on the other hand, is very similar to that of~\res 3, with the sweetspot resonance frequency being reduced by~\qty{\sim 10}{\mega\hertz} and the linewidth increasing by~\qty{\sim 40}{\mega\hertz}, though its absolute value is~\qty{\sim 20}{\mega\hertz} larger, probably due to a lower constriction~$T_{\mathrm c}$~\cite{uhl2024}.
The flux responsivity, on the other hand, is considerably higher than for the \nD 2~constriction resonator, a typical result for \nD 3~constrictions~\cite{uhl2024} due to their reduced critical current, making them better suited for sensing and parametric coupling applications.
Here, the responsivity is larger by a factor of~\num{\sim 2} both with and without applied magnetic field, i.e.~the larger responsivity is maintained even at large fields.

\begin{figure}
  \includegraphics{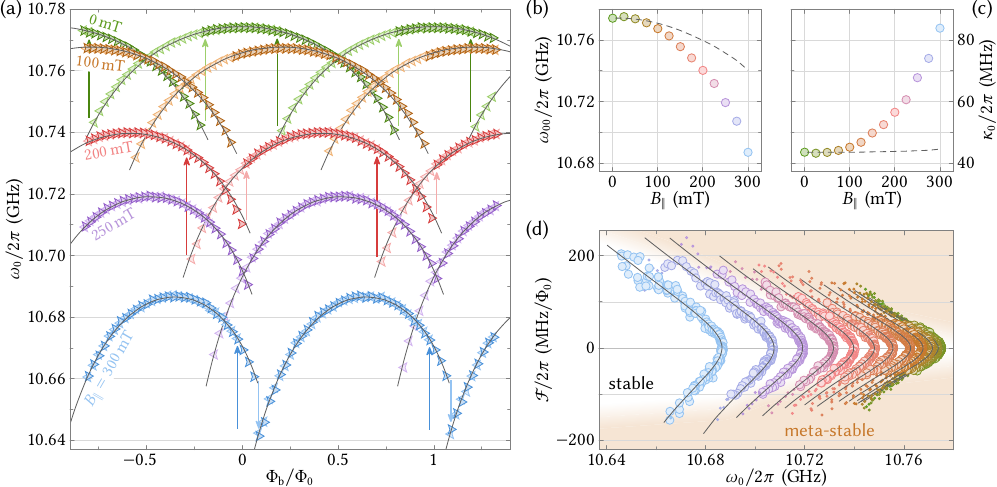}
  \titlecaption{Skewed flux-response and enhanced flux-responsivity in large magnetic in-plane fields for resonator~\res 4}{%
    All data correspond to those shown in Fig.~\ref{fig:field-response} for resonator~\res 3.
    \sublabel{a}~Bias flux response~$\omega_0(\Phi_{\mathrm b})$ of the circuit for five different magnetic in-plane fields between \qty{0}{\milli\tesla} and~\qty{300}{\milli\tesla}; labels next to the datasets denote~$B_\parallel$.
    Data combine flux up-sweep (dark kites pointing right) and flux down-sweep (light kites pointing left); lines are fits.
    Note that the flux arcs skew in the opposite direction compared to those of resonator~\res 3 presented in Fig.~\subref{fig:field-response}{a}, with all other features of the field response present here as well.
    \sublabel{b},~\sublabel{c}~Sweetspot resonance frequency~$\omega_{00}$ and linewidth~$\kappa_0$ as a function of~$B_\parallel$.
    Symbols are data extracted from the flux arcs, colors denote the value of~$B_\parallel$, as in all panels, and the dashed lines show the expected behavior in the absence of the constrictions as inferred from reference circuits.
    Just like for resonator~\res 3, about two thirds of the $\omega_{00}$~decrease and almost all of the $\kappa_0$~increase can be attributed to the constrictions.
    \sublabel{d}~Flux responsivity $\mathcal F = \idell{\omega_0}{\Phi_{\mathrm b}}$ for various~$B_\parallel$ and plotted vs~$\omega_0$ as a figure of merit for sensing and parametric coupling applications.
    Large symbols correspond to values on stable branches, small symbols to those on metastable branches, lines are derivatives of the arc fits.
    Typical for \nD 3~constrictions, the responsivity here is larger than that of the \nD 2~resonator~\res 3 shown in Fig.~\subref{fig:field-response}{d}, with a factor of~\num{\sim 2} between them both with and without a magnetic field.
  }
  \label{fig:suppl-res4-field-response}
\end{figure}

\begin{figure}[p]
  \includegraphics{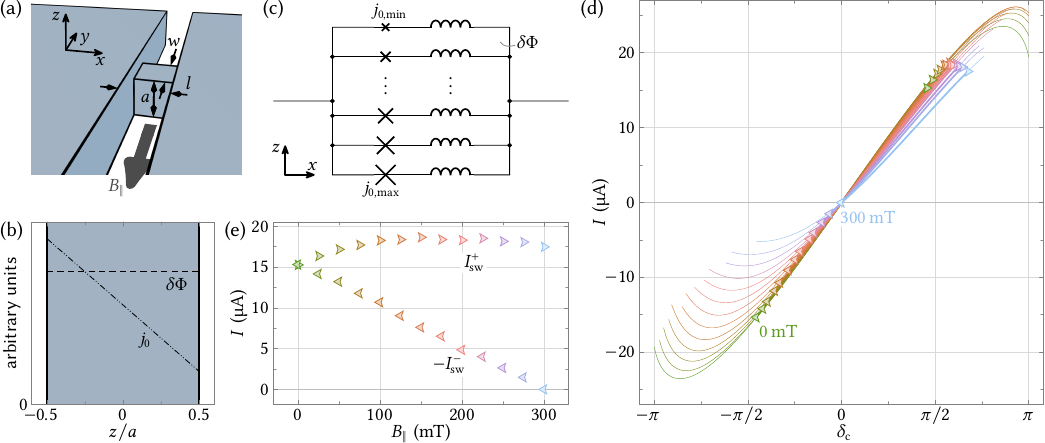}
  \titlecaption{Simplified Josephson-diode model for resonator~\res 4}{%
    All data correspond to those shown in Fig.~\ref{fig:cpr-response} for resonator~\res 3.
    \sublabel{a}~Schematic of a nano-constriction defining its width~$w$, length~$l$ and height~$a$ as well as the direction of~$B_\parallel$.
    Unlike with resonator~$\res 3$, here we have a \nD 3~constriction and the magnetic field enters the junction from the opposite side (when viewed from the inside of the SQUID loop).
    \sublabel{b}~To model the skewed flux response of this resonator in accordance with the measurement data, no gradient in flux per height $\delta\Phi = B_\parallel l_{\mathrm{eff}}$ is necessary, so we only include the $j_0$\nobreakdash-gradient for the analysis here, i.e.~$b = 0$.
    \sublabel{c}~Circuit equivalent of the Josephson-diode model consisting of a multi-loop parallel arrangement of infinitesimal constrictions whose inductive Josephson contribution grows with~$z$ due to a decreasing critical current density.
    In accordance with the constant $\delta\Phi$\nobreakdash-value used, in contrast to the $z$\nobreakdash-gradient illustrated in Fig.~\subref{fig:cpr-response}{c}, here the loop length is independent of~$z$.
    \sublabel{d}~Constriction current-phase relation~$I(\delta_{\mathrm c})$ as a function of the applied field~$B_\parallel$ obtained from fits to the flux arc data (lines in Fig.~\subref{fig:suppl-res4-field-response}{a}).
    Thin lines show the full range for which our numerical algorithm yields a result, thick line segments show the parts that correspond to the experimental flux arcs in between the discontinuous jumps, symbols mark the corresponding switching currents/phases.
    The magnetic-field response is much the same as that for resonator~\res 3 presented in Fig.~\subref{fig:cpr-response}{d} but with inverted polarity.
    Here, the positive critical currents decrease quickly while the negative ones mostly retain their magnitude.
    As before as well as in the next panel, colors encode~$B_\parallel$.
    \sublabel{e}~Switching currents $I_{\mathrm{sw}}^+$ and~$-I_{\mathrm{sw}}^-$ as derived from the flux arc discontinuities and the CPR model.
    For $B_\parallel = 0$, positive and negative switching currents are equal in magnitude, but with increasing~$B_\parallel$ the difference between the two values grows, indicating an increasing diode effect.
    Positive switching currents stay larger than the initial value for all $B_\parallel > 0$, negative switching currents decrease in magnitude nearly linearly.
    For the largest field, the latter value reaches zero, so just as for resonator~\res 3, the current flows in the same direction in the SQUID ring for any applied bias flux~$\Phi_{\mathrm b}$, though this point is reached at slightly higher fields here.
  }
  \label{fig:suppl-res4-cpr-response}
\end{figure}

\begin{figure}[p]
  \includegraphics{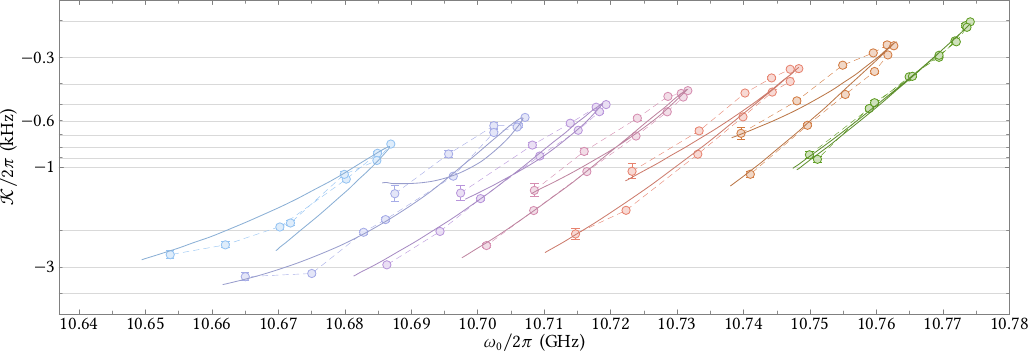}
  \titlecaption{Bimodel Kerr anharmonicity for resonator~\res 4}{%
    The data correspond to those shown in Fig.~\subref{fig:kerr-response}{d} for resonator~\res 3.
    Again, the in-plane fields shown are $B_\parallel \in \qtyset{0; 125; 175; 225; 250; 275; 300}{\milli\tesla}$, indicated by color as before, symbols are data, connected with dashed lines as guides to the eye, solid lines are fit curves based on the CPRs shown in Supplemental Fig.~\ref{fig:suppl-res4-cpr-response} plus a small polynomial correction.
    The bimodality is much weaker here than for resonator~\res 3, but still present with all the same qualitative features, confirming the diode effect for this resonator as well.
  }
  \label{fig:suppl-res4-kerr-response}
\end{figure}

For \res 3, we used a $z$\nobreakdash-gradient of both $j_0$ and~$B_\parallel l_{\mathrm{eff}}$ (i.e.~the fit parameters $\epsilon$ and~$b$, cf.~Eqs.~\eqref{eq:j0-gradient} and~\eqref{eq:leff-gradient}) to fit the flux arcs.
Indeed, both of them were necessary to achieve good agreement with the data, though the effect of~$b$ is much weaker than that of~$\epsilon$ for the parameters found.
Here, we are able to fit the flux arcs without the gradient in~$B_\parallel l_{\mathrm{eff}}$, i.e.~we use $b = 0$.
This matches the smaller constriction height, as we can expect less non-trivial parameter variation to occur in the smaller space.
Supplemental Fig.~\subref{fig:suppl-res4-cpr-response}{a} through~\subref{fig:suppl-res4-cpr-response}*{c} illustrate this slightly simplified model behavior and Supplemental Fig.~\subref{fig:suppl-res4-cpr-response}{d} and~\subref{fig:suppl-res4-cpr-response}*{e} show the CPRs and switching currents extracted from the flux-arc fits.
As expected, just like the flux arcs, the CPRs deform in the opposite direction and the reactions of the switching currents $I_{\mathrm{sw}}^\pm$ to~$B_\parallel$ are swapped with respect to those of~\res 3 presented in Fig.~\ref{fig:cpr-response}.
The fits yield a field $B_0 = \qty{361}{\milli\tesla}$ for coupling one flux quantum into the junction, corresponding to an effective area $A_{\mathrm{eff}} = \qty{5.7e-15}{\square\meter}$ and an effective constriction length $l_{\mathrm{eff}}(0) = \qty{76}{\nano\meter}$, slightly larger than that of the \res 3~constrictions, which is consistent with a suppressed~$T_{\mathrm c}$ and a correspondingly increased~$\lambda_{\mathrm L}$ in the \nD 3~constrictions.
See Supplemental Note~\ref{sec:suppl-fit-parameters} for a discussion of all parameters resulting from the fits.

Moving on to the Kerr anharmonicity, we again find a bimodal distribution for non-zero in-plane fields that is shown in Supplemental Fig.~\ref{fig:suppl-res4-kerr-response}, though it is significantly weaker than that of~\res 3 presented in Fig.~\subref{fig:kerr-response}{d}.
For small fields, the short leg of the fit curve is at a lower value of~$\abs{\mathcal K}$ than the long leg, while this is inverted at the highest field.
The transition occurs at around~\qty{250}{\milli\tesla}.
In the two-tone measurement data, inversion seems to be happening right at~\qty{300}{\milli\tesla}, making the bimodality disappear at that particular field.
The same transition was also present in Fig.~\subref{fig:kerr-response}{d}, but at lower fields (\qty{\sim 125}{\milli\tesla} for the fit curves and \qty{\sim 250}{\milli\tesla} for the two-tone data), making its appearance less pronounced.
Quantitatively, the Kerr anharmonicity of the \nD 3~constriction resonator is again larger than that of~\res 3 by a factor of~\num{\sim 2} due to the lower critical currents, similarly to the increased flux responsivity discussed above.

\clearpage

\vspace*{0pt plus 0fil}

\section{Reversing the in-plane field}\label{sec:suppl-reversed-inpl-field}
Let us compare this inversion of the circuit flux response for resonators with opposite orientation on the chip to the change of flux response of a resonator upon inverting the in-plane field.
In Supplemental Fig.~\ref{fig:suppl-reversed-inpl-field} we present resonance frequency flux arcs for the same two circuits \res 3 and~\res 4 at $\abs{B_\parallel} = \qty{250}{\milli\tesla}$ for both positive and negative field orientation.
Just like for the inverted resonator orientation and just as predicted by our constriction model, reversing the field orientation flips the skewing and shifting direction of the flux arcs; the arcs of~\res 3 at positive~$B_\parallel$ skew in the same way as the arcs of~\res 4 at negative~$B_\parallel$.
This result re-affirms that the diode-effect causing the distorted CPR (and thus the skewed bias-flux response) is indeed induced by the magnetic field.

\begin{figure}[t]
  \includegraphics{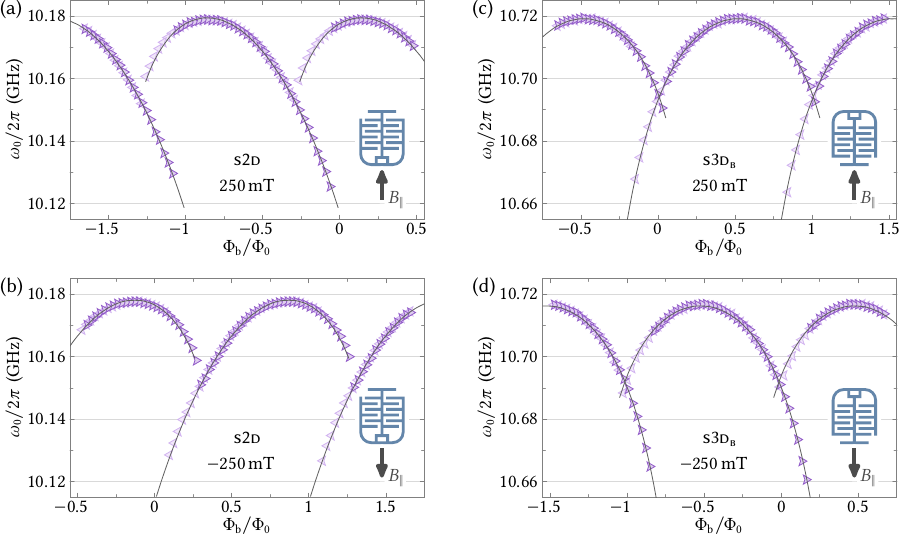}
  \titlecaption{Reversing the in-plane field reverses the flux arc skewing}{%
    Bias flux response~$\omega_0(\Phi_{\mathrm b})$ of resonators~\res 3 and~\res 4 at $B_\parallel = \qty{\pm 250}{\milli\tesla}$.
    The resonator as well as the $B_\parallel$ values are indicated in the panels, a small sketch additionally illustrates the orientation of resonator and field.
    Kite symbols are measurement data with the pointing direction indicating the flux-sweep direction, lines are fits.
    For the fits, the model parameters $j_0(0)$, $\epsilon$, $l_{\mathrm{eff}}(0)$ and~$b$ are identical for both field orientations, only $\ell_{\mathrm{lin}}$ being allowed to differ.
    Both in the data and in the fits, reversing the in-plane field also reverses the skewing of the flux arcs as well as the sweetspot-shift from $\Phi_{\mathrm b} = 0$.
  }
  \label{fig:suppl-reversed-inpl-field}
\end{figure}

The discerning reader may notice that the sweetspot frequencies of the two measurements for each resonator are not exactly identical, but differ by~\qty{\sim 2}{\mega\hertz}.
This is due to a small drift present in our setup, shifting all frequencies to slightly lower values over the course of the measurements (ca.~one week).
We attribute this to the pressure in the vacuum compartment slowly increasing, raising the cooling power reaching the sample assembly, which in turn causes the temperature controller to provide a higher heating power in order to keep the temperature sensor at the temperature setpoint $T_{\mathrm s} = \qty{2.8}{\kelvin}$ (cf.~Supplemental Fig.~\ref{fig:suppl-setup}).
Since the sample diode, although directly attached to the copper housing surrounding the sample, will never have exactly the same temperature as the resonator circuits, this can cause an upwards creep of the circuit temperature, increasing its kinetic inductance and thus lowering its resonance frequency.
This effect can also be observed in the sweetspot frequencies of the compensation angle measurements presented in Fig.~\subref{fig:meth-field-alignment}{b}; those were taken at a much lower pressure (before settling on the final measurement conditions) and accordingly show a noticeably higher resonance frequency at the optimum compensation angle.
Here, we account for this drift in the fit curves by allowing~$\ell_{\mathrm{lin}}$ to vary between fits for positive and for negative field values, with small shifts in that inductance (here~\qty{\lesssim 3}{\percent}) effectively shifting the resulting flux arcs without changing their shape.
All other model parameters $j_0(0)$, $\epsilon$, $l_{\mathrm{eff}}(0)$ and~$b$ are left unchanged for the negative-field fits, ensuring that the observed reversal of the skewing direction can indeed be attributed to the field reversal and not to changing parameter values.

\clearpage

\section{Model parameters resulting from the fits}\label{sec:suppl-fit-parameters}
As described in Appendix~\ref{sec:meth-arcfit}, we obtain the model parameters $I_{00}$, $\epsilon$, $b$, $\delta_B = 2 \pi B_\parallel / B_0$ and~$\delta_\ell(B_\parallel)$ from the flux arc fits.
From this we can directly calculate the effective junction area $A_{\mathrm{eff}} = \Phi_0 / B_0$ threaded by the magnetic in-plane field.
While the inductive phase~$\delta_\ell(B_\parallel)$ is field dependent, at zero field we can calculate the equivalent linear inductance $L_{\mathrm{lin}} = \Phi_0 \delta_\ell(0) / 2 \pi I_{00}$ of a simple constriction without gradients, as discussed in the context of Fig.~\ref{fig:flux-response}.
Given the constriction dimensions $a$ and~$w$, we can furthermore calculate the effective junction length $l_{\mathrm{eff}}(0) = A_{\mathrm{eff}} / a$ as well as the critical current density in the center of the junction $j_0(0) = I_{00} / w a$.
All of these parameters are displayed in Supplemental Table~\ref{tab:suppl-fit-parameters}, except for the field-dependent~$\delta_\ell$ which is shown in Supplemental Fig.~\ref{fig:suppl-fit-inductance}.
As discussed in Supplemental Note~\ref{sec:suppl-reversed-inpl-field}, a resonator that is rotated on the chip by~\ang{180} exhibits the same behavior when applying a positive field as one that is not rotated when applying a negative field, i.e.~the fit results in a negative~$\delta_B$ when not accounting for the rotation elsewhere in the formalism.
We account for the rotation after the fact by removing the sign when calculating $B_0 = 2 \pi \abs{B_\parallel / \delta_B}$ for Supplemental Table~\ref{tab:suppl-fit-parameters}, leaving the fit scripts rotation-agnostic.

\begin{table}[b]
  \titlecaption{Model parameters resulting from the fits and derived quantities}{
    We show the CPR model parameters $I_{00}$, $\epsilon$, $b$ and $B_0 = 2 \pi \abs{B_\parallel / \delta_B}$ extracted from the flux arc fits as described in Appendix~\ref{sec:meth-arcfit}, as well as the effective constriction area~$A_{\mathrm{eff}}$ implied by these results.
    The field dependent parameter~$\delta_\ell$ is depicted in plot form in Supplemental Fig.~\ref{fig:suppl-fit-inductance}, here we only show the equivalent linear inductance at zero field~$L_{\mathrm{lin}}$.
    Furthermore, we state the constriction dimensions $a$~(estimated from the NIM beam dose~$D_2$, cf.~Supplemental Table~\ref{tab:suppl-resonator-parameters}), $w$ and~$l$ as well as the effective constriction length~$l_{\mathrm{eff}}$ and the critical current density in the center of the junction~$j_0(0)$ resulting from these together with the fit results.
    Finally, we include the parameters $\omega_{\mathrm{ref}}$, $\zeta_0$ and~$\zeta_1$ obtained from the Kerr fits as described in Appendix~\ref{sec:meth-kerrfit} that define the pump-line damping behavior~$\zeta(\omega_{\mathrm p})$.
  }
  \label{tab:suppl-fit-parameters}
  \vskip -.4\belowcaptionskip
  \setlength\tabcolsep{3pt}
  \def\extraspace{\hspace{.8em}}
  \begin{tabular}[b]{l !{\extraspace}
                     S[table-format=2] S[table-format=1.2] S[table-format=1] S[table-format=3] !{\extraspace}
                     S[table-format=1.1e-2] S[table-format=2] !{\extraspace}
                     S[table-format=3] *{2}{S[table-format=2]} !{\extraspace}
                     S[table-format=2] S[table-format=1.1e1] !{\extraspace}
                     S[table-format=2.2] S[table-format=1.1e-1] S[table-format=-1.1e-1]} \toprule
    {} & {$I_{00}$} & {$\epsilon$} & {$b$} & {$B_0$} & {$A_{\mathrm{eff}}$} & {$L_{\mathrm{lin}}$} & {$a$} & {$w$} & {$l$} & {$l_{\mathrm{eff}}(0)$} & {$j_0(0)$} & {$\omega_{\mathrm{ref}} / 2 \pi$} & {$\zeta_0$} & {$2 \pi \zeta_1$} \\
    {} & {in \unit{\micro\ampere}} & {} & {} & {in \unit{\milli\tesla}} & {in \unit{\square\meter}} & {in \unit{\pico\henry}} & {in \unit{\nano\meter}} & {in \unit{\nano\meter}} & {in \unit{\nano\meter}} & {in \unit{\nano\meter}} & {in \unit{\ampere\per\square\meter}} & {in \unit{\giga\hertz}} & {} & {in \unit{\per\mega\hertz}} \\\midrule
    \res 3 & 35 & 0.78 & 1 & 305 & 6.8e-15 & 12 & 100 & 40 & 40 & 68 & 8.7e+09 & 10.25 & 3.9e-05 & -7.4e-03 \\
    \res 4 & 26 & 0.82 & 0 & 361 & 5.7e-15 & 15 & 75 & 40 & 40 & 76 & 8.8e+09 & 10.80 & 7e-05 & -1.9e-02 \\\bottomrule
  \end{tabular}
\end{table}

Overall, the values we find are quite sensible and match the typical parameter range for similar devices.
Note again that, unlike in a standard junction experiencing a magnetic field, the presence of the $j_0$\nobreakdash-gradient represented by~$\epsilon$ has the effect that the CPR no longer becomes zero when $B_\parallel = B_0$, cf.~Supplemental Note~\ref{sec:suppl-cpr-parameters}.
Thus, despite the value $B_0 = \qty{305}{\milli\tesla}$ for~\res 3 being barely above the highest measured field of~\qty{300}{\milli\tesla}, we do not expect the critical currents to nearly disappear, and indeed our estimate of~\qtyrange[range-units=bracket]{350}{400}{\milli\tesla} derived from Fig.~\subref{fig:cpr-response}{e} was considerably larger.
The value $b = 1$ characterizing the $B_\parallel l_{\mathrm{eff}}(z)$\nobreakdash-gradient, however, is at the very edge of its permissible parameter range~$[0, 1]$.
While the resulting CPRs fit the data quite well, this indicates that the model is reaching its limits here and more sophisticated dependencies than those introduced in Eqs.~\eqref{eq:j0-gradient} and~\eqref{eq:leff-gradient} are needed for a more complete description of the circuit response, such as a nonlinear~$j_0(z)$ or a $z$-dependent~$\ell_{\mathrm{lin}}$.

\begin{figure}[b]
  \begin{minipage}[b]{.35\textwidth}
    \includegraphics[trim={0pt 10.5pt 0pt -1ex}]{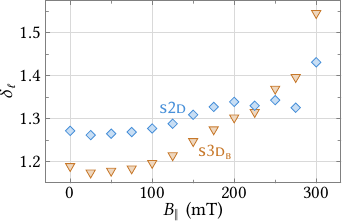}
  \end{minipage}\hfill
  \begin{minipage}[b]{.6\textwidth}
    \titlecaption{Field-dependent specific linear constriction inductance extracted from the flux-response fits}{%
      Symbols are the normalized specific linear constriction inductance~$\delta_\ell(B_\parallel)$ for resonators~\res 3 and~\res 4 resulting from the flux arc fits described in Appendix~\ref{sec:meth-arcfit}.
      The non-normalized value~$\ell_{\mathrm{lin}}$ can be obtained via multiplication with the specific Josephson inductance in the center of the constriction $\ell_{\mathrm J 0} = \Phi_0 / 2 \pi j_0(0)$, which is approximately~\qty{3.8e-14}{\pico\henry\square\meter} for both resonators, cf.~Supplemental Table~\ref{tab:suppl-fit-parameters}.
      \label{fig:suppl-fit-inductance}
    }
  \end{minipage}
\end{figure}

The $B_\parallel$\nobreakdash-dependence of~$\delta_\ell$ also seems reasonable for the most part.
We already discussed the expected increase of kinetic inductance due to the magnetic field that should thus lead to a higher $\delta_\ell$\nobreakdash-value at larger fields, which is indeed what we see.
The absolute values $\delta_\ell > 1$ indicate that the current density in the center of the junction~$j(0, \delta_{\mathrm c})$ is not single-valued (cf.~Supplemental Note~\ref{sec:suppl-cpr-parameters}), similar to $L_{\mathrm{lin}} / L_{\mathrm J 0} > 1$ resulting in an over-hanging CPR for the simpler case discussed in the context of Fig.~\ref{fig:flux-response}.
We can gain a more intuitive comparison with this field-free, gradient-free model by considering the equivalent linear inductance~$L_{\mathrm{lin}}$, finding a larger value for the \nD 3~constrictions, which makes intuitive sense.
Reassuringly, the value of~\qty{12}{\pico\henry} we find for~\res 3 matches the value we found using the simpler fit.
Note that the zero-field value of $\delta_\ell \propto L_{\mathrm{lin}} I_{00}$ itself is smaller for the \nD 3~constrictions than for the \nD 2 variants, corresponding to a less skewed CPR.
While some features of~$\delta_\ell(B_\parallel)$, like the large jump at $B_\parallel = \qty{300}{\milli\tesla}$ or the plateau in the \res 3\nobreakdash-values right before it, offer no intuitive physical explanation and might simply be artifacts of model insufficiencies, the larger value at $B_\parallel = 0$ can be explained by the small value drift discussed in Supplemental Note~\ref{sec:suppl-reversed-inpl-field}:
While the other measurements were taken in order -- larger fields were measured later during the measurement period -- the zero-field value was measured last, so the difference between it and the next data point represents the accumulated value drift during the entire measurement process, demonstrating its smallness.

For completeness, Supplemental Table~\ref{tab:suppl-fit-parameters} also lists the parameters $\omega_{\mathrm{ref}}$, $\zeta_0$ and~$\zeta_1$ resulting from the Kerr anharmonicity fits described in Appendix~\ref{sec:meth-kerrfit}.

\end{supplsetup}


\begin{thebibliography}{90}
\bibitem [{met()}]{metastable-transition-note}%
  \BibitemOpen
  \href@noop {} {}\bibinfo {note} {We assume the transition to the metastable
  state on a flux arc branch to be located in the center between the two jumps
  (arrows in Fig.~\subref{fig:field-response}{a}) to and from that branch,
  which at $B_\parallel = 0$ corresponds to the arc crossing point but shifts
  right of it for larger fields. While we believe that this is the most
  reasonable value to use, our conclusions would not be changed much if the
  transition were in fact located somewhere else between the
  jumps.}\BibitemShut {Stop}%
\end{thebibliography}
\end{document}